\documentclass[twocolumn]{aastex7}

\usepackage{graphicx}
\usepackage{amsmath}


\shorttitle{The 2175\AA\ Bump}
\shortauthors{Guo et al.}

\submitjournal{ApJ}

\begin{document}

\title{Mapping Dust Attenuation at Kiloparsec Scales. III. The 2175\AA\ Bump}

\correspondingauthor{Cheng Li}
\email{cli2015@tsinghua.edu.cn}

\author[0009-0008-4962-665X]{Ruonan Guo}
\affiliation{Department of Astronomy, Tsinghua University, Beijing 100084, China}
\email{grn19@mails.tsinghua.edu.cn}

\author[0000-0002-8711-8970]{Cheng Li}
\affiliation{Department of Astronomy, Tsinghua University, Beijing 100084, China}
\email{cli2015@tsinghua.edu.cn}

\author[0009-0004-6271-4321]{Tao Jing}
\affiliation{Department of Astronomy, Tsinghua University, Beijing 100084, China}
\email{jingt20@mails.tsinghua.edu.cn}

\author[0000-0002-8999-6814]{Shuang Zhou}
\affiliation{INAF–Osservatorio Astronomico di Brera, via Brera, 28, 20159 Milano, Italy}
\email{shuang.zhou@inaf.it}

\author[0000-0002-0656-075X]{Niu Li}
\affiliation{National Astronomical Observatories, Chinese Academy of Sciences, Beijing 100101, China}
\affiliation{Department of Astronomy, Tsinghua University, Beijing 100084, China}
\email{liniu126@126.com}

\author[0009-0001-5437-410X]{Zhuo Cheng}
\affiliation{Department of Physics, The Chinese University of Hong Kong, Sha Tin, NT, Hong Kong, China}
\affiliation{Department of Astronomy, Tsinghua University, Beijing 100084, China}
\email{zhuocheng@cuhk.edu.hk}

\begin{abstract}
We combine the Swift/UVOT+MaNGA catalog with 2MASS $K_s$ imaging to map the 2175\AA\ attenuation bump at kiloparsec scales in nearby galaxies. We use two complementary estimators: an ultraviolet-to-near-infrared attenuation-curve method, yielding $A_{bump}^{UOIR}$ and $B$ for 2487 high-continuum-S/N spaxels, and the NUV-only method of \citet{2025PASA...42...22B}, yielding $A_{bump}^{NUV}$ and $k_{bump}$ for 7934 spaxels. The two absolute bump estimates agree well where they overlap. We compare bump strength with local stellar-population, emission-line, attenuation-curve, and geometric diagnostics after separating star-forming (SF) and non-SF regions. The strongest bumps occur at low specific H$\alpha$ surface brightness, $\Sigma_{\text{H}\alpha}/\Sigma_\ast$, especially in non-SF regions, where this ratio traces ionized-gas emission per unit stellar mass rather than sSFR. The bump also weakens with EW(H$\alpha$) and strengthens with $D_n4000$ and stellar age. In contrast, metallicity, inclination, galactocentric radius, $A_V$, and optical attenuation-curve slope are secondary predictors. The absolute strength $A_{bump}^{NUV}$ increases with $\Sigma_{\text{H}\alpha}$ and $\Sigma_\ast$, while the relative strengths $k_{bump}$ and $B$ do not, indicating that absolute bump amplitude partly follows dust column whereas normalized strengths better trace effective bump prominence. These results support local radiation-field processing of the 2175\AA\ carriers.
\end{abstract}


\section{Introduction}\label{sec:introduction}

The broad absorption feature near 2175\AA\ is one of the most prominent ultraviolet signatures of interstellar dust. It is strong in the average Milky Way extinction curve, weaker or more variable in the Large Magellanic Cloud, and weak or absent along many Small Magellanic Cloud sightlines; recent Local Group surveys have expanded the SMC sightline sample and extended ultraviolet extinction-curve measurements to M31 and M33, reinforcing the diversity of UV extinction curves in nearby galaxies \citep{1986ApJ...307..286F,1989ApJ...345..245C,2003ApJ...594..279G,2024ApJ...970...51G,2025ApJ...989...61C,2026ApJ..1001..221G}. In contrast, the empirical attenuation curve of local starburst galaxies is nearly featureless around 2175\AA\ \citep{1994ApJ...429..582C,2000ApJ...533..682C}. Because attenuation curves also depend on dust-star geometry, scattering, and stellar-population mixing, a weak bump does not necessarily imply absent carrier grains \citep[e.g.][]{1997ApJ...487..625G,2000ApJ...528..799W,2007MNRAS.375..640P,2016ApJ...833..201S,2020ARA&A..58..529S}.

Integrated-galaxy studies have found 2175\AA\ bumps with strengths that vary with inclination, spectral type, stellar mass, and star-formation activity \citep[e.g.][]{2010ApJ...718..184C,2011MNRAS.417.1760W,2017ApJ...851...90B,2018ApJ...859...11S}. The feature has also been detected or constrained at intermediate and high redshift, including LEGA-C galaxies at $z\sim0.8$ and JWST/JADES galaxies in the first billion years \citep[e.g.][]{2013ApJ...775L..16K,2020ApJ...903..146B,2021ApJ...909..213K,2022MNRAS.514.1886S,2023Natur.621..267W,2025arXiv250221119O}. Resolved work with Swift/UVOT and infrared data now makes it possible to connect the bump to local dust and PAH-sensitive emission within nearby galaxies \citep{2019MNRAS.486..743D,2023ApJ...953...54B,2025PASA...42...22B}. These measurements are especially useful because integrated attenuation curves average over regions with different radiation fields, dust columns, and geometries.

The feature is usually associated with small carbonaceous grains, including graphitic grains and PAH-like aromatic material \citep[e.g.][]{2003ARA&A..41..241D,2001ApJ...554..778L,2023MNRAS.525.2380L,2025A&A...694A..84L}. The abundance and size distribution of these grains can evolve through growth, shattering, coagulation, and destruction by intense radiation fields or shocks \citep[e.g.][]{2020MNRAS.492.3779H}. Radiative-transfer effects can also dilute the apparent bump even when the underlying extinction curve contains the feature \citep{2000ApJ...528..799W,2011A&A...533A.117F}. The key observational task is therefore to identify which local conditions best predict the effective bump strength.

This is the third paper in a series on kiloparsec-scale dust attenuation in nearby galaxies using Swift/UVOT NUV imaging, MaNGA spectroscopy, and 2MASS near-infrared imaging. Paper I \citep{2023ApJ...957...75Z} developed the resolved attenuation-curve method and found that the 2175\AA\ bump weakens toward higher specific star formation rate. Paper II \citep{2025arXiv250314028G} extended the analysis to SwiM\_v4.2 and focused on opacity and attenuation-curve slopes. Here we return to the bump, asking whether the Paper I trend persists in the larger sample, whether it differs between star-forming (SF) and non-SF regions, and whether absolute and normalized bump strengths point to the same physical drivers.

We use two complementary bump estimators. The first derives the ultraviolet-to-near-infrared attenuation curve and measures both the absolute excess attenuation around the {\tt uvm2} band, $A_{bump}^{UOIR}$, and the relative bump strength $B$. This method is physically direct but requires high-quality optical continua and significant attenuation. The second uses only the three UVOT NUV bands and the nebular color excess, following \citet{2025PASA...42...22B}, and yields $A_{bump}^{NUV}$ and $k_{bump}$ for a larger sample. Comparing the two methods where they overlap lets us exploit the larger NUV-selected sample while keeping the interpretation anchored to the attenuation-curve measurement.

Section~\ref{sec:data} describes the data and samples, Section~\ref{sec:methodology} defines the bump measurements, Section~\ref{sec:results} presents the empirical trends, Section~\ref{sec:discussion} discusses their implications, and Section~\ref{sec:summary} summarizes the conclusions. We assume $\Omega_{m}=0.3$, $\Omega_{\Lambda}=0.7$, and $H_{0}=70~\text{km~s}^{-1}~\text{Mpc}^{-1}$.

\section{Data}\label{sec:data}

\subsection{SwiM\_v4.2, MaNGA and 2MASS}

We use SwiM\_v4.2, the Swift/UVOT+MaNGA (SwiM) Value-added Catalog version described by \citet{2023ApJS..268...63M} and used in Paper II, publicly available from SDSS\footnote{\url{https://www.sdss4.org/dr17/data_access/value-added-catalogs/?vac_id=swift-manga-value-added-catalog}}. The catalog contains 559 MaNGA galaxies with Swift/UVOT imaging in {\tt uvw2} and {\tt uvw1}; 490 also have {\tt uvm2}. The three UVOT bands are centered at 1928\AA, 2246\AA, and 2600\AA, with {\tt uvm2} lying close to the 2175\AA\ feature. SwiM provides matched UVOT images, SDSS {\tt ugriz} images, and MaNGA emission-line and spectral-index maps on the {\tt uvw2} grid, whose PSF has FWHM $2.^{\prime\prime}92$ and pixel size $1^{\prime\prime}$.

MaNGA \citep{2015ApJ...798....7B,2017AJ....154...28B} provides optical integral-field spectroscopy for 10,010 SDSS-IV galaxies over $0.01<z<0.15$ and $5\times10^{8}M_{\odot} \leq M_{\ast} \leq 3\times10^{11}M_{\odot}$ \citep{2017AJ....154...86W}. The datacubes have $0.5^{\prime\prime}$ spaxels, spectral coverage 3622--10354\AA, and effective spatial resolution about $2.5^{\prime\prime}$ \citep{2015AJ....150...19L}. We use the DR17 DRP and DAP products \citep{2016AJ....152...83L,2019AJ....158..231W,2019AJ....158..160B,2022ApJS..259...35A}, resampled and PSF-matched to the UVOT grid by SwiM.

We retrieve 2MASS $K_s$-band atlas images \citep{2006AJ....131.1163S} for the 559 SwiM\_v4.2 galaxies. The $1^{\prime\prime}$ pixels and $2.5^{\prime\prime}$--$3.5^{\prime\prime}$ resolution are comparable to the UVOT data. Following Paper I, which found that additional UVOT--2MASS convolution changes pixel fluxes by less than 1\%, we resample the $K_s$ images to the UVOT grid without an extra resolution correction.

\begin{figure*}
\centering
    \includegraphics[width=0.32\textwidth]{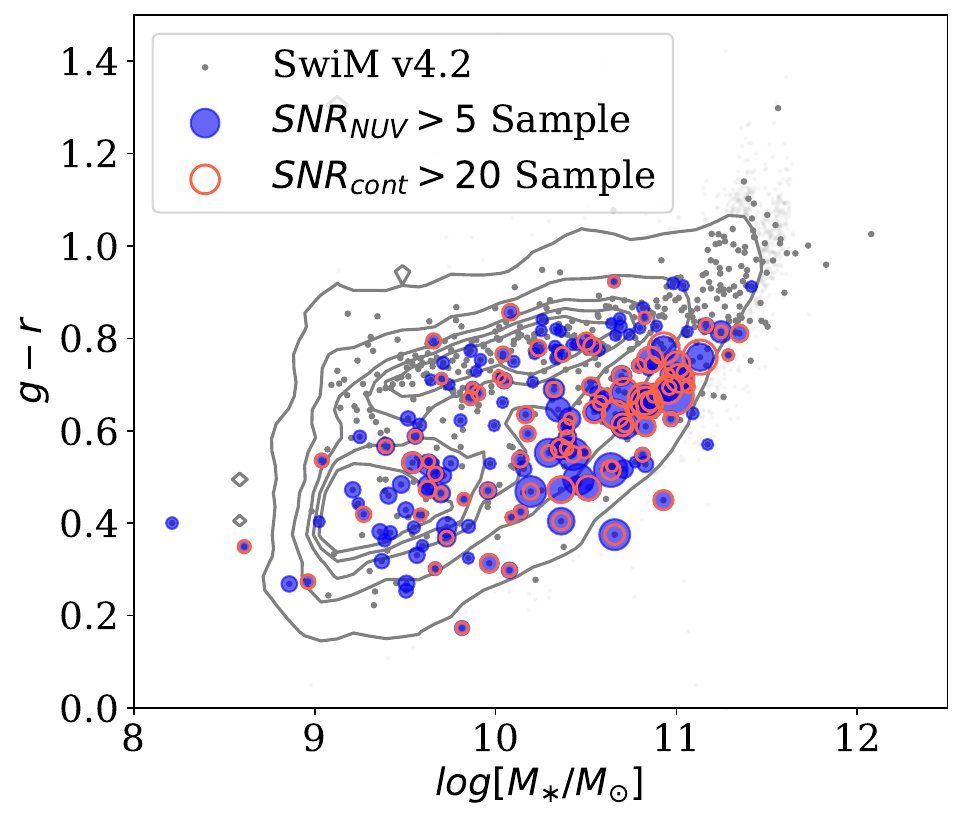}
    \includegraphics[width=0.32\textwidth]{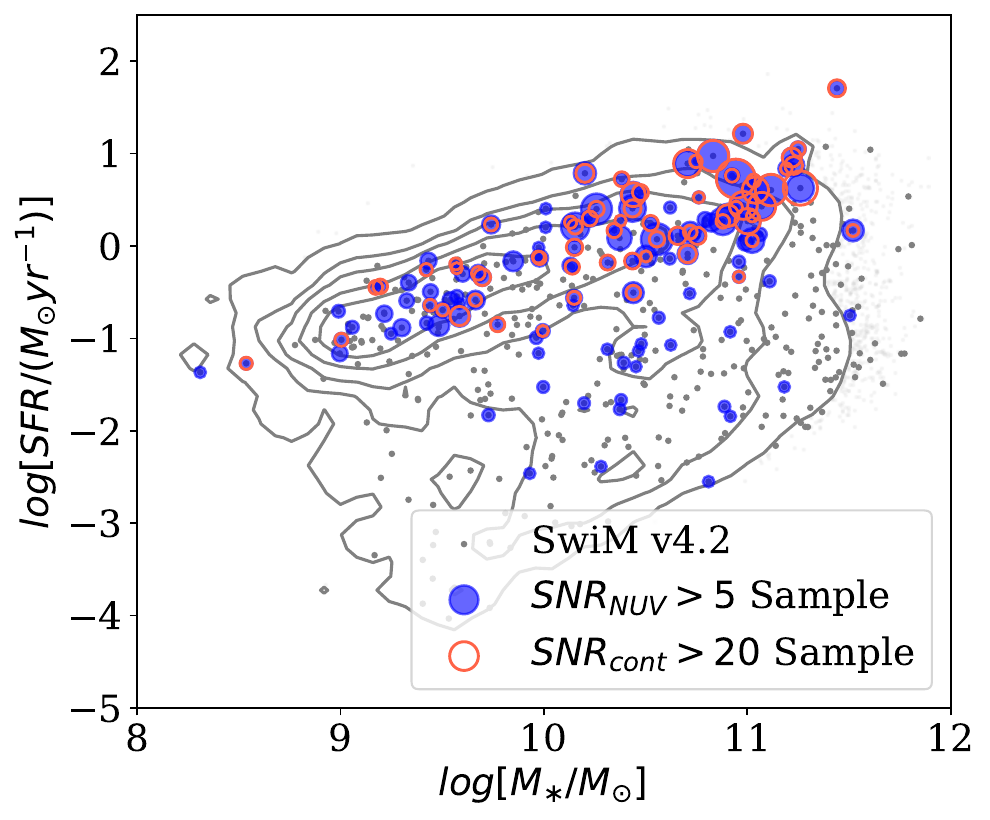}
    \includegraphics[width=0.32\textwidth]{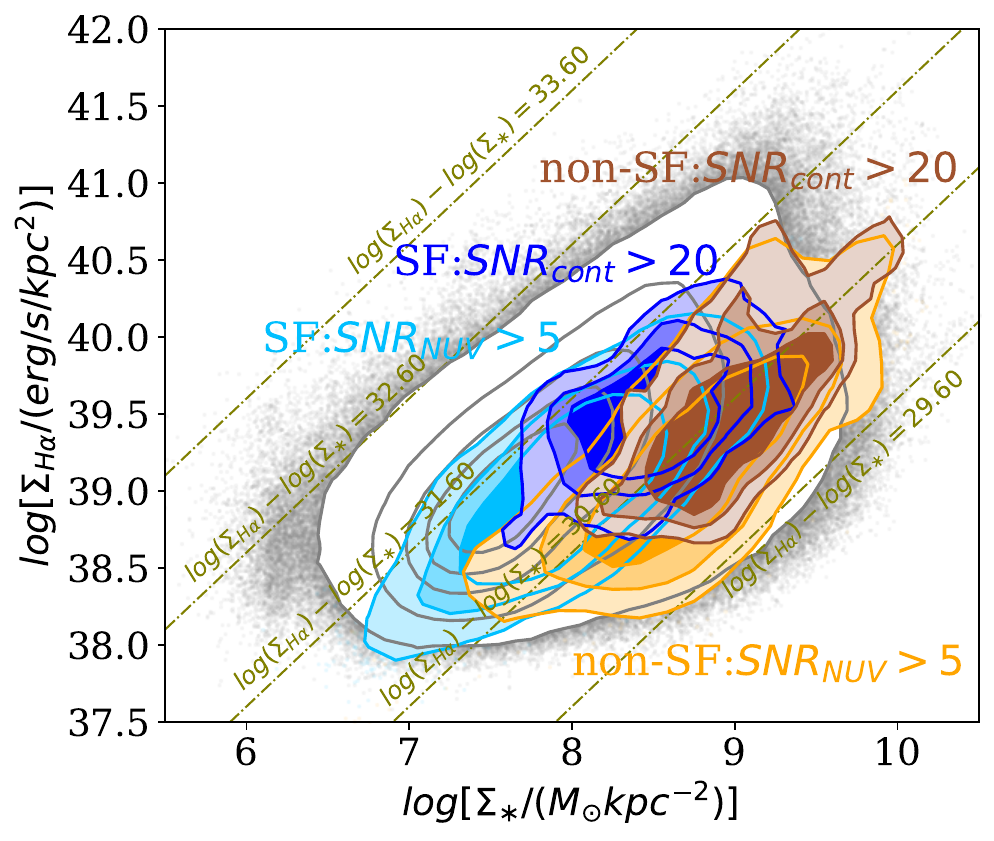}
    \caption{Selected galaxies and spaxels relative to the parent MaNGA/SwiM samples. Left: global $g-r$ color versus stellar mass. Middle: global SFR versus stellar mass. Blue filled circles and red open circles mark galaxies with at least one selected spaxel in the $SNR_{NUV}>5$ and $SNR_{cont}>20$ samples, respectively; symbol size scales with the number of selected spaxels. Right: selected spaxels in the $\Sigma_{\text{H}\alpha}$--$\Sigma_\ast$ plane, with dashed lines of constant $\Sigma_{\text{H}\alpha}/\Sigma_\ast$. Contours distinguish SF and non-SF regions in the two analysis samples; gray contours show the parent MaNGA comparison sample with continuum S/N $>3$.}
\label{fig:globel_and_classify}
\end{figure*}

\subsection{Sample selection and classification}

Because the two bump measurements have different data requirements, we use two analysis samples drawn from the same SwiM\_v4.2 data products as Paper II. For quantities derived from the UOIR attenuation curve, we adopt exactly the Paper II sample selection: continuum S/N $>20$, measured around 5500\AA, and $A_V>0.25$. This sample contains 2487 spaxels in 91 galaxies, and we denote it as the $SNR_{cont}>20$ sample. The continuum S/N used here is distinct from the photometric S/N in the individual UVOT bands. The NUV-only estimator can also be applied to a lower-continuum-S/N subset of the same parent data set. This lower-continuum-S/N subset is newly defined for the present work from the same SwiM\_v4.2 parent catalog. We denote this subset as the $SNR_{NUV}>5$ sample; it requires continuum S/N $>5$, the three UVOT NUV bands, and the H$\alpha$ and H$\beta$ emission-line fluxes, but not the high-continuum-S/N and 2MASS $K_s$ normalization required by the UOIR attenuation-curve measurement. It contains 7934 spaxels. The label $SNR_{NUV}>5$ identifies the lower-continuum-S/N sample used for the NUV-only bump estimator; it does not denote the photometric S/N in the UVOT NUV bands.

The final analysis masks also include conservative quality cuts on the UV attenuation measurements: positive fitted attenuations in the three UVOT bands, fitted NUV continuum slope $\beta>-3$, $\log_{10}\chi^2_{\nu}\leq -3.1$, and a monotonic optical-to-NUV attenuation shape that avoids pathological interpolations. Separately from the continuum-S/N thresholds above, we require photometric S/N $>3$ in each of the three UVOT bands. The $SNR_{NUV}>5$ sample additionally requires $E(B-V)_{gas}>0.01$ so that $k_{bump}$ is well defined. These extra cuts define the sample used in the figures and statistical tests; they do not change the photometric definition of $A_{bump}^{NUV}$ itself.

The $SNR_{cont}>20$ sample is used whenever $A_{bump}^{UOIR}$ or $B$ is required; the $SNR_{NUV}>5$ sample provides better statistical leverage for $A_{bump}^{NUV}$ and $k_{bump}$.

Following Paper II, we classify spaxels into star-forming (SF) and non-star-forming (non-SF) regions using the diagnostic diagram of \citet{2020MNRAS.499.5749J}. We compute the Ji \& Yan $P1$ and $P2$ coordinates from $N2=\log_{10}([{\rm NII}]\lambda6583/{\rm H}\alpha)$, $S2=\log_{10}(([{\rm SII}]\lambda6717+[{\rm SII}]\lambda6731)/{\rm H}\alpha)$, and $R3=\log_{10}([{\rm OIII}]\lambda5007/{\rm H}\beta)$; spaxels on the HII-region side of their $P1$--$P2$ demarcation are labeled SF, and the rest are grouped as non-SF. The $SNR_{cont}>20$ sample contains 1468 SF spaxels and 1019 non-SF spaxels, while the $SNR_{NUV}>5$ sample contains 5800 SF spaxels and 2134 non-SF spaxels.

The SF/non-SF labels are empirical spectral classifications. Non-SF spaxels may include diffuse ionized gas, evolved stellar populations, shocks, weak nuclear activity, or mixtures of these components; $\Sigma_{\text{H}\alpha}/\Sigma_\ast$ is therefore interpreted as an sSFR-like quantity only for SF spaxels.

\autoref{fig:globel_and_classify} places the selected galaxies and spaxels in context. The selected galaxies are biased toward blue, star-forming, and relatively massive systems compared with the parent MaNGA sample, although the $SNR_{NUV}>5$ sample includes more low-mass galaxies than the $SNR_{cont}>20$ sample. In the $\Sigma_{\text{H}\alpha}$--$\Sigma_\ast$ plane, SF and non-SF regions occupy partly distinct but overlapping loci; the larger NUV-selected sample extends mainly toward lower surface densities, not to a substantially different range of $\Sigma_{\text{H}\alpha}/\Sigma_\ast$.

\section{Methodology}\label{sec:methodology}

\subsection{Deriving attenuation curves and local diagnostics}

We use the attenuation-curve method developed in Paper I and the opacity and slope measurements discussed in Paper II. In brief, we derive a relative optical attenuation curve from each MaNGA spectrum \citep{2020ApJ...896...38L}, fit the attenuation-corrected spectrum with BIGS \citep{2019MNRAS.485.5256Z} to obtain a dust-free stellar model, and normalize that model with the observed 2MASS $K_s$ flux. Comparing the normalized model with the MaNGA spectrum and UVOT photometry gives the absolute attenuation curve, including $A_V$, $A_B/A_V$, $A_{w2}/A_{w1}$, and the UOIR bump quantities below.

We measure local stellar and nebular diagnostics from the dust-corrected MaNGA spectra following \citet{2021ApJ...917...72L}. These include $\Sigma_\ast$, luminosity- and mass-weighted ages and metallicities, $D_n4000$ \citep{1999ApJ...527...54B}, EW(H$\delta_A$), the Balmer-decrement color excess $E(B-V)_{gas}$, $\Sigma_{\text{H}\alpha}$, $\Sigma_{\text{H}\alpha}/\Sigma_\ast$, EW(H$\alpha$), oxygen abundance from O3N2 and R23 \citep{2013A&A...559A.114M,2022ApJS..262....3N}, and N2S2, $\log_{10}([{\rm NII}]\lambda6583/[{\rm SII}]\lambda6717)$.

\subsection{UOIR attenuation-curve bump measurement}

Our first bump measurement follows Paper I and uses the ultraviolet-to-optical-to-near-infrared (UOIR) attenuation curve. It provides an absolute bump strength, $A_{bump}^{UOIR}$, and a relative bump strength, $B$, for the smaller $SNR_{cont}>20$ sample. Both estimators below draw a smooth bump-free NUV continuum between the {\tt uvw2} and {\tt uvw1} bands and measure the attenuation excess, or flux deficit, at {\tt uvm2}.

We define $A_{bump}^{UOIR}$ as the excess attenuation near 2175\AA\ relative to a smooth attenuation curve without the bump. Because the UVOT {\tt uvm2} band is centered close to this feature, we use the {\tt uvm2} attenuation, $A_{m2}$, as a practical proxy for $A_{2175}$. We denote the smooth, bump-free attenuation expected at the same wavelength as $A_{m2}^{\prime}$. The absolute bump strength is therefore

\begin{equation}
A_{bump}^{UOIR}\equiv{A}_{2175}-A_{2175}^{\prime}\simeq{A}_{m2}-A_{m2}^{\prime}
\end{equation}

To estimate $A_{m2}^{\prime}$, we approximate the bump-free NUV attenuation curve as a power law, $A(\lambda)\propto\lambda^{-\beta}$. The {\tt uvw2} and {\tt uvw1} bands bracket {\tt uvm2}; we use their attenuations to determine the NUV slope and interpolate the bump-free attenuation at {\tt uvm2}. We then normalize the excess attenuation by $A_{m2}^{\prime}$:

\begin{equation}
B\equiv{A}_{bump}^{UOIR}/A_{m2}^{\prime}
\end{equation}

With this definition, $B=0$ for a Calzetti-like curve without a UV bump \citep{1994ApJ...429..582C,2000ApJ...533..682C}, while $B=0.26$ for the Milky Way CCM curve \citep{1989ApJ...345..245C}. We apply empirical K-corrections derived from mock spectra; Appendix~\ref{sec:appendix_mock_kcorr} shows that the corrected $A_{bump}^{UOIR}$ and $B$ have median residuals close to zero.

\subsection{NUV-only bump measurement}

Our second bump measurement adopts the Swift/UVOT-only estimator of \citet{2025PASA...42...22B}. It uses the three UVOT NUV bands and the Balmer-decrement color excess, but does not require the high-S/N optical continuum or 2MASS normalization used in the UOIR attenuation-curve method; it can therefore be applied to the larger, lower-continuum-S/N $SNR_{NUV}>5$ sample.

Following \citet{2025PASA...42...22B}, we assume that the bump-free UV flux density is locally described by a power law. The {\tt uvw2} and {\tt uvw1} flux densities define the expected bump-free {\tt uvm2} flux density, $f_{m2}^{\prime}$, and the observed {\tt uvm2} flux density is $f_{m2}$. The NUV-only absolute bump strength is then

\begin{equation}
A_{bump}^{NUV}\equiv-2.5\log_{10}(f_{m2}/f_{m2}^{\prime})
\end{equation}

and the normalized bump strength is

\begin{equation}
k_{bump}\equiv{A}_{bump}^{NUV}/E(B-V)_{gas}. 
\end{equation}

The two absolute quantities are directly comparable because both express, in magnitudes, the \texttt{uvm2} flux deficit relative to a locally smooth NUV baseline defined by \texttt{uvw2} and \texttt{uvw1}. They are not algebraically identical, however: the UOIR method constructs this baseline from the absolute attenuation curve after estimating the dust-free stellar spectrum, whereas the NUV-only method constructs it directly from the observed NUV flux densities. Their normalized counterparts are likewise analogous but not interchangeable: $B$ divides $A_{bump}^{UOIR}$ by the bump-free \texttt{uvm2} attenuation $A_{m2}^{\prime}$, whereas $k_{bump}$ divides $A_{bump}^{NUV}$ by the nebular color excess $E(B-V)_{gas}$. We therefore compare $A_{bump}^{NUV}$ and $A_{bump}^{UOIR}$ directly, but assess consistency between $k_{bump}$ and $B$ through their trends rather than equality of their numerical values. The superscripts ``NUV'' and ``UOIR'' label the method, not distinct physical bump components.

We apply the same mock-based K-correction strategy to $A_{bump}^{NUV}$ and $k_{bump}$; unless otherwise stated, all bump strengths below are K-corrected.

\begin{figure}[!tbp]
\centering
        \includegraphics[width=0.9\columnwidth]{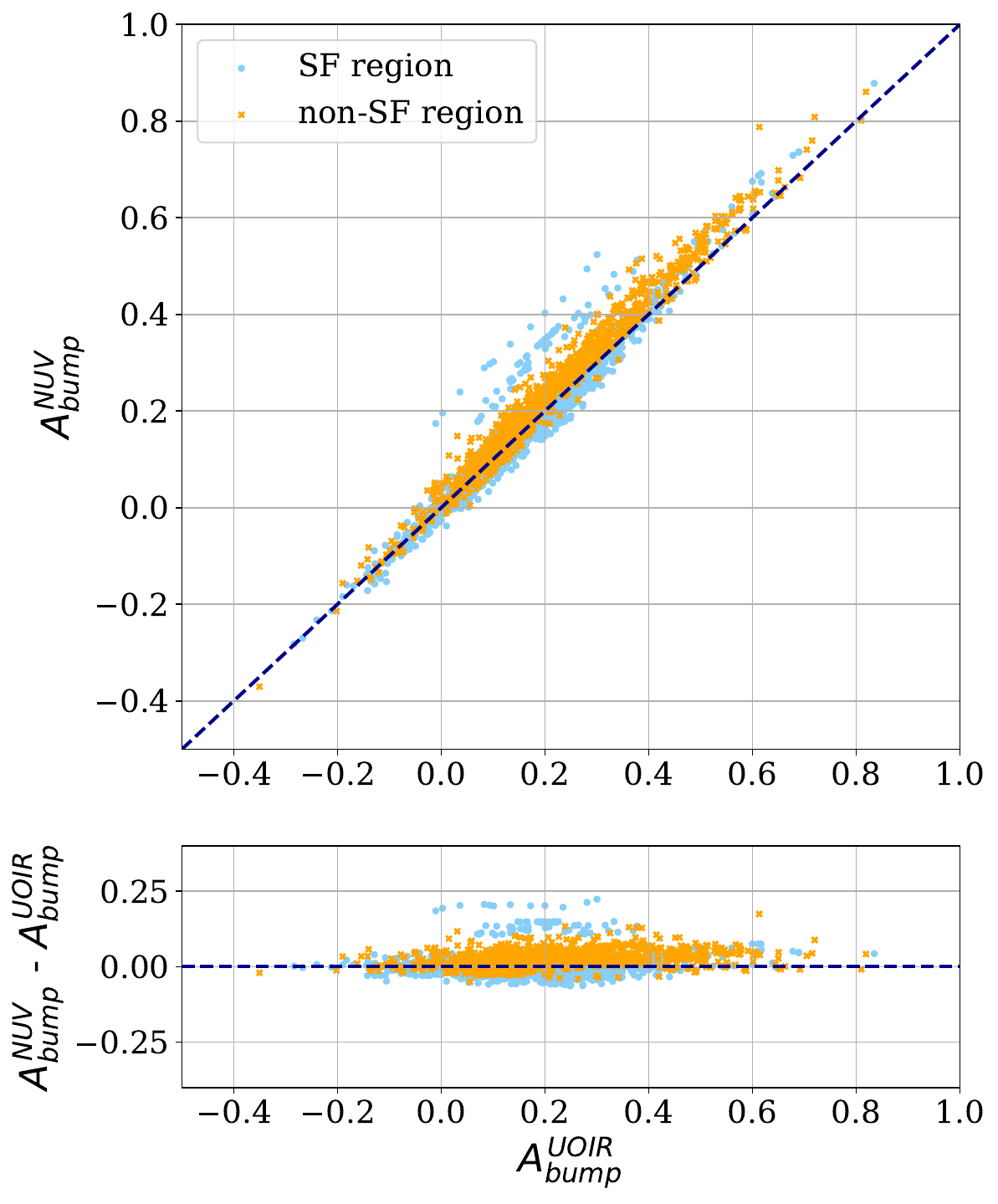}
                \caption{Comparison between $A_{bump}^{NUV}$ and $A_{bump}^{UOIR}$ in the overlapping $SNR_{cont}>20$ sample. The upper panel shows the spaxel-by-spaxel comparison, and the lower panel shows $A_{bump}^{NUV}-A_{bump}^{UOIR}$ versus $A_{bump}^{UOIR}$. SF and non-SF regions are shown in blue and orange.\label{fig:Abump_one_by_one}}
\end{figure}  

\begin{figure*}
\centering
    \includegraphics[width=0.3\textwidth]{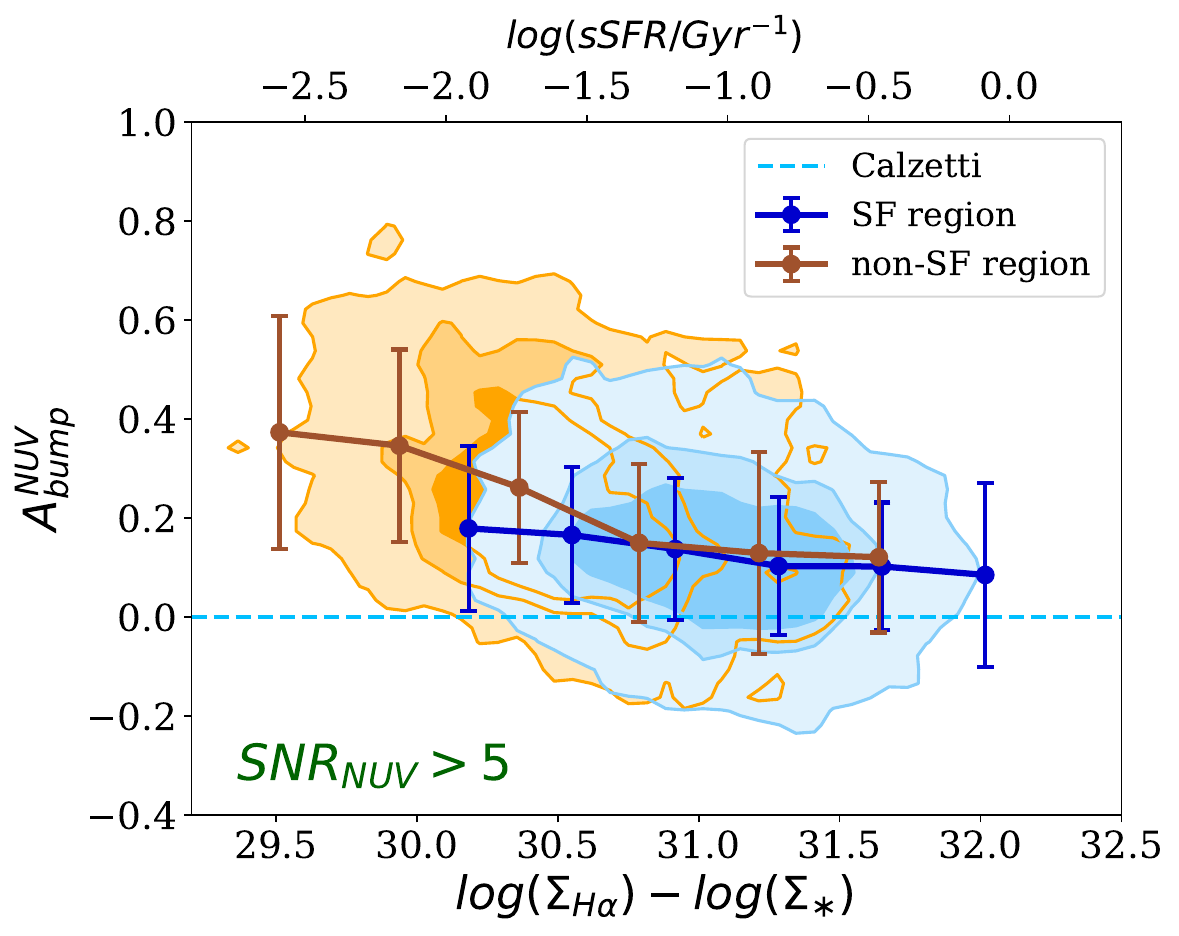}
    \includegraphics[width=0.3\textwidth]{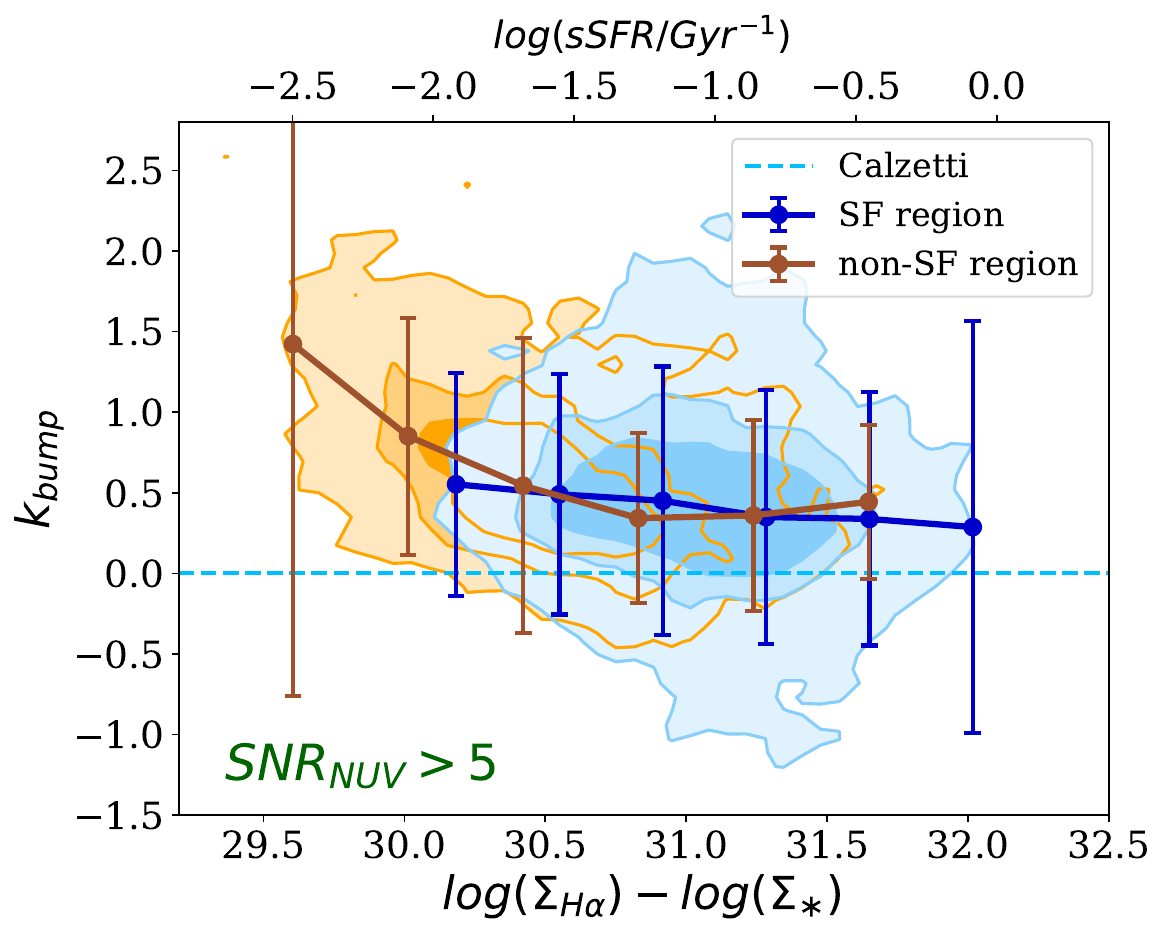}
    \includegraphics[width=0.3\textwidth]{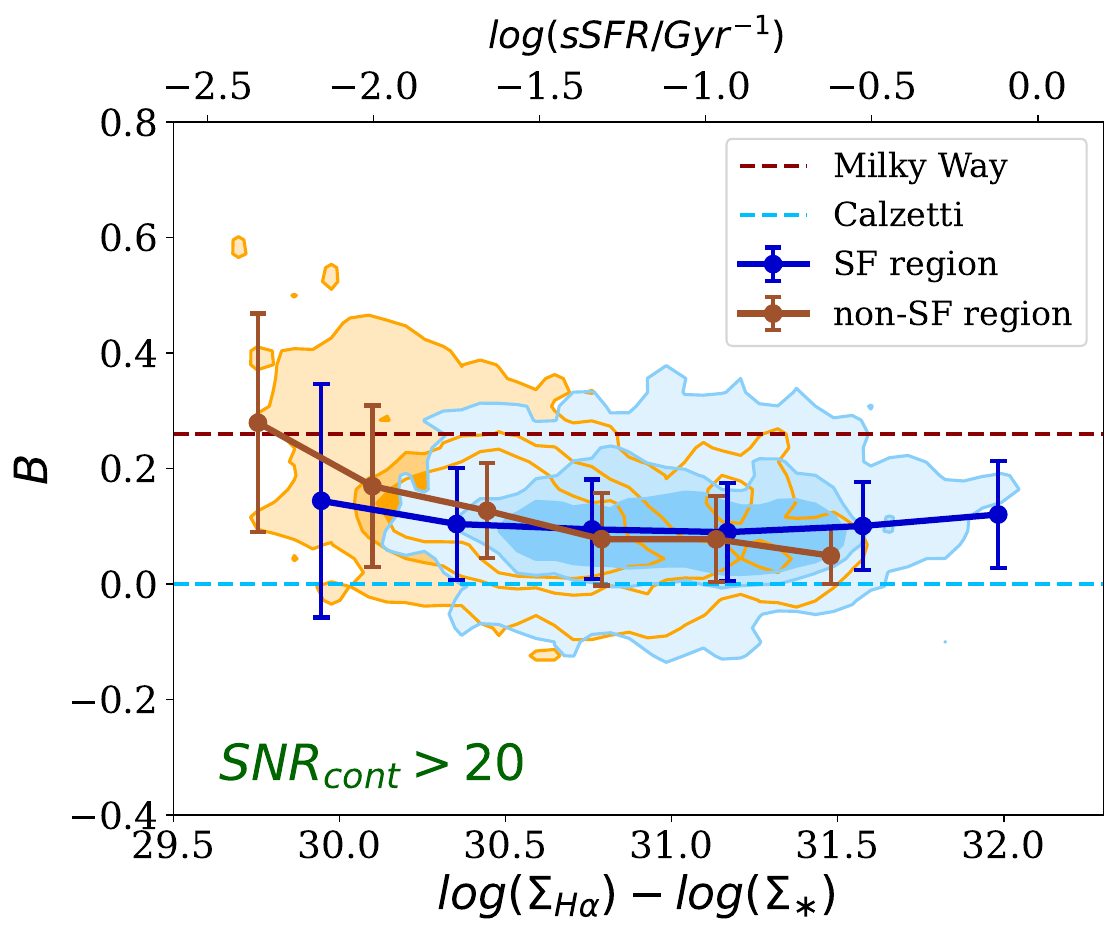}
    \includegraphics[width=0.3\textwidth]{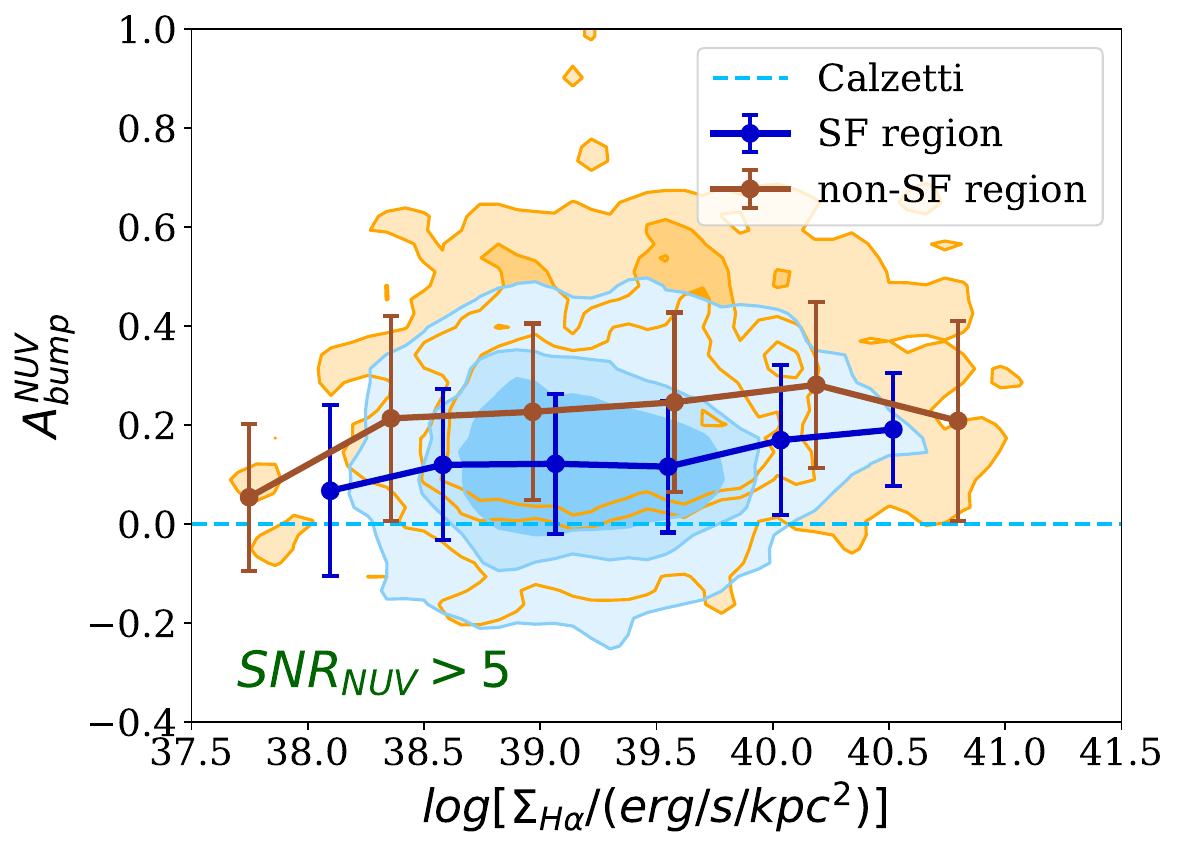}
    \includegraphics[width=0.3\textwidth]{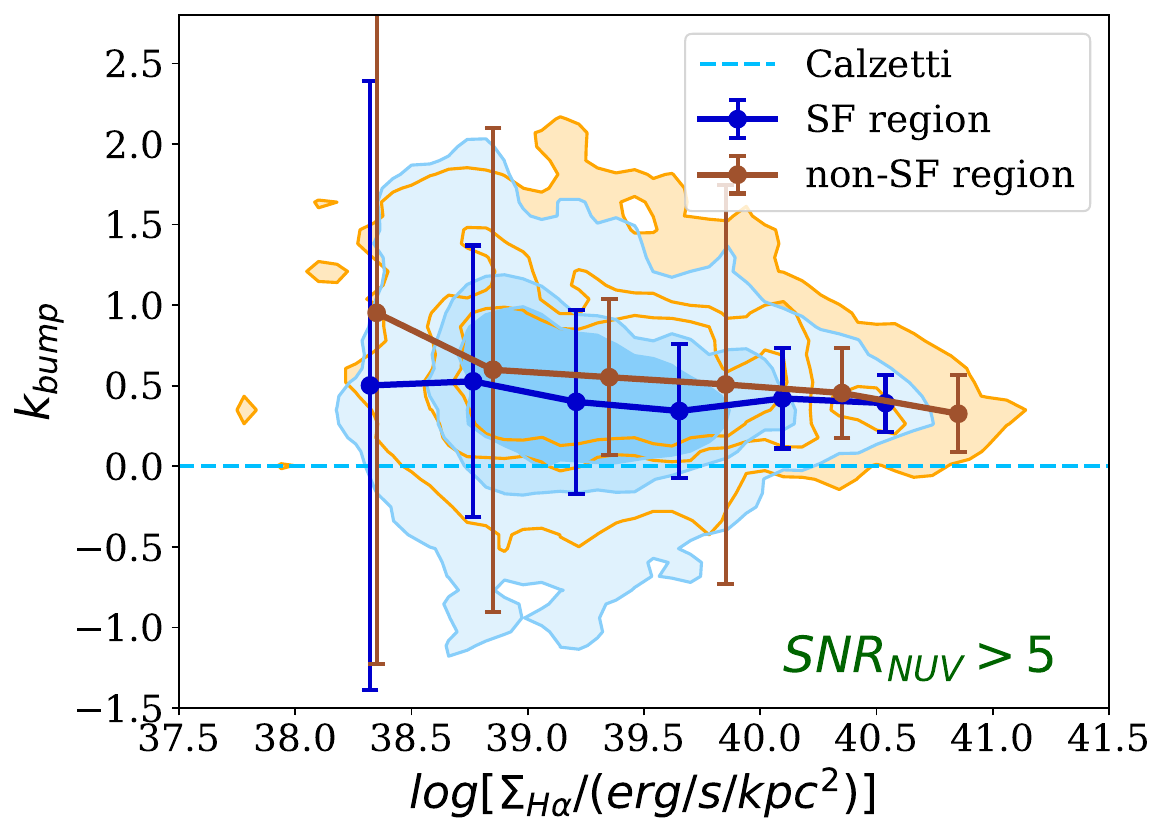}
    \includegraphics[width=0.3\textwidth]{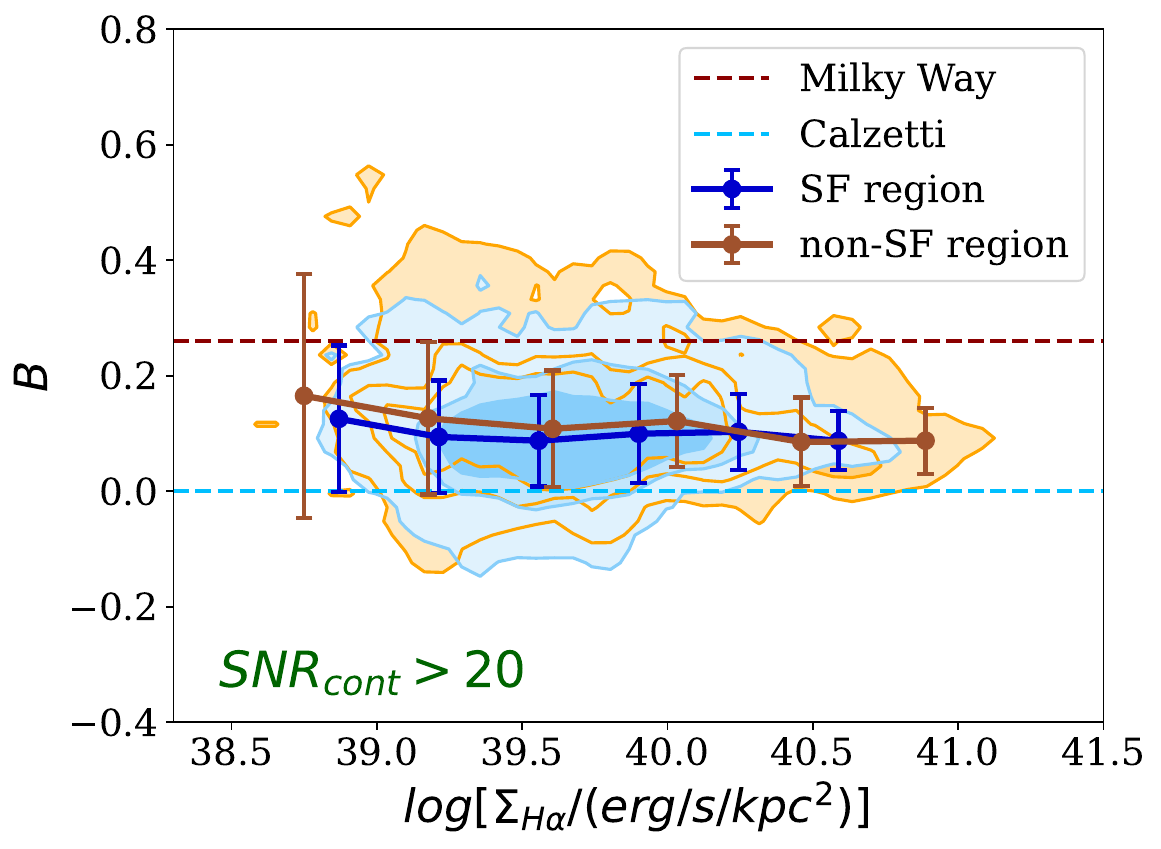}
    \includegraphics[width=0.3\textwidth]{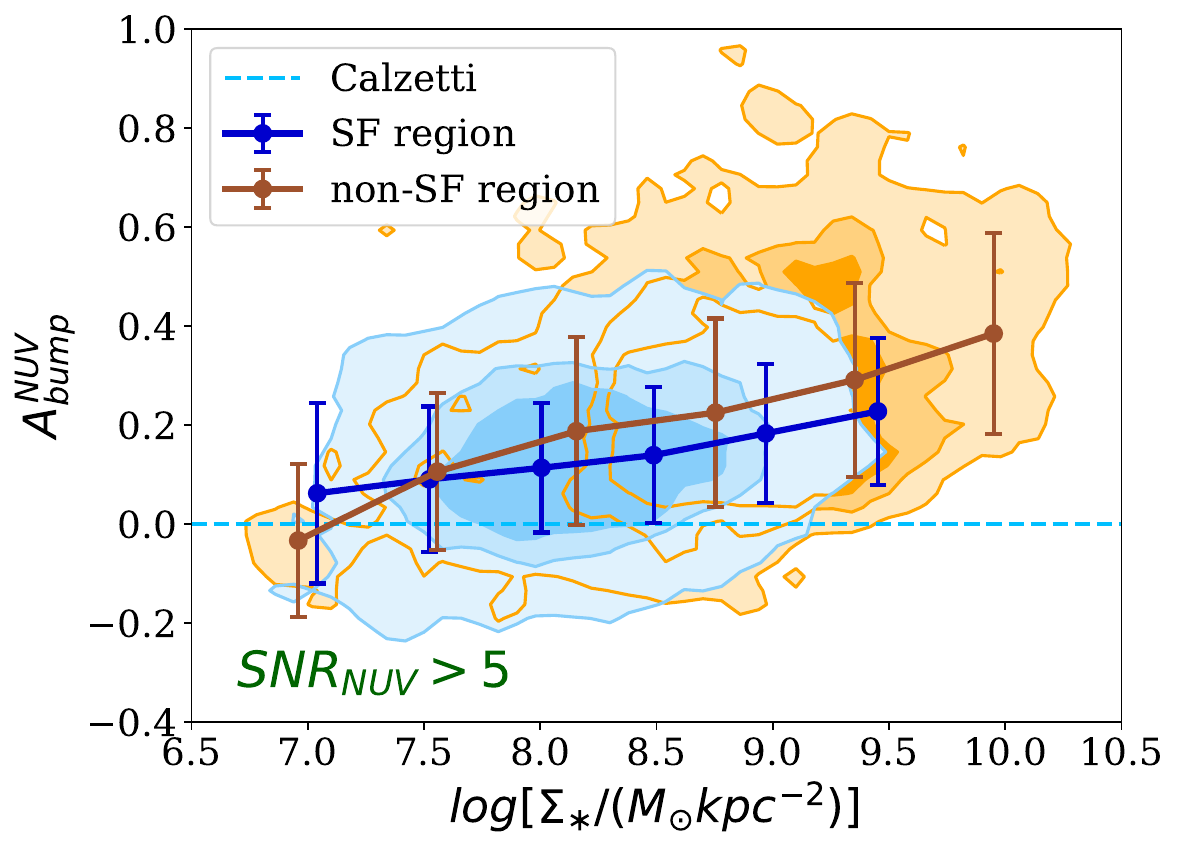}
    \includegraphics[width=0.3\textwidth]{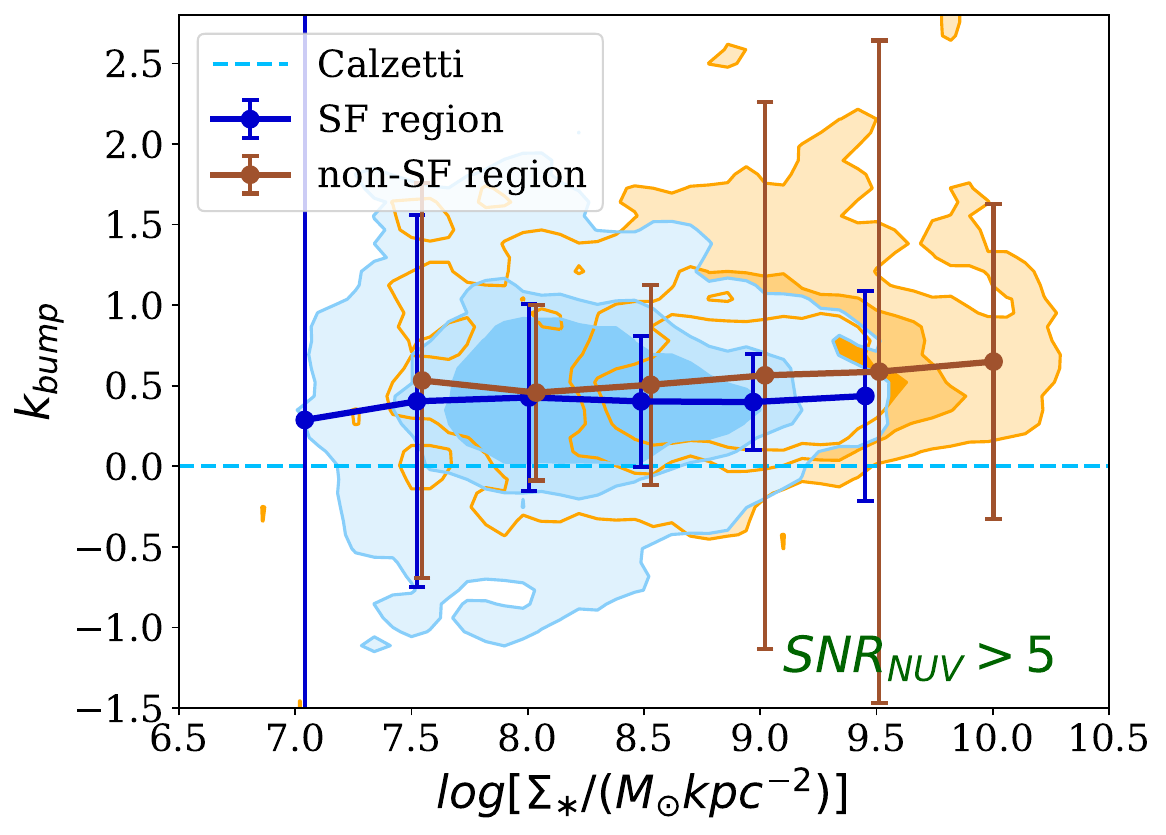}
    \includegraphics[width=0.3\textwidth]{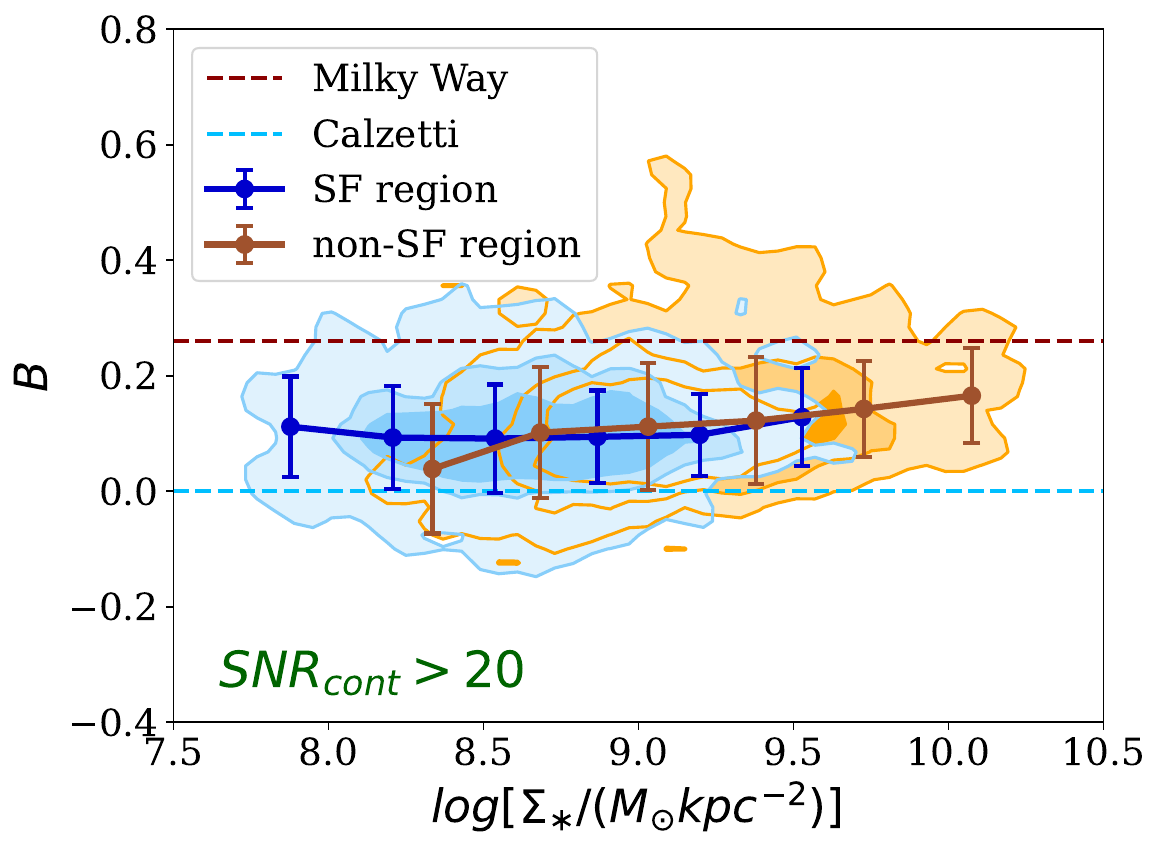}
    \caption{Bump strength versus $\Sigma_{\text{H}\alpha}/\Sigma_\ast$ (top row), $\Sigma_{\text{H}\alpha}$ (middle row), and $\Sigma_\ast$ (bottom row). Columns show $A_{bump}^{NUV}$, $k_{bump}$, and $B$. The first two columns use the $SNR_{NUV}>5$ sample; the $B$ column uses the $SNR_{cont}>20$ sample. Contours and median relations distinguish SF and non-SF regions, with dashed lines marking Calzetti-like and Milky-Way-type values.\label{fig:sSFR_bump}}
\end{figure*}

\subsection{Comparison of the two absolute bump measurements}

\autoref{fig:Abump_one_by_one} compares $A_{bump}^{NUV}$ and $A_{bump}^{UOIR}$ in the overlapping $SNR_{cont}>20$ sample. The two estimates agree well in both SF and non-SF regions, so we use $A_{bump}^{NUV}$ as the fiducial absolute bump strength in the larger NUV-selected sample, while retaining $B$ as the primary attenuation-curve-based relative strength.

Because both estimates use the UVOT NUV photometry, this comparison is not fully independent; instead, it tests whether the NUV-only estimator is consistent with the full attenuation-curve normalization in the overlap sample.

Quantitatively, the two absolute estimators are very tightly correlated in the overlap sample. With galaxy-bootstrap uncertainties, the full overlap has $\rho_S=0.97^{+0.01}_{-0.01}$ and a median offset $A_{bump}^{NUV}-A_{bump}^{UOIR}=0.01$ mag. The agreement remains similarly strong for SF and non-SF spaxels separately, with $\rho_S=0.96^{+0.02}_{-0.02}$ and $0.99^{+0.00}_{-0.00}$ and median offsets of $0.00$ and $0.02$ mag, respectively.

\begin{figure*}
\centering
    \includegraphics[width=0.3\textwidth]{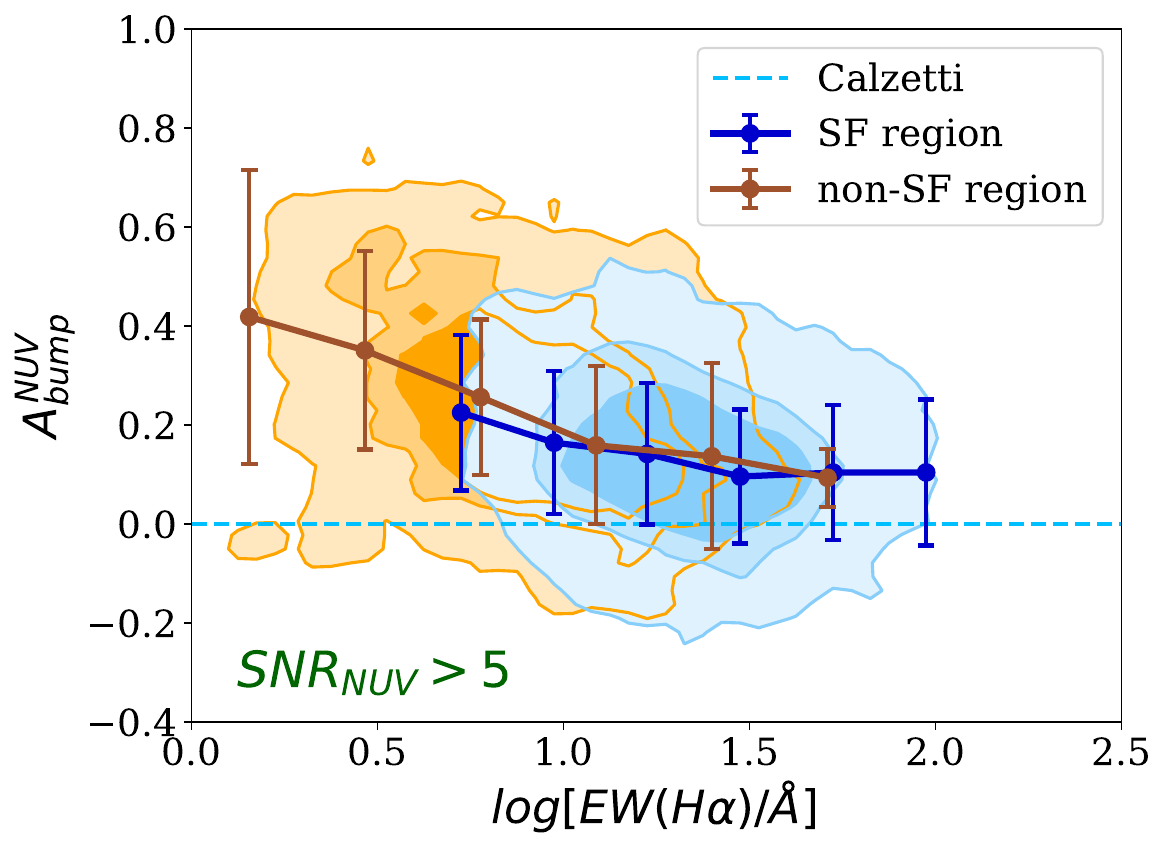}
    \includegraphics[width=0.3\textwidth]{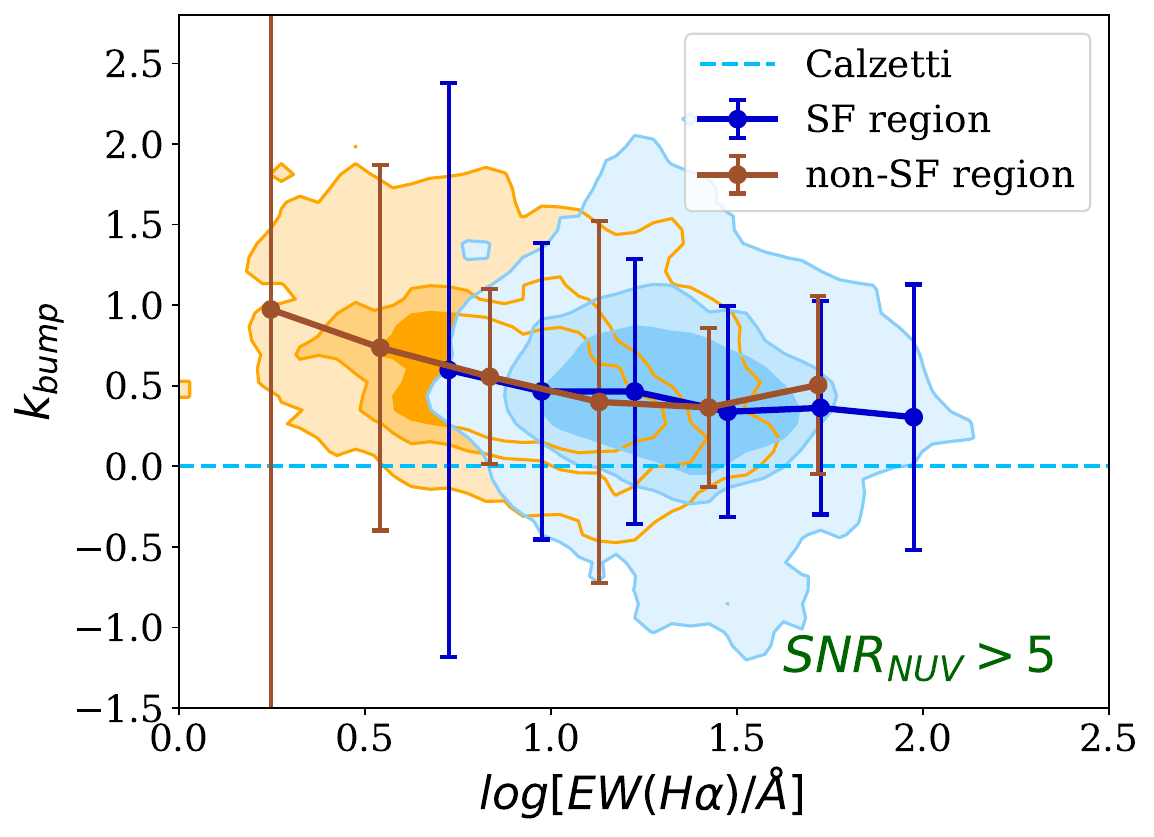}
    \includegraphics[width=0.3\textwidth]{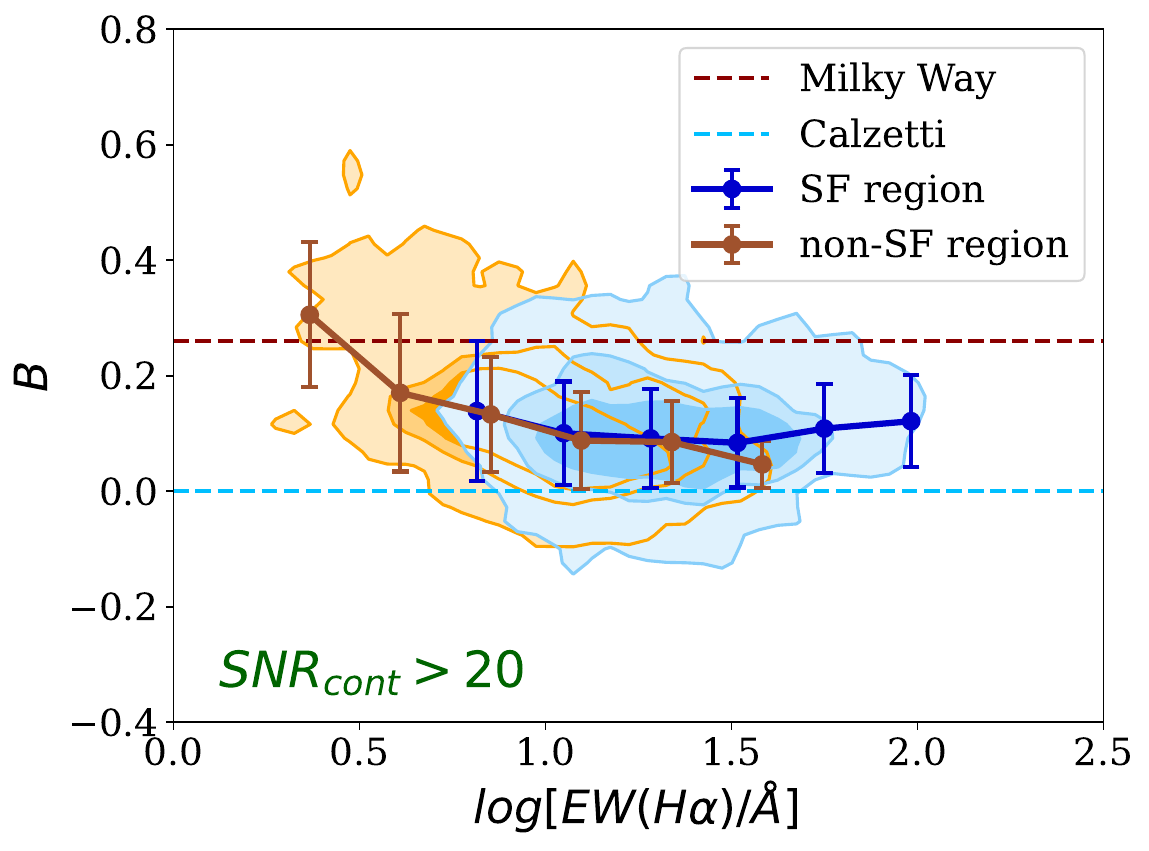}
    \includegraphics[width=0.3\textwidth]{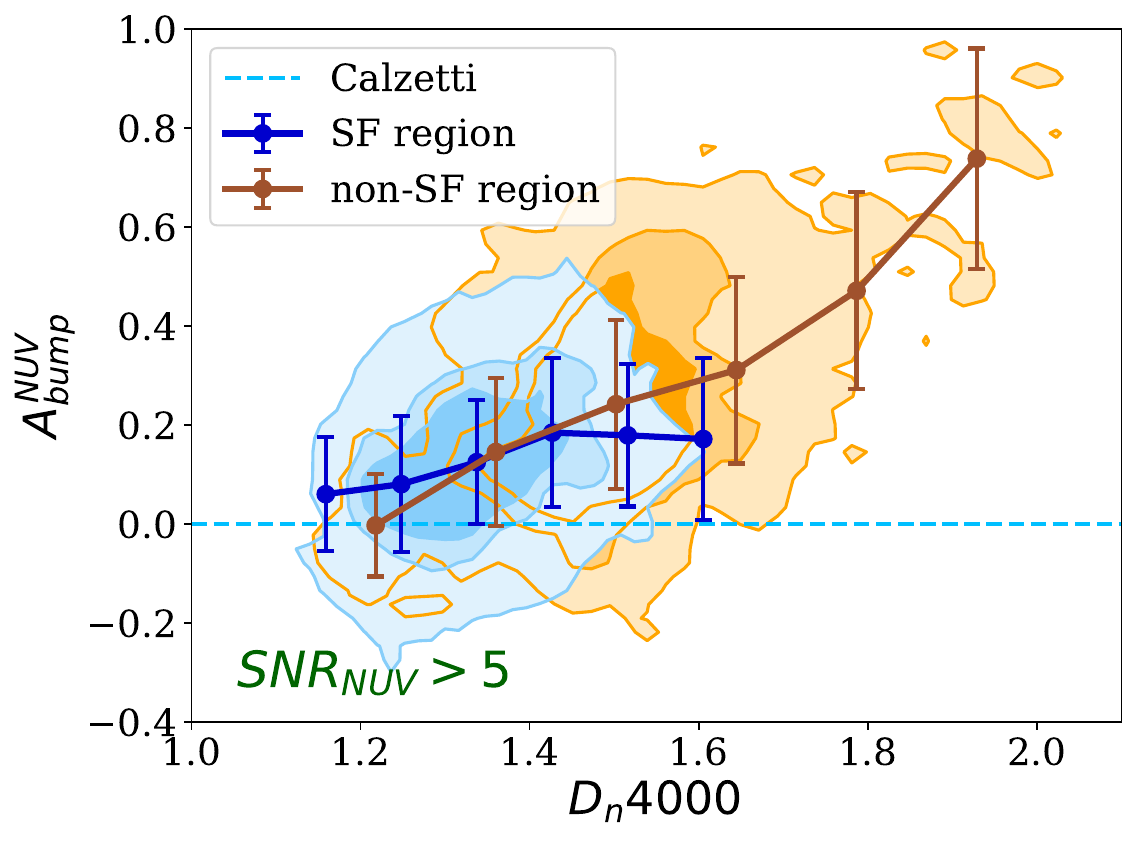}
    \includegraphics[width=0.3\textwidth]{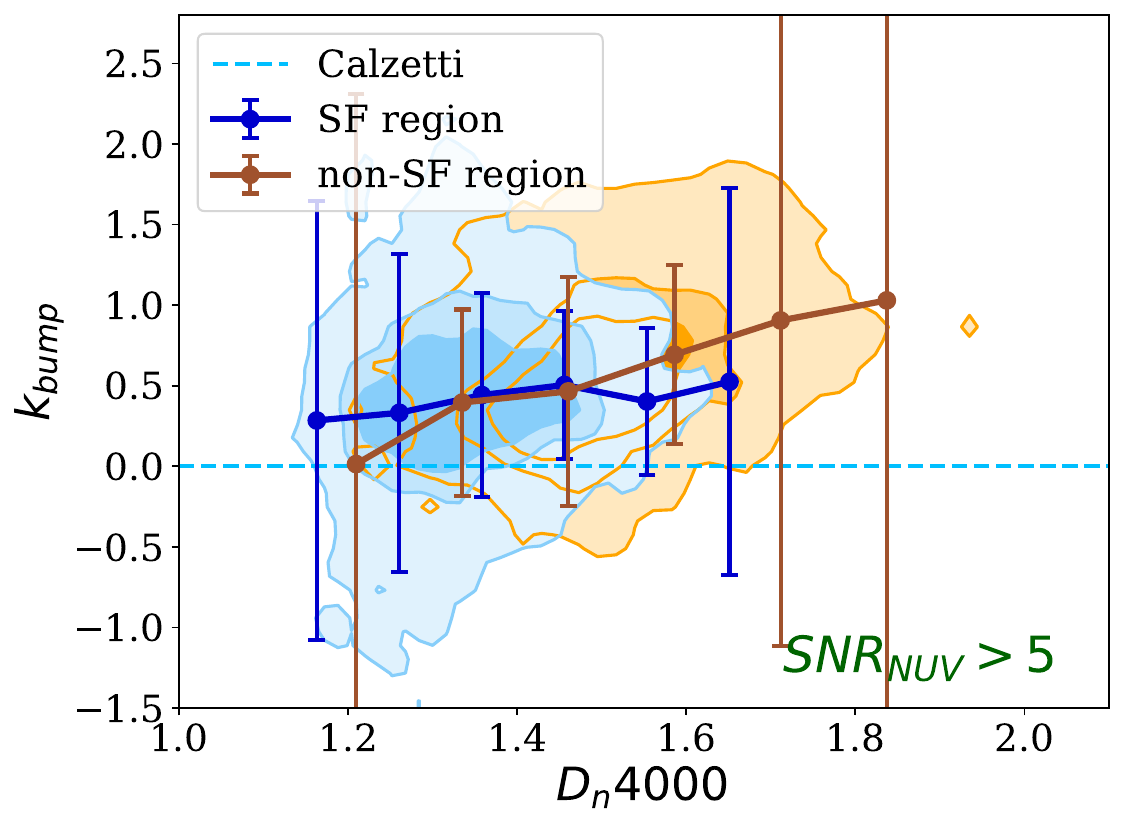}
    \includegraphics[width=0.3\textwidth]{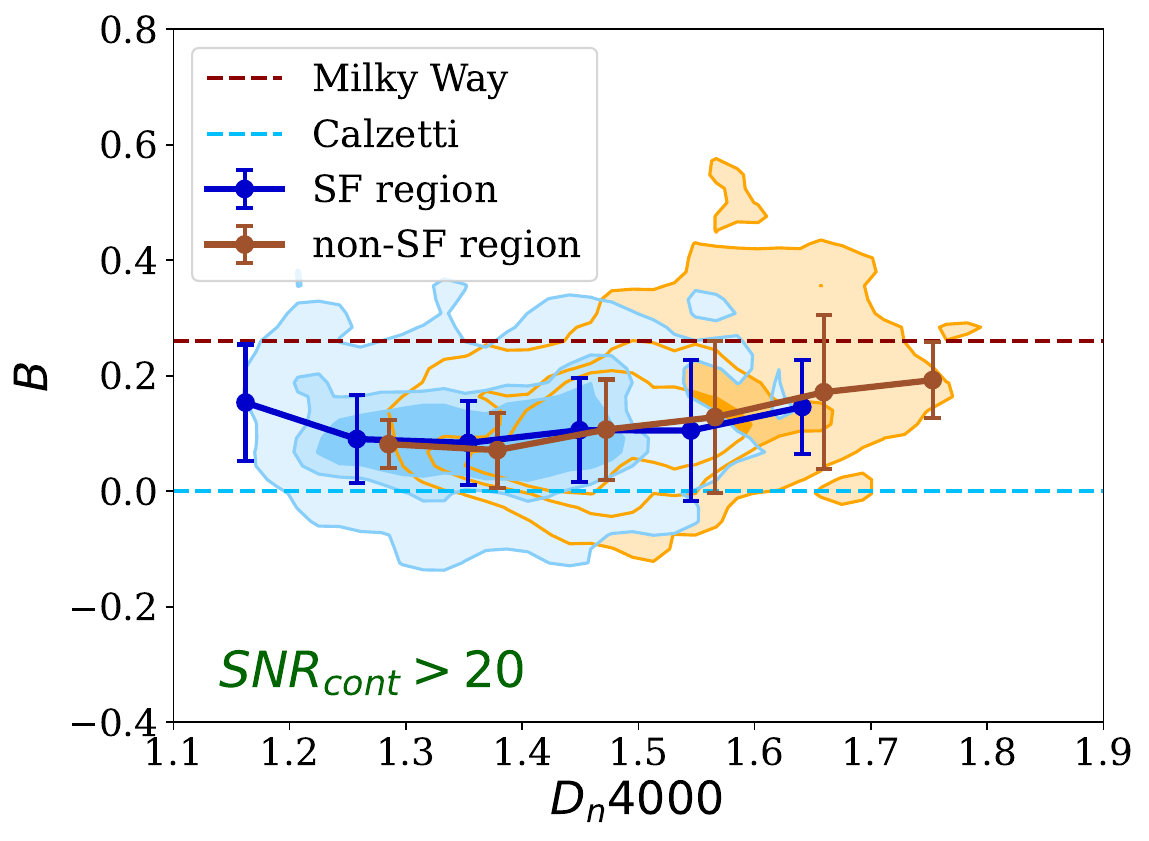}
    \includegraphics[width=0.3\textwidth]{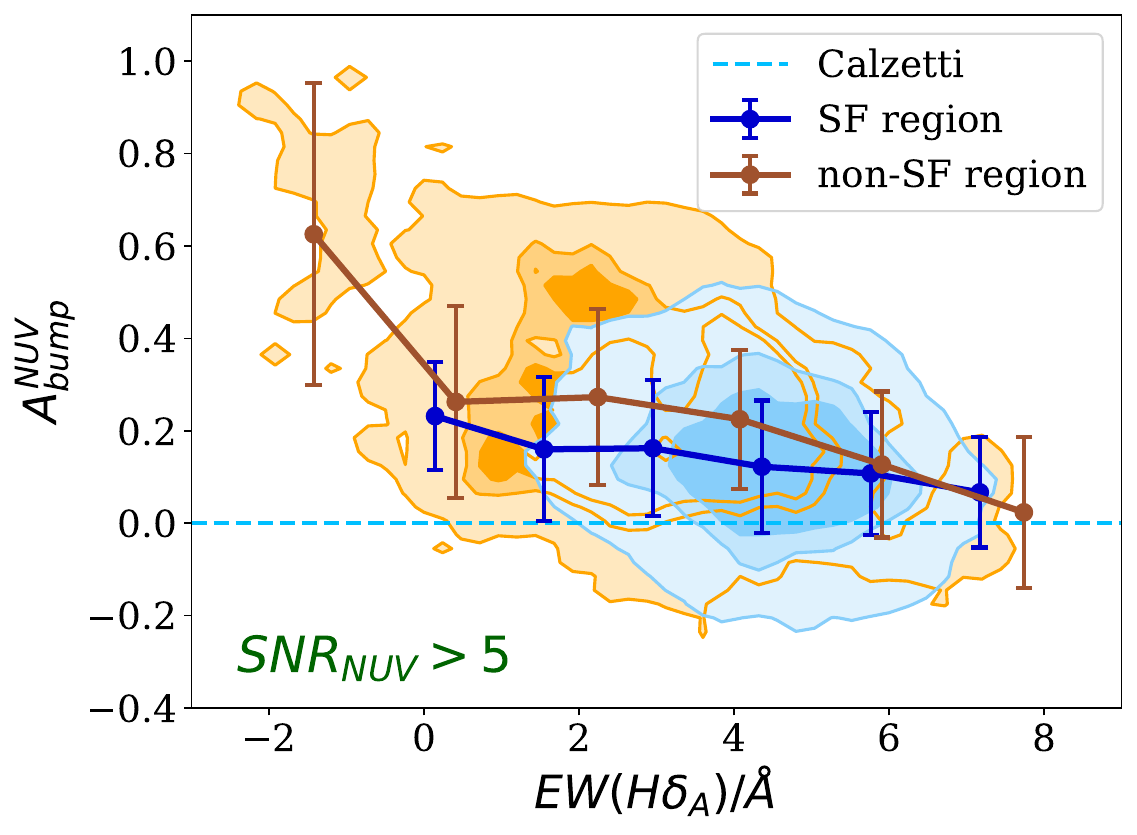}
    \includegraphics[width=0.3\textwidth]{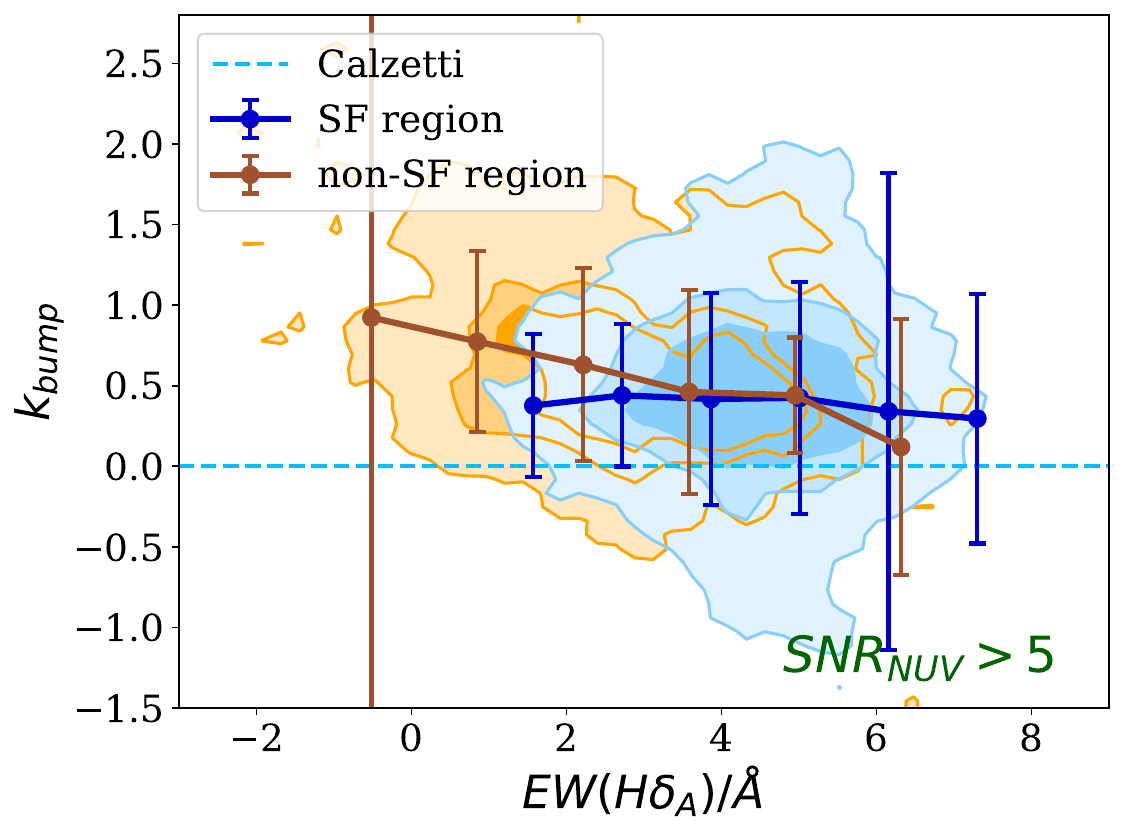}
    \includegraphics[width=0.3\textwidth]{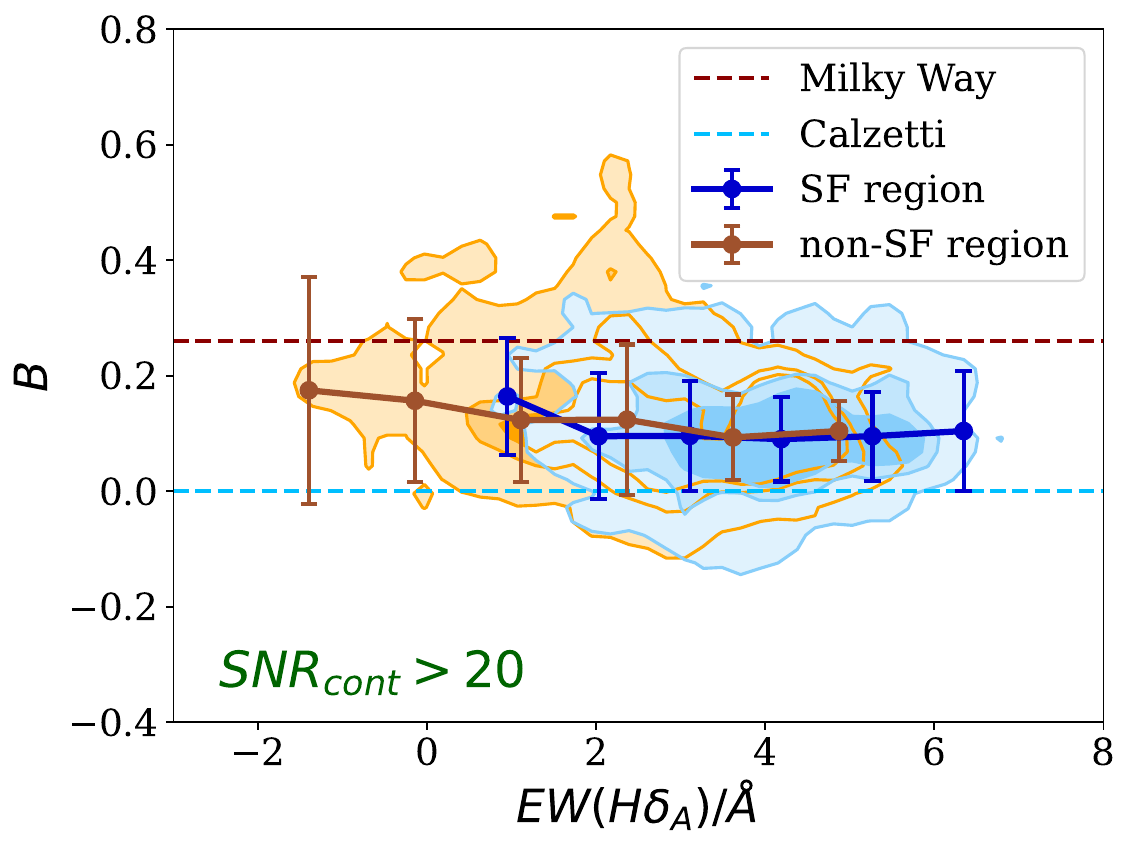}
    \caption{Bump strength versus recent-SFH indicators: EW(H$\alpha$), $D_n4000$, and EW(H$\delta_A$). Columns and symbols are as in \autoref{fig:sSFR_bump}.\label{fig:SFH_bump}}
\end{figure*}

\section{Results}\label{sec:results}

Here we describe the empirical trends before turning to their physical interpretation in Section~\ref{sec:discussion}. We focus on whether each trend appears in absolute and/or relative bump strength, whether SF and non-SF regions behave differently, and whether secondary correlations remain important compared with the dependence on $\Sigma_{\text{H}\alpha}/\Sigma_\ast$.

To quantify the visual trends, \autoref{tab:bootstrap_trends} lists Spearman rank coefficients and the difference in median bump strength between the highest and lowest predictor quartiles. The uncertainties are estimated by resampling host galaxies with replacement and retaining all selected spaxels from each resampled galaxy, so the quoted ranges account for the fact that neighboring spaxels and spaxels within the same galaxy are not independent.

\begin{table*}
\centering
\caption{Galaxy-bootstrap trend statistics for representative bump correlations.}
\label{tab:bootstrap_trends}
\scriptsize
\begin{tabular}{llllrrr}
\hline
Bump measure & Region & Predictor & Sample & $N_{\rm gal}$ & $\rho_S$ & $\Delta y_{\rm Q4-Q1}$ \\
\hline
$A_{bump}^{NUV}$ & SF & $\log(\Sigma_{\mathrm{H}\alpha}/\Sigma_*)$ & NUV & 114 & -0.21$^{+0.06}_{-0.05}$ & -0.07$^{+0.02}_{-0.02}$ \\
$A_{bump}^{NUV}$ & SF & EW(H$\alpha$) & NUV & 114 & -0.21$^{+0.07}_{-0.05}$ & -0.07$^{+0.03}_{-0.02}$ \\
$A_{bump}^{NUV}$ & SF & $D_n4000$ & NUV & 114 & 0.35$^{+0.06}_{-0.07}$ & 0.12$^{+0.02}_{-0.03}$ \\
$A_{bump}^{NUV}$ & non-SF & $\log(\Sigma_{\mathrm{H}\alpha}/\Sigma_*)$ & NUV & 70 & -0.39$^{+0.08}_{-0.07}$ & -0.20$^{+0.05}_{-0.05}$ \\
$A_{bump}^{NUV}$ & non-SF & EW(H$\alpha$) & NUV & 97 & -0.41$^{+0.08}_{-0.09}$ & -0.27$^{+0.07}_{-0.06}$ \\
$A_{bump}^{NUV}$ & non-SF & $D_n4000$ & NUV & 97 & 0.45$^{+0.07}_{-0.06}$ & 0.23$^{+0.07}_{-0.05}$ \\
$k_{bump}$ & SF & $\log(\Sigma_{\mathrm{H}\alpha}/\Sigma_*)$ & NUV & 114 & -0.18$^{+0.04}_{-0.04}$ & -0.16$^{+0.05}_{-0.05}$ \\
$k_{bump}$ & SF & EW(H$\alpha$) & NUV & 114 & -0.16$^{+0.05}_{-0.04}$ & -0.14$^{+0.06}_{-0.05}$ \\
$k_{bump}$ & SF & $D_n4000$ & NUV & 114 & 0.19$^{+0.05}_{-0.07}$ & 0.19$^{+0.04}_{-0.08}$ \\
$k_{bump}$ & non-SF & $\log(\Sigma_{\mathrm{H}\alpha}/\Sigma_*)$ & NUV & 70 & -0.47$^{+0.07}_{-0.07}$ & -0.56$^{+0.09}_{-0.09}$ \\
$k_{bump}$ & non-SF & EW(H$\alpha$) & NUV & 70 & -0.37$^{+0.07}_{-0.06}$ & -0.42$^{+0.09}_{-0.08}$ \\
$k_{bump}$ & non-SF & $D_n4000$ & NUV & 70 & 0.40$^{+0.06}_{-0.07}$ & 0.45$^{+0.06}_{-0.10}$ \\
$B$ & SF & $\log(\Sigma_{\mathrm{H}\alpha}/\Sigma_*)$ & UOIR & 63 & -0.00$^{+0.05}_{-0.06}$ & -0.00$^{+0.01}_{-0.01}$ \\
$B$ & SF & EW(H$\alpha$) & UOIR & 63 & -0.04$^{+0.08}_{-0.09}$ & -0.01$^{+0.01}_{-0.02}$ \\
$B$ & SF & $D_n4000$ & UOIR & 63 & 0.07$^{+0.10}_{-0.08}$ & 0.02$^{+0.02}_{-0.02}$ \\ 
$B$ & non-SF & $\log(\Sigma_{\mathrm{H}\alpha}/\Sigma_*)$ & UOIR & 46 & -0.44$^{+0.13}_{-0.06}$ & -0.10$^{+0.04}_{-0.04}$ \\
$B$ & non-SF & EW(H$\alpha$) & UOIR & 46 & -0.44$^{+0.09}_{-0.07}$ & -0.10$^{+0.03}_{-0.05}$ \\
$B$ & non-SF & $D_n4000$ & UOIR & 46 & 0.36$^{+0.08}_{-0.09}$ & 0.10$^{+0.02}_{-0.04}$ \\
\hline
\end{tabular}
\tablecomments{The uncertainties are the 16th--84th percentile ranges from galaxy-level bootstrap resampling. $\rho_S$ is the Spearman rank coefficient. $\Delta y_{\rm Q4-Q1}$ is the difference between the median bump strength in the highest and lowest quartiles of the predictor; negative values therefore indicate weaker bumps at larger predictor values.}
\end{table*}

\subsection{Recent star formation and ionized-gas activity}

The overlap sample shows that $A_{bump}^{NUV}$ and $A_{bump}^{UOIR}$ have consistent behavior with $\Sigma_{\text{H}\alpha}/\Sigma_{\ast}$, including the stronger decline toward larger values in non-SF regions. For relative bump strength, the two normalized quantities should be read together: $k_{bump}$ is normalized by $E(B-V)_{gas}$, while $B$ is normalized by the bump-free attenuation $A_{m2}^{\prime}$. A trend shared by $k_{bump}$ and $B$ is therefore the strongest evidence for a change in relative bump prominence.

The top row of \autoref{fig:sSFR_bump} shows the main result. In SF regions, where $\Sigma_{\text{H}\alpha}/\Sigma_{\ast}$ is sSFR-like, $A_{bump}^{NUV}$ decreases with increasing $\Sigma_{\text{H}\alpha}/\Sigma_{\ast}$, while $k_{bump}$ and $B$ change more weakly. In non-SF regions, all three measurements decrease more clearly, and the strongest bumps occur at the lowest $\Sigma_{\text{H}\alpha}/\Sigma_{\ast}$. Thus the dominant common trend is a weakening of relative bump prominence in non-SF regions with larger H$\alpha$ emission per unit stellar mass.

The galaxy-bootstrap statistics support this interpretation. For non-SF regions, the correlations with $\log_{10}(\Sigma_{\text{H}\alpha}/\Sigma_\ast)$ are $\rho_S=-0.39^{+0.08}_{-0.07}$ for $A_{bump}^{NUV}$, $-0.47^{+0.07}_{-0.07}$ for $k_{bump}$, and $-0.44^{+0.13}_{-0.06}$ for $B$. The corresponding high-minus-low-quartile changes are $-0.20$, $-0.56$, and $-0.10$, respectively. In SF regions the same trend is weaker, and $B$ is consistent with no correlation.

The middle and bottom rows of \autoref{fig:sSFR_bump} separate the ratio into its two surface-density components. $A_{bump}^{NUV}$ increases with both $\Sigma_{\text{H}\alpha}$ and $\Sigma_{\ast}$, while $k_{bump}$ and $B$ show weak or no dependence on either quantity alone. This contrast indicates that absolute bump amplitude is partly sensitive to attenuation scale or dust column, whereas normalized quantities better isolate relative bump prominence.

\autoref{fig:SFH_bump} shows the same picture using spectral indices. $A_{bump}^{NUV}$ and $k_{bump}$ decrease with EW(H$\alpha$), especially in non-SF regions, and all bump strengths increase with $D_n4000$. EW(H$\delta_A$) gives weaker trends but does not contradict this behavior. Recent-SFH indicators are therefore more coherent predictors of bump strength than $\Sigma_{\text{H}\alpha}$ or $\Sigma_{\ast}$ alone.

The same table shows that the non-SF trends with EW(H$\alpha$) and $D_n4000$ are comparable in strength to the $\Sigma_{\text{H}\alpha}/\Sigma_\ast$ trend. For example, in non-SF regions $A_{bump}^{NUV}$ has $\rho_S=-0.41^{+0.08}_{-0.09}$ with EW(H$\alpha$) and $\rho_S=0.45^{+0.07}_{-0.06}$ with $D_n4000$, while $k_{bump}$ has $\rho_S=-0.37^{+0.07}_{-0.06}$ and $0.40^{+0.06}_{-0.07}$ for the same two predictors.

\subsection{Secondary dependences}

The age trends are consistent with the recent-SFH picture. The bump strength increases with luminosity-weighted stellar age more clearly than with mass-weighted age, indicating that quantities sensitive to the recent light-weighted population are more predictive than those weighted by the accumulated stellar mass.

Metallicity is a secondary correlation. $A_{bump}^{NUV}$ shows weak positive trends with stellar metallicity and, in SF regions, with gas-phase oxygen abundance, but the relative strengths $k_{bump}$ and $B$ show little dependence on either stellar metallicity or oxygen abundance. The N2S2 ratio varies differently in non-SF regions, where it is better treated as a gas-excitation sequence than as a pure metallicity tracer \citep{2020ApJ...888...88L}. These results argue against metallicity as the dominant driver of relative bump prominence.

Geometry is also secondary. Using SDSS $r$-band axis ratio $b/a$ as an inclination proxy and $R_c/R_e$ as normalized galactocentric radius, we find only weak residual trends once $\Sigma_{\text{H}\alpha}/\Sigma_\ast$ is considered. Similarly, $B$ is largely independent of $A_V$ and the optical attenuation-curve slope $A_B/A_V$, but increases with the NUV slope $A_{w2}/A_{w1}$. Thus the bump behaves as a distinct NUV-shape component rather than as a simple consequence of optical opacity, optical slope, projection, or radius. The corresponding diagnostic panels are shown in Appendix~\ref{sec:appendix_diagnostic_figures}.

\section{Discussion}\label{sec:discussion}

\subsection{Radiation-field processing versus dust-column dependence}

The main result is that the 2175\AA\ bump is most strongly connected to recent star-formation and ionized-gas diagnostics, rather than to metallicity, inclination, radius, optical opacity, or optical slope alone. The conclusion is clearest when the absolute and relative measurements are considered together. $A_{bump}^{NUV}$ responds to both bump prominence and total attenuation, while $k_{bump}$ and $B$ reduce part of this column dependence. The fact that all three quantities show their strongest common trend with $\Sigma_{\text{H}\alpha}/\Sigma_\ast$, especially in non-SF regions, argues that the trend is not simply a dust-column effect.

This extends Paper I, where the bump was found to weaken with increasing sSFR. In SF regions, $\Sigma_{\text{H}\alpha}/\Sigma_\ast$ is sSFR-like, so the anti-correlation naturally suggests weaker bump features in stronger young-star radiation fields. The trends with EW(H$\alpha$), $D_n4000$, and stellar age point in the same direction: regions dominated by recent star formation have weaker bumps, whereas older or more quiescent regions preserve a stronger feature. We use ``radiation-field processing'' empirically here; the data identify the environments associated with strong or weak bumps, but do not uniquely distinguish carrier abundance, grain charge state, size distribution, shielding, scattering, or unresolved star--dust geometry.

Non-SF regions are essential because they populate the low-$\Sigma_{\text{H}\alpha}/\Sigma_\ast$ regime where the strongest bumps are found. In these regions, however, $\Sigma_{\text{H}\alpha}/\Sigma_\ast$ should not be interpreted as sSFR: the H$\alpha$ emission may come from diffuse ionized gas, hot evolved stars, shocks, weak AGN activity, or mixtures of these sources. We therefore treat it as ionized-gas emission per unit stellar mass.

Under this interpretation, the non-SF trend suggests that bump carriers are sensitive to radiation fields or gas-excitation conditions beyond classical HII-region star formation. The strongest bumps occur where ionized-gas emission per unit stellar mass is lowest, consistent with easier survival of small carbonaceous grains in relatively quiescent diffuse environments. The apparent continuity between SF and non-SF regions should therefore not be read as evidence for one ionizing source; it more likely reflects a common balance between grain survival and processing.

The non-SF trend in the NUV-selected sample is robust against cuts designed to remove low-quality Balmer decrements, low UVOT S/N, AGN host galaxies, and central regions. Requiring $E(B-V)_{gas}>0.05$ or $>0.10$ gives nearly unchanged correlations with $\log_{10}(\Sigma_{\text{H}\alpha}/\Sigma_\ast)$: $\rho_S\simeq-0.38$ for $A_{bump}^{NUV}$ and $\rho_S\simeq-0.47$ for $k_{bump}$. Requiring S/N $>5$ in all three UVOT bands gives $\rho_S=-0.46$ and $-0.52$, and requiring H$\alpha$ and H$\beta$ S/N $>5$ gives $\rho_S=-0.38$ and $-0.47$. Excluding AGN hosts, restricting to $R_c/R_e\geq0.5$, or applying both cuts still leaves negative correlations, with the combined cut giving $\rho_S=-0.29$ and $-0.40$.

The same conclusion holds when each host galaxy is given equal weight. Using one median point per host galaxy, the correlations with $\log_{10}(\Sigma_{\text{H}\alpha}/\Sigma_\ast)$ remain negative for $A_{bump}^{NUV}$, $k_{bump}$, and $B$, with $\rho_S=-0.41$, $-0.57$, and $-0.44$, respectively.

\subsection{Absolute and relative bump strengths}

The different behavior of $A_{bump}^{NUV}$, $k_{bump}$, and $B$ clarifies what each quantity measures. The positive correlations of $A_{bump}^{NUV}$ with $\Sigma_{\text{H}\alpha}$ and $\Sigma_\ast$ likely reflect, at least in part, larger dust columns and stronger stellar continua in denser regions. By contrast, $k_{bump}$ and $B$ show little dependence on either surface-density component alone. Thus increasing surface density can raise the absolute attenuation excess around 2175\AA, while a larger ionized-gas output per unit stellar mass can still correspond to a weaker bump relative to the local continuum attenuation.

This distinction also explains why metallicity is not the dominant axis. The absolute bump strength shows weak positive trends with stellar metallicity and oxygen abundance in SF regions, but the relative strengths show little or no metallicity dependence. Likewise, the weak dependence on inclination, radius, $A_V$, and optical slope argues against a primarily geometric or optical-opacity explanation. The correlation between $B$ and the NUV slope $A_{w2}/A_{w1}$ instead links the bump to the detailed NUV shape of the attenuation curve, as expected if it traces a specific grain population whose abundance or survival changes from region to region.

\subsection{Context and caveats}

Our results connect the classical sightline view of the 2175\AA\ feature with galaxy-scale attenuation measurements. The resolved attenuation curves do not reduce to either a Milky-Way-like strong-bump case or a Calzetti-like featureless case; substantial variation exists within the same nearby-galaxy sample and even within individual galaxies. Integrated correlations with inclination, stellar mass, or star-formation activity \citep[e.g.][]{2010ApJ...718..184C,2011MNRAS.417.1760W,2017ApJ...851...90B,2018ApJ...859...11S,2013ApJ...775L..16K,2021ApJ...909..213K,2022MNRAS.514.1886S} may therefore reflect changing mixtures of local environments.

The weakening of the bump in regions with stronger recent-star-formation indicators is qualitatively consistent with radiation-field processing of small carbonaceous or aromatic grains, as also suggested by resolved UV/infrared and PAH-related studies \citep{2019MNRAS.486..743D,2023ApJ...953...54B,2025PASA...42...22B,2022MNRAS.514.1886S,2023ApJ...944L..16E}. However, all bump measurements here are effective attenuation measurements, not direct extinction curves along single sightlines. They include dust-star geometry, scattering, and population mixing within each kiloparsec-scale spaxel. The agreement between $A_{bump}^{NUV}$ and $A_{bump}^{UOIR}$ supports the main absolute-strength trends, while conclusions about relative bump prominence should rely primarily on trends shared by $k_{bump}$ and $B$. This is particularly important in non-SF regions, where $k_{bump}$ is normalized by $E(B-V)_{gas}$; our interpretation of the non-SF relative-bump trend therefore relies on both the agreement with $B$ and the robustness checks against stricter Balmer-decrement and S/N cuts.

We therefore interpret the correlations as evidence for changes in the effective abundance, survival, or processing state of bump carriers, not as a unique inversion of carrier chemistry or grain-size distribution. Detailed dust modeling is presented in Paper IV \citep{2026arXiv260713868G}.

\section{Summary}\label{sec:summary}

This is the third paper in a series using the SwiM catalog to study dust attenuation at kiloparsec scales in nearby galaxies. We combine Swift/UVOT NUV imaging, MaNGA optical integral-field spectroscopy, and 2MASS $K_s$ imaging to measure the 2175\AA\ attenuation bump with two complementary methods: an attenuation-curve method that gives $A_{bump}^{UOIR}$ and $B$ for a high-continuum-S/N sample, and a NUV-only photometric method that gives $A_{bump}^{NUV}$ and $k_{bump}$ for a larger sample. We classify spaxels into SF and non-SF regions and compare the bump strengths with stellar-population, emission-line, attenuation-curve, and geometric properties.

Our main conclusions can be summarized as follows:

\begin{enumerate}
  \item In the overlapping high-continuum-S/N sample, the absolute bump strengths $A_{bump}^{NUV}$ and $A_{bump}^{UOIR}$ agree well for most spaxels. We therefore use $A_{bump}^{NUV}$ to exploit the larger NUV-selected sample, while using $B$ as the most direct relative bump measurement in the attenuation-curve sample.
  \item The bump strength is anti-correlated with specific H$\alpha$ surface brightness, $\Sigma_{\text{H}\alpha}/\Sigma_{\ast}$. The strongest bumps occur at the lowest values of $\Sigma_{\text{H}\alpha}/\Sigma_{\ast}$ and are found primarily in non-SF regions. In SF regions this quantity is closely related to sSFR, while in non-SF regions it is better interpreted as ionized-gas emission per unit stellar mass. The trend is weaker in SF regions, especially for the relative strengths $k_{bump}$ and $B$.
  \item The bump is weaker in regions with stronger recent star-formation indicators. It decreases with EW(H$\alpha$), increases with $D_n4000$, and is generally stronger in older stellar populations. These trends are most pronounced in non-SF regions. This suggests that the bump carriers are affected by radiation-field processing in both young-star-dominated and diffuse older-stellar environments.
  \item The absolute strength $A_{bump}^{NUV}$ increases with $\Sigma_{\text{H}\alpha}$ and $\Sigma_\ast$, while the relative strengths $k_{bump}$ and $B$ show weak or no dependence on these quantities. This indicates that absolute bump strength is partly affected by dust column or surface density, whereas the relative measures better trace the prominence of the bump carriers.
  \item Metallicity is not the dominant driver of the relative bump strength. The relative quantities show weak or no dependence on stellar metallicity and gas-phase oxygen abundance. The weak positive trends seen in $A_{bump}^{NUV}$ are secondary to the stronger correlations with recent-SFH and age indicators.
  \item The bump strength shows little dependence on $A_V$, optical slope $A_B/A_V$, host-galaxy inclination, or galactocentric radius, while $B$ correlates with the NUV slope $A_{w2}/A_{w1}$. Thus, the resolved bump variations are more closely linked to the local NUV attenuation shape and radiation-field environment than to global geometry or optical attenuation amplitude. The results support a picture in which the 2175\AA\ feature is shaped mainly by local radiation-field processing of small carbonaceous grains on kiloparsec or smaller scales.
\end{enumerate}

Taken together, these results extend the bump--sSFR connection found in Paper I into a resolved picture in which the 2175\AA\ feature is weakest where radiation-field processing appears strongest, while metallicity, dust column, and geometry act mainly as secondary modifiers. This empirical interpretation does not uniquely identify carrier chemistry or grain-size distribution. Paper IV \citep{2026arXiv260713868G} builds on this framework with detailed dust modeling of the bump carriers and their response to local radiation-field conditions.

\begin{acknowledgments}

This work is supported by the National Science Foundation of China (grant No. 12433003) and the National Key R\&D Program of China (grant No. 2018YFA0404502).

Funding for SDSS-IV has been provided by the Alfred P. Sloan Foundation and Participating Institutions. Additional funding towards SDSS-IV has been provided by the US Department of Energy Office of Science. SDSS-IV acknowledges support and resources from the Centre for High-Performance Computing at the University of Utah. The SDSS website is \url{https://www.sdss.org}.

We acknowledge the Tsinghua Astrophysics High-Performance Computing platform at Tsinghua University for providing computational and data storage resources that have contributed to the research results reported within this paper.

\end{acknowledgments}

\begin{contribution}
RG performed the data analysis, produced the figures, and prepared the initial manuscript. CL initiated the project, supervised the analysis, and edited the manuscript. TJ, SZ, NL, and ZC contributed to the interpretation of the results and reviewed the manuscript.
\end{contribution}

\section*{Data Availability}

The MaNGA data products used in this work are publicly available as part of SDSS DR17. The Swift/UVOT+MaNGA value-added catalog is publicly available from the SDSS value-added catalog archive, and the 2MASS imaging data are publicly available from the NASA/IPAC Infrared Science Archive. Derived measurements and analysis products underlying this article are available from Zenodo at \dataset[doi:10.5281/zenodo.22151440]{https://doi.org/10.5281/zenodo.22151440}.

\appendix
\onecolumngrid

\section{Mock test of K-correction for bump strength measurement}\label{sec:appendix_mock_kcorr}

\begin{figure}[!htbp]
\centering
        \includegraphics[width=0.3\textwidth]{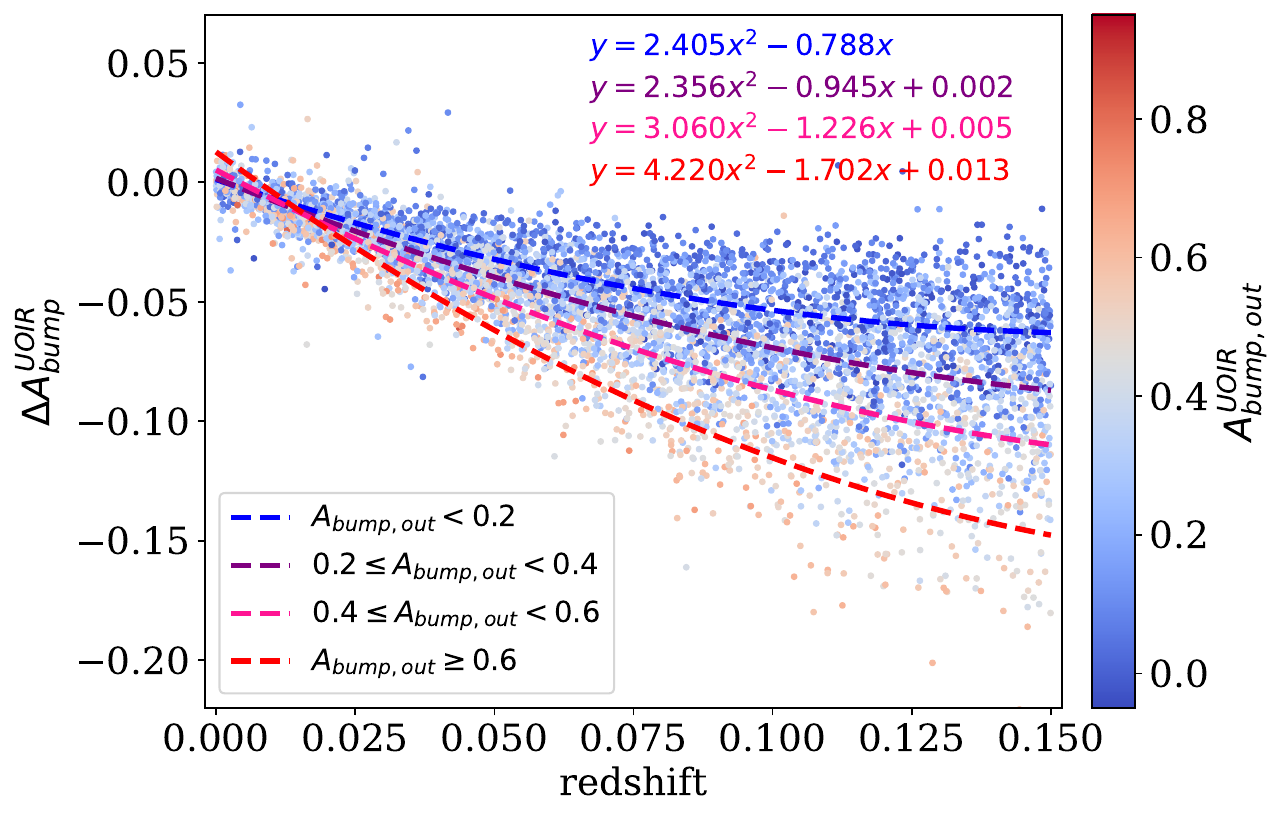}
    \includegraphics[width=0.3\textwidth]{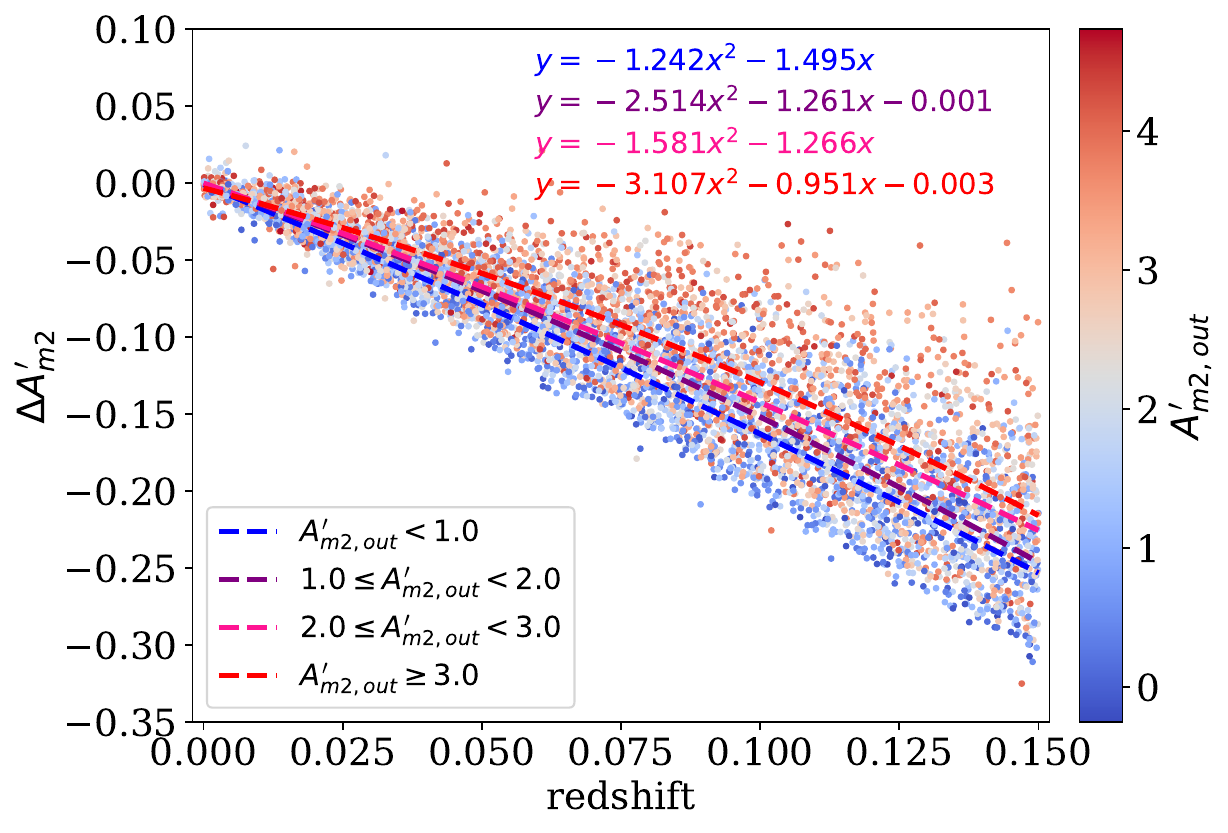}
    \includegraphics[width=0.3\textwidth]{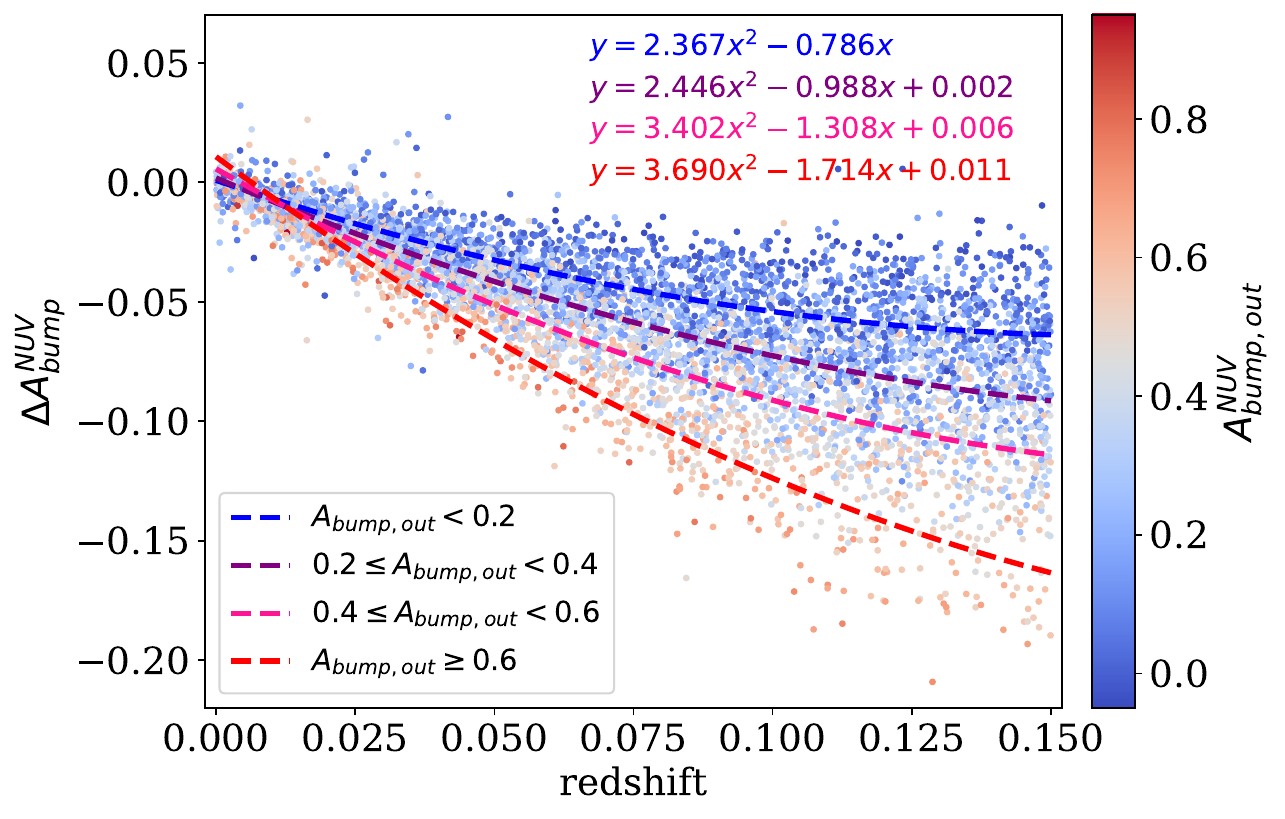}
                \caption{Mock-test K-corrections for the attenuation-curve method (left and middle panels) and the NUV-only method (right panel). The left and right panels show the deviations between output and input absolute bump strengths, $\Delta A_{bump}^{UOIR}$ and $\Delta A_{bump}^{NUV}$, as functions of redshift. The middle panel shows the deviation in the bump-free {\tt uvm2} attenuation, $\Delta A_{m2}^{\prime}$. Points are color-coded by the measured output quantity. In each panel, the mock spectra are divided into four output-value bins, and the redshift dependence in each bin is fit with a quadratic polynomial. The fitted relations are used to correct the observed bump measurements.\label{fig:Abump_Kcorr}}
\end{figure}

To test the bump measurements and derive K-corrections, we construct 6000 mock spectra. The intrinsic spectra are dust-free model spectra selected from different galaxy regions. We add noise corresponding to random S/N values between 3 and 50 and assign redshifts over the range covered by our sample, $0<z<0.15$. We then attenuate each spectrum with a CCM extinction curve \citep{1989ApJ...345..245C} using a range of $E(B-V)$ and $R_V$ values. For each input attenuation curve, we compute the intrinsic $A_{bump}^{UOIR}$ and $A_{m2}^{\prime}$ using the definitions in Section~\ref{sec:methodology}. We then remeasure these quantities after applying the redshift and noise effects, and define the measurement deviations as output minus input. The left and middle panels of \autoref{fig:Abump_Kcorr} show that $\Delta A_{bump}^{UOIR}$ and $\Delta A_{m2}^{\prime}$ depend on both redshift and the measured quantity. We therefore divide the mock sample into four bins of measured $A_{bump}^{UOIR}$ or $A_{m2}^{\prime}$ and fit the redshift dependence in each bin with a quadratic polynomial. We apply the same procedure to $A_{bump}^{NUV}$ for the NUV-only method, as shown in the right panel of \autoref{fig:Abump_Kcorr}. The corrected $A_{bump}^{UOIR}$ and $A_{m2}^{\prime}$ are then used to compute the corrected relative bump strength $B$.

\begin{figure}
\centering
        \includegraphics[width=0.9\textwidth]{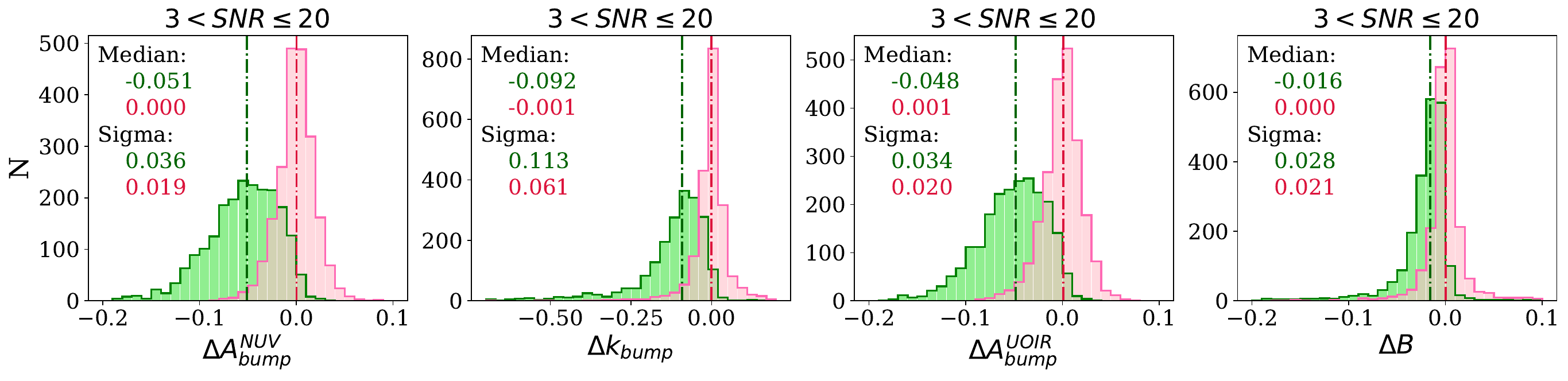}
    \includegraphics[width=0.9\textwidth]{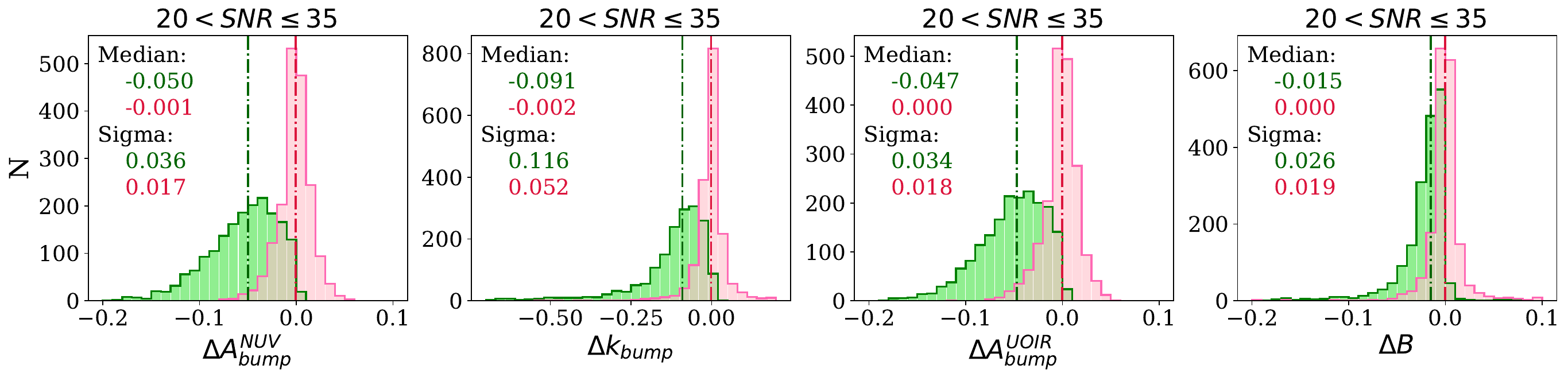}
    \includegraphics[width=0.9\textwidth]{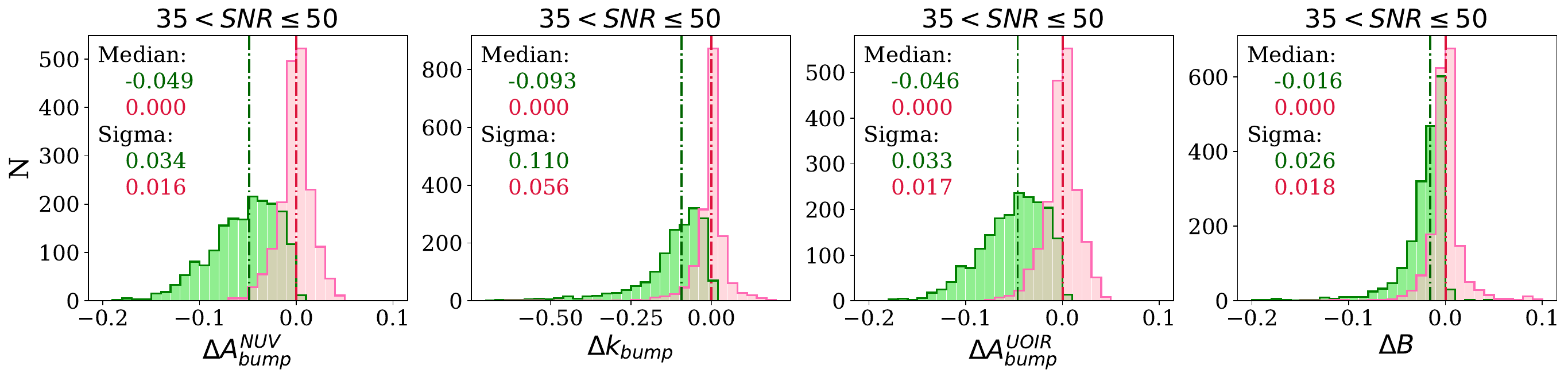}
                \caption{Mock-test deviations before and after K-correction in three S/N bins. Columns show, from left to right, $\Delta A_{bump}^{NUV}$, $\Delta k_{bump}$, $\Delta A_{bump}^{UOIR}$, and $\Delta B$. Rows show increasing S/N bins from top to bottom. Green histograms show output-minus-input deviations before K-correction; pink histograms show the deviations after K-correction. Vertical dashed lines mark the corresponding medians, and the median and scatter values are listed in each panel.\label{fig:Delta_Abump_SNRbin}}
\end{figure}

\autoref{fig:Delta_Abump_SNRbin} shows the residual measurement deviations in three S/N bins for, from left to right, $\Delta A_{bump}^{NUV}$, $\Delta k_{bump}$, $\Delta A_{bump}^{UOIR}$, and $\Delta B$. The green histograms show the deviations before K-correction, and the pink histograms show the deviations after K-correction. The corresponding vertical dashed lines mark the medians, with median and scatter values listed in each panel. After correction, the median deviations are close to zero in all three S/N bins. This test suggests that the photometric and attenuation-curve bump measurements can be corrected reliably for spectra or photometry with S/N $>3$. The test does not include the additional uncertainty associated with deriving the full attenuation curve, so we retain the stricter $SNR_{cont}>20$ requirement for the UOIR attenuation-curve sample.

\section{Additional diagnostic figures}\label{sec:appendix_diagnostic_figures}

This appendix shows the secondary diagnostic panels summarized in Section~\ref{sec:results}. These figures support the conclusion that age and recent-SFH indicators are more predictive of relative bump strength than metallicity, geometry, optical opacity, or optical attenuation-curve slope.

\begin{figure}[!htbp]
\centering
    \includegraphics[width=0.285\textwidth]{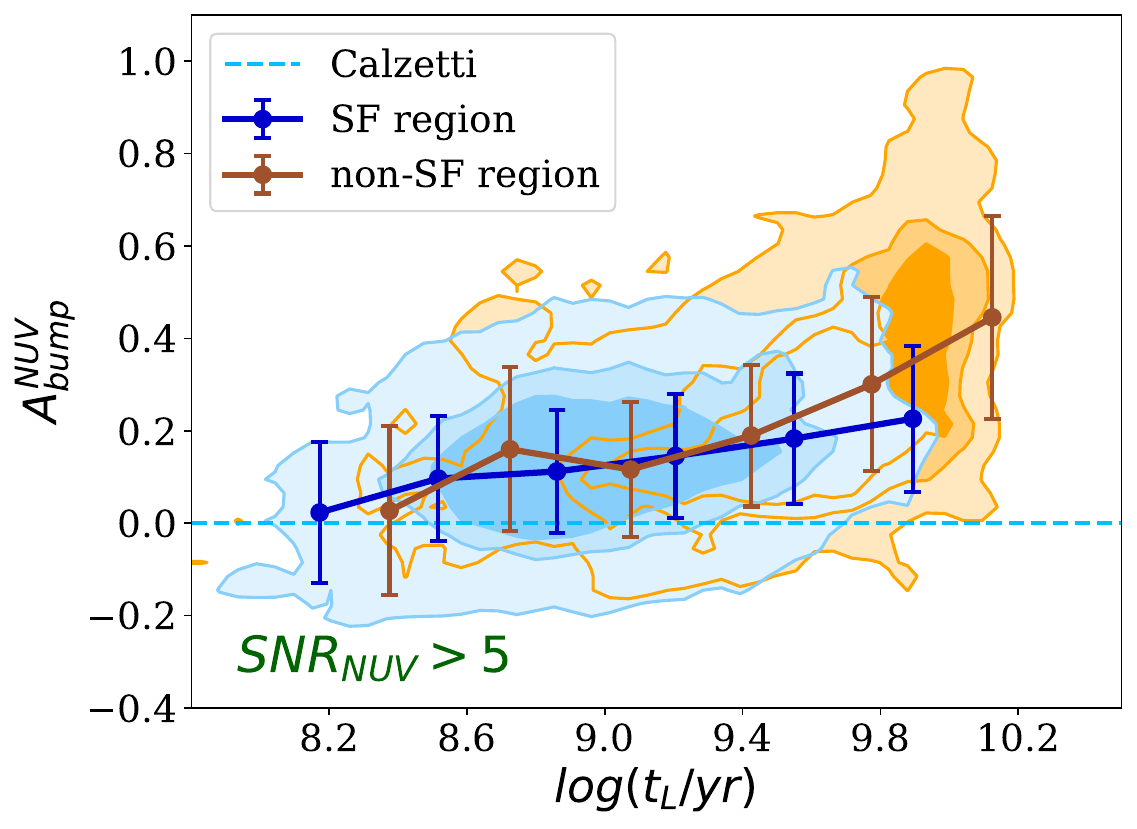}
    \includegraphics[width=0.285\textwidth]{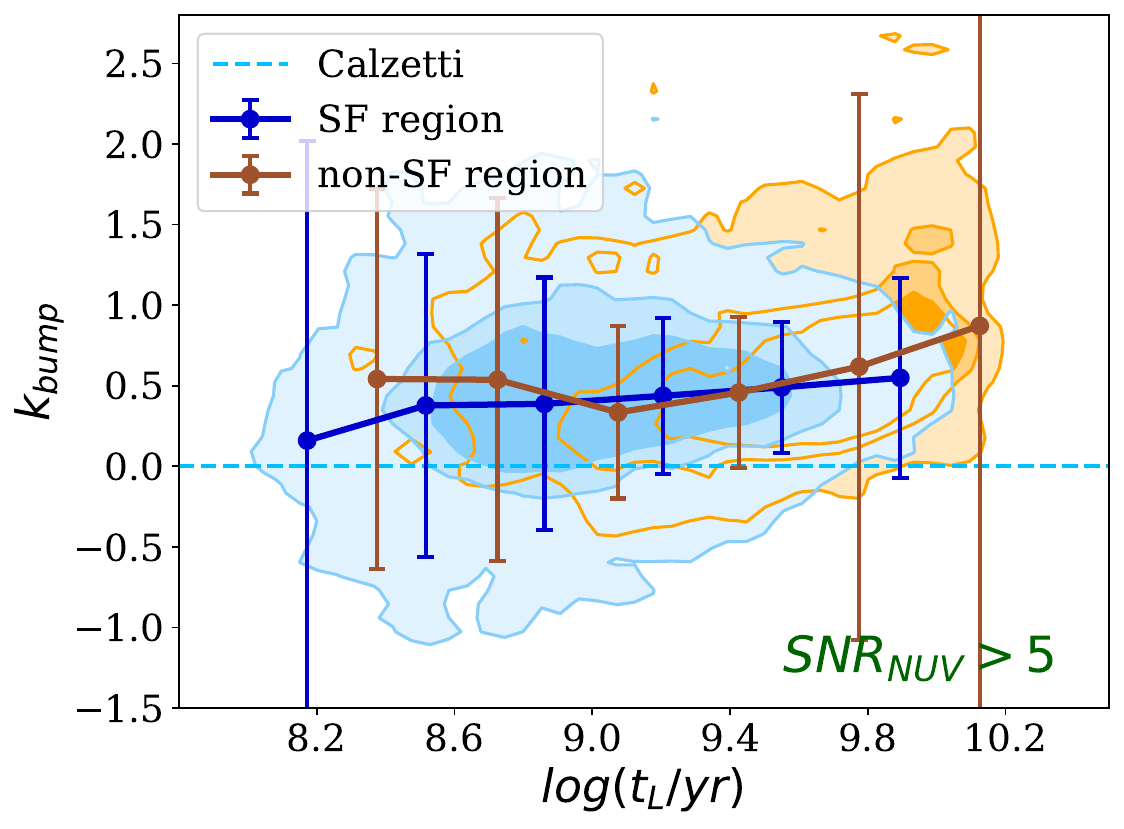}
    \includegraphics[width=0.285\textwidth]{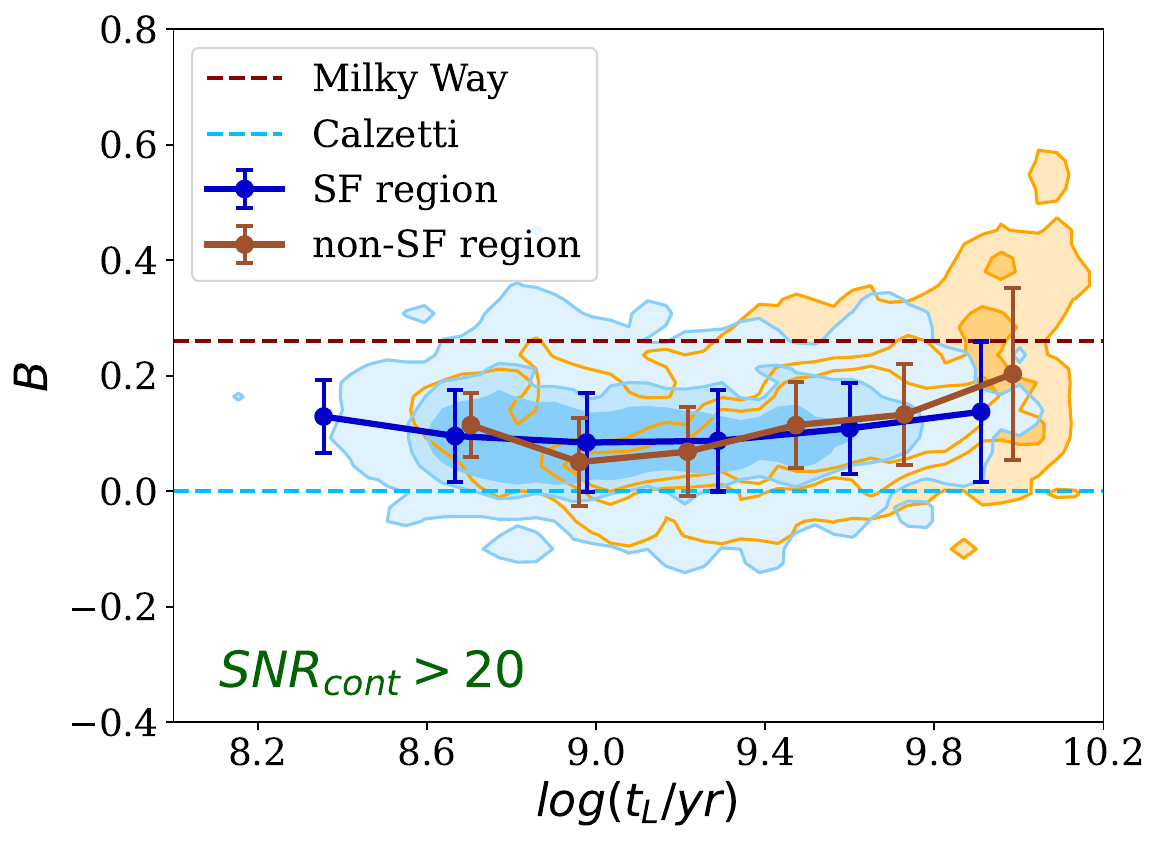}
    \includegraphics[width=0.285\textwidth]{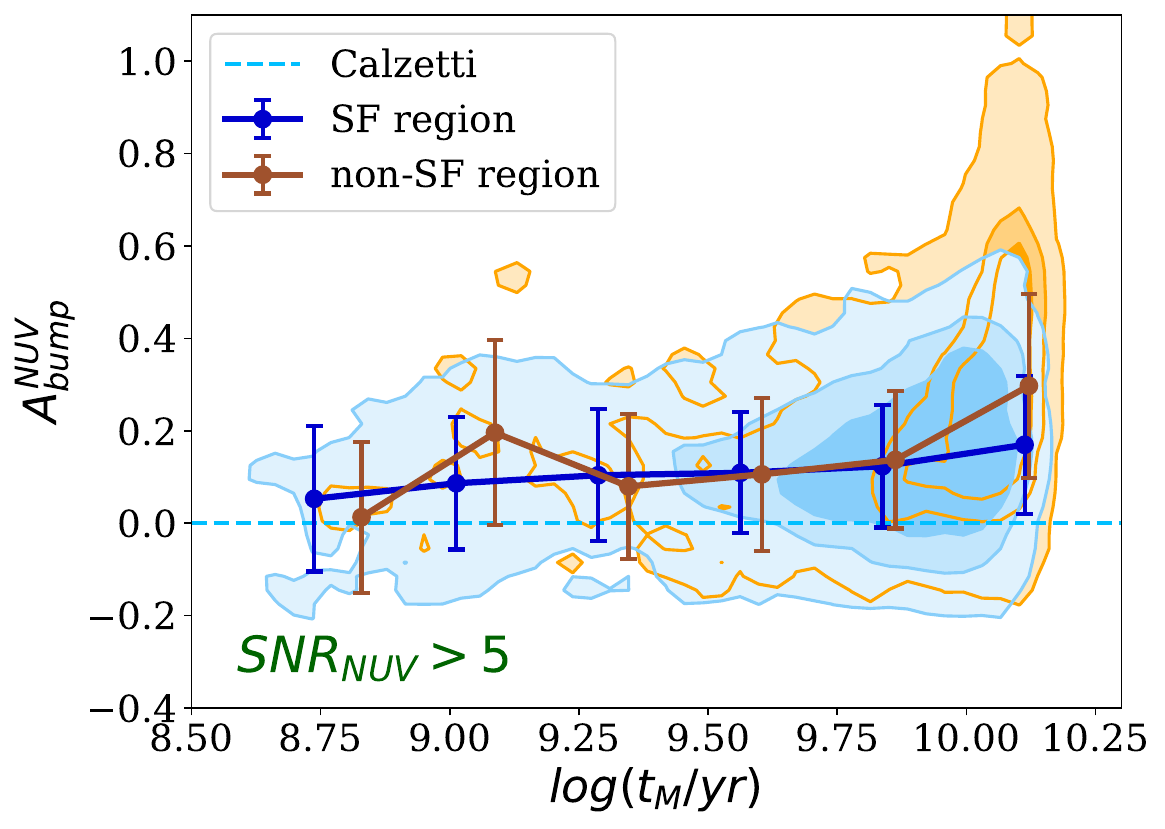}
    \includegraphics[width=0.285\textwidth]{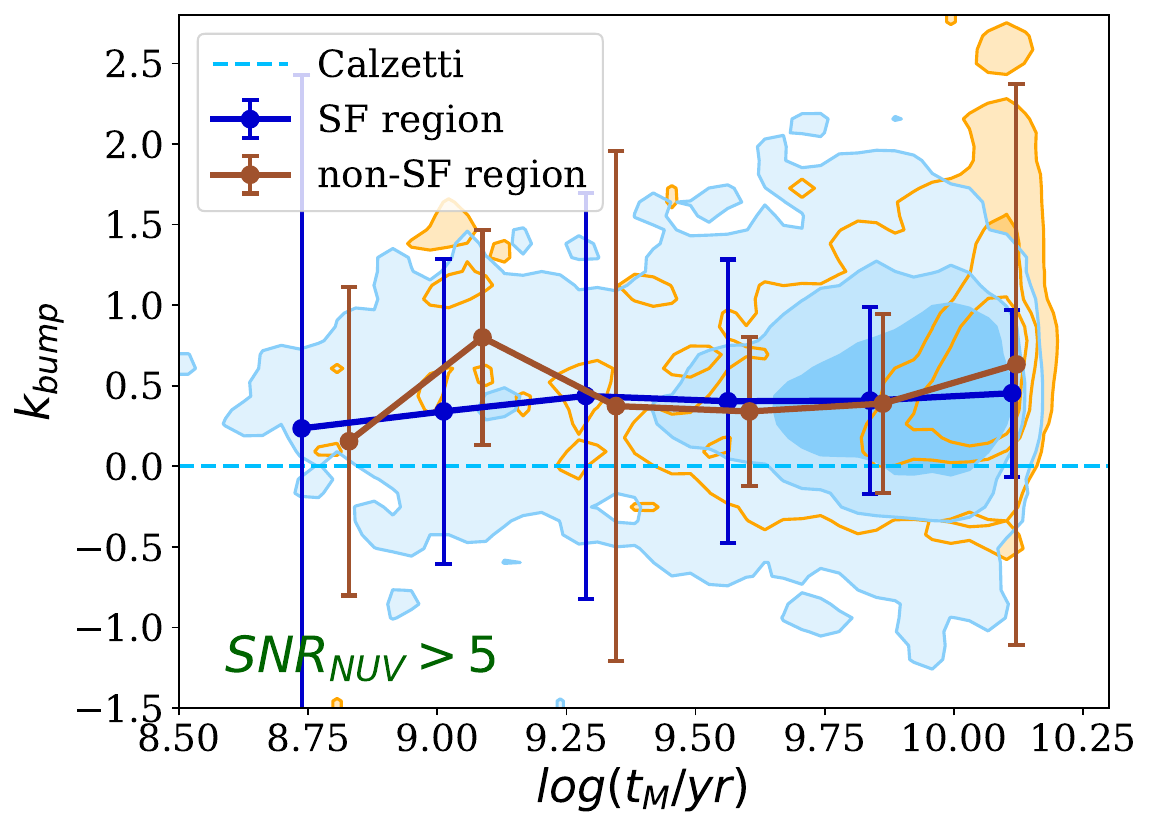}
    \includegraphics[width=0.285\textwidth]{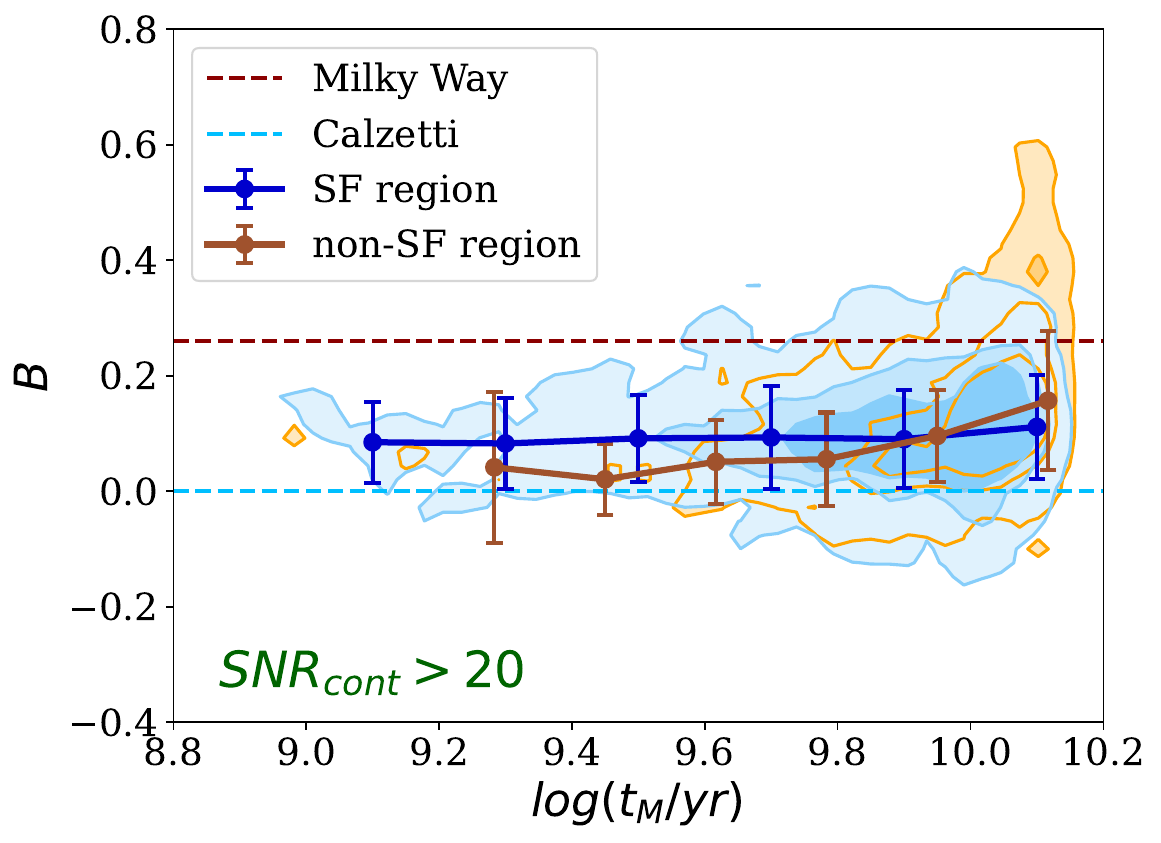}
    \caption{Bump strength versus luminosity-weighted stellar age (top row) and mass-weighted stellar age (bottom row). Columns show $A_{bump}^{NUV}$, $k_{bump}$, and $B$ from left to right. Symbols and colors are as in \autoref{fig:sSFR_bump}.\label{fig:stellar_age_bump}}
\end{figure}

\begin{figure}[!htbp]
\centering
    \includegraphics[width=0.275\textwidth]{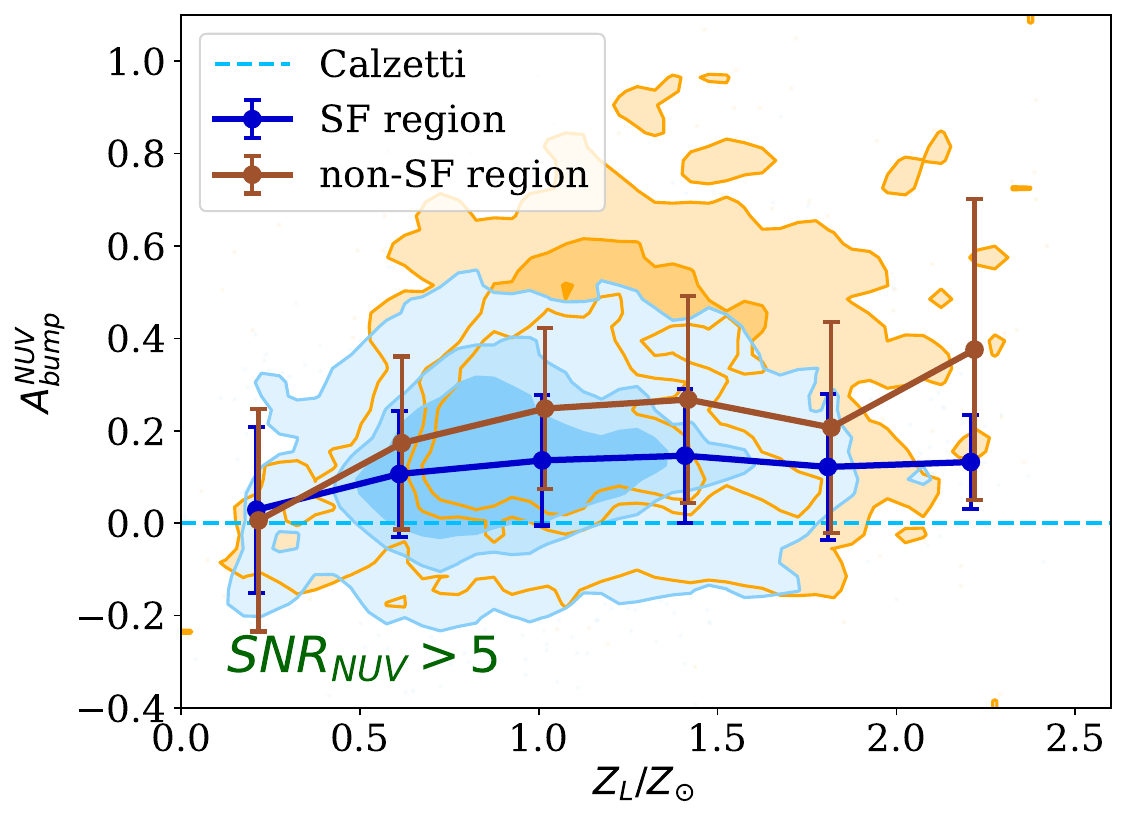}
    \includegraphics[width=0.275\textwidth]{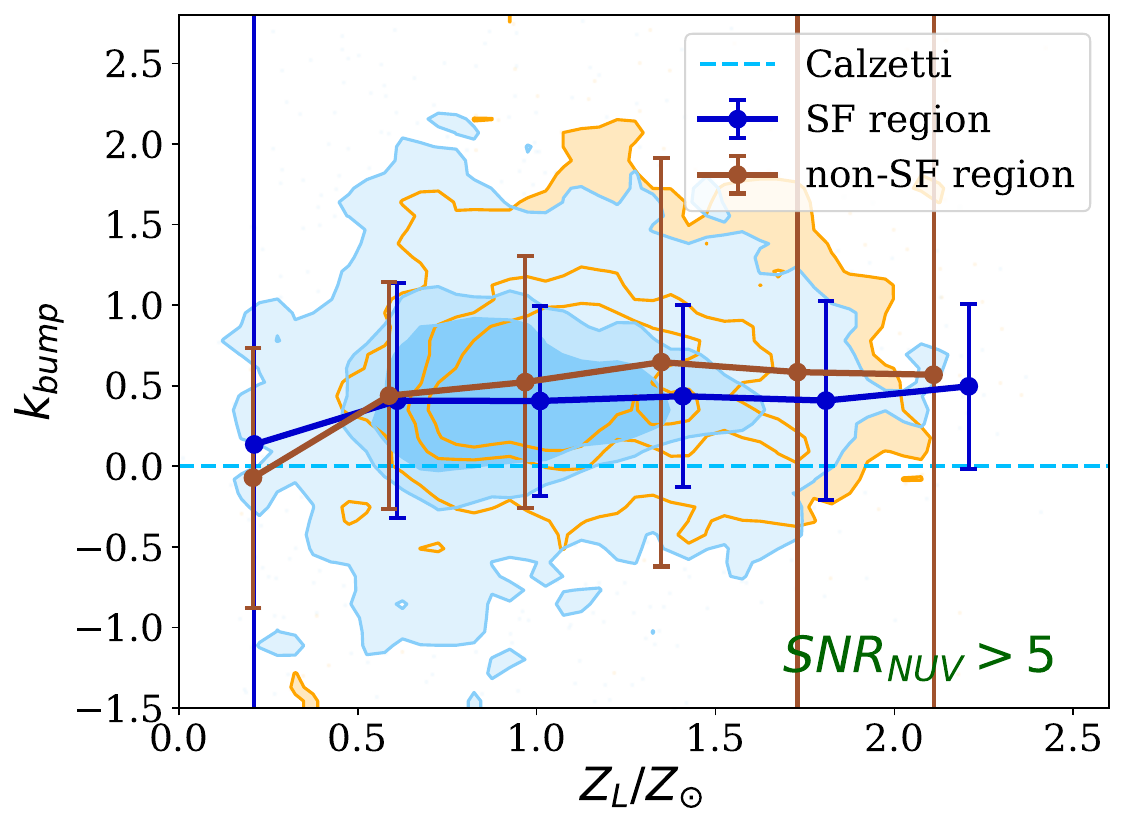}
    \includegraphics[width=0.275\textwidth]{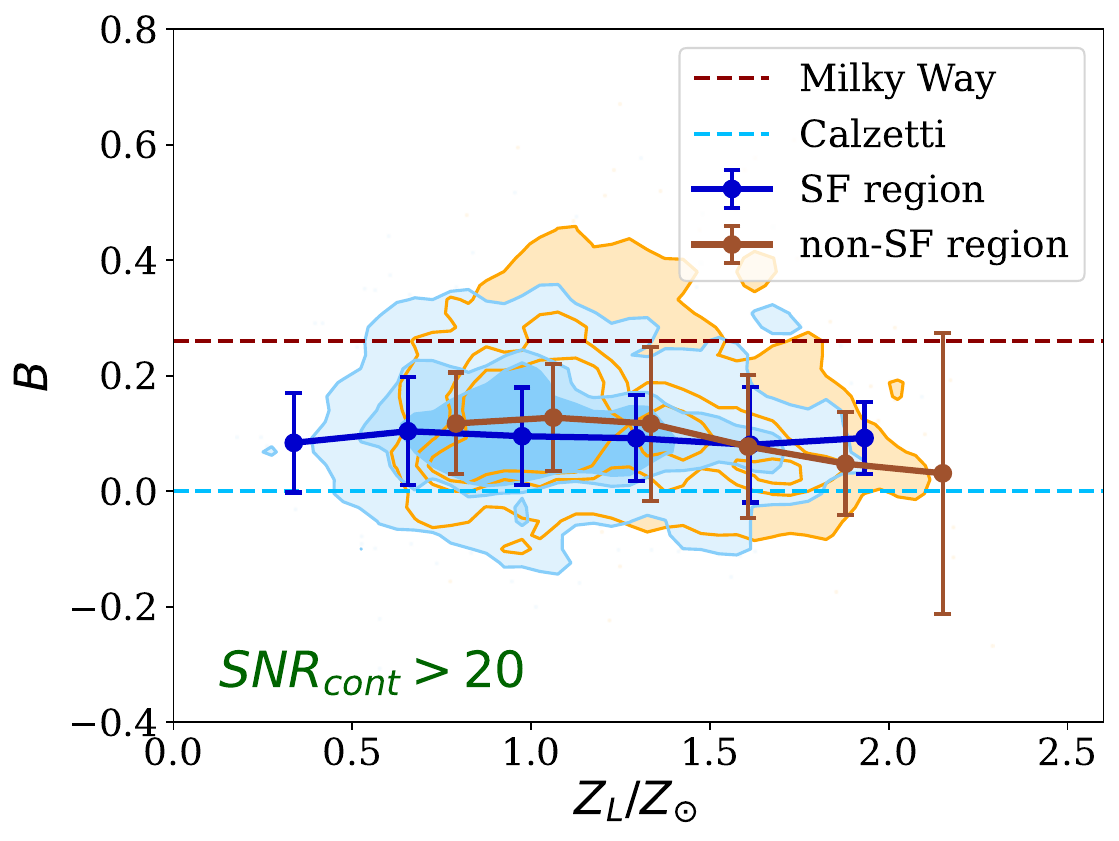}
    \includegraphics[width=0.275\textwidth]{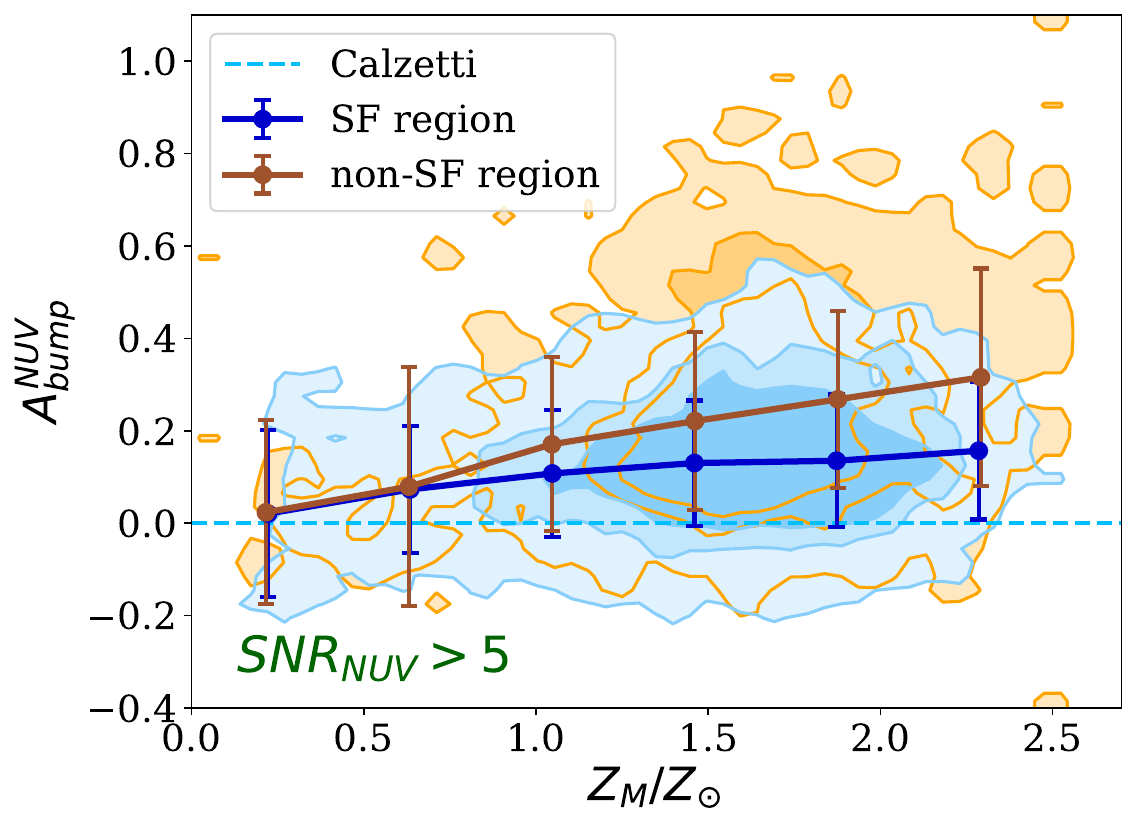}
    \includegraphics[width=0.275\textwidth]{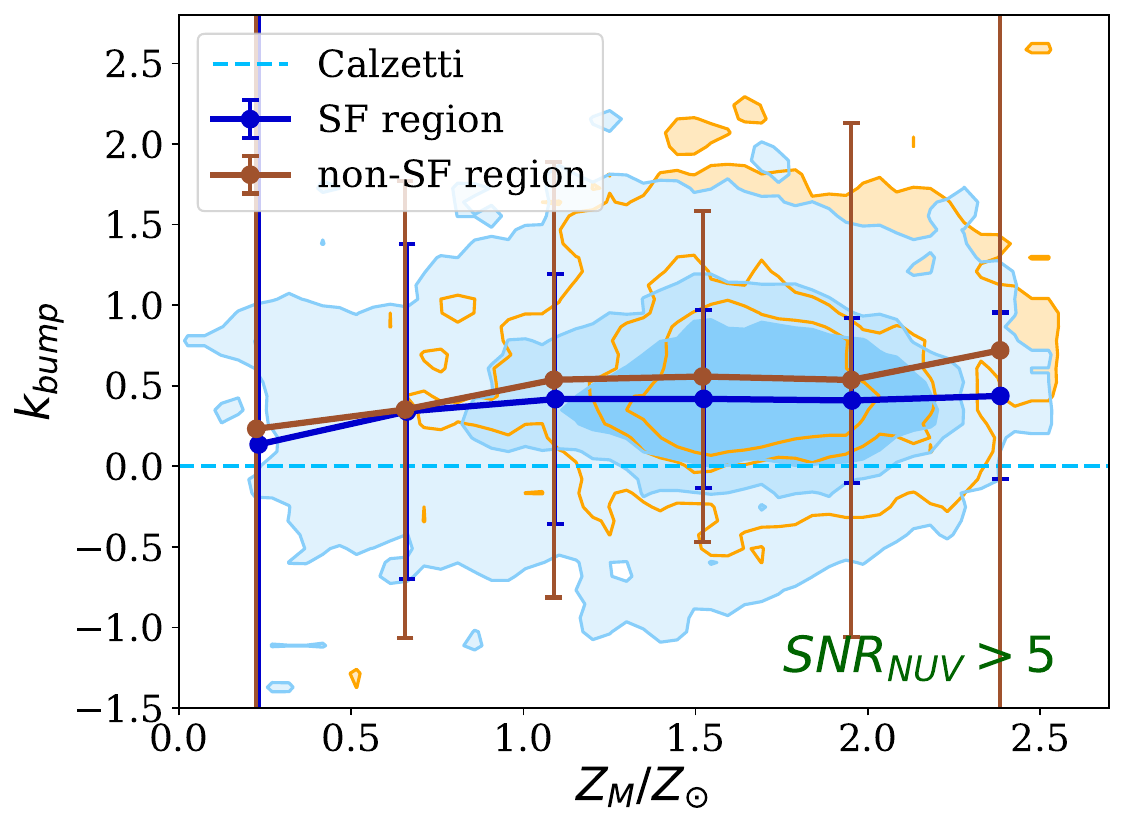}
    \includegraphics[width=0.275\textwidth]{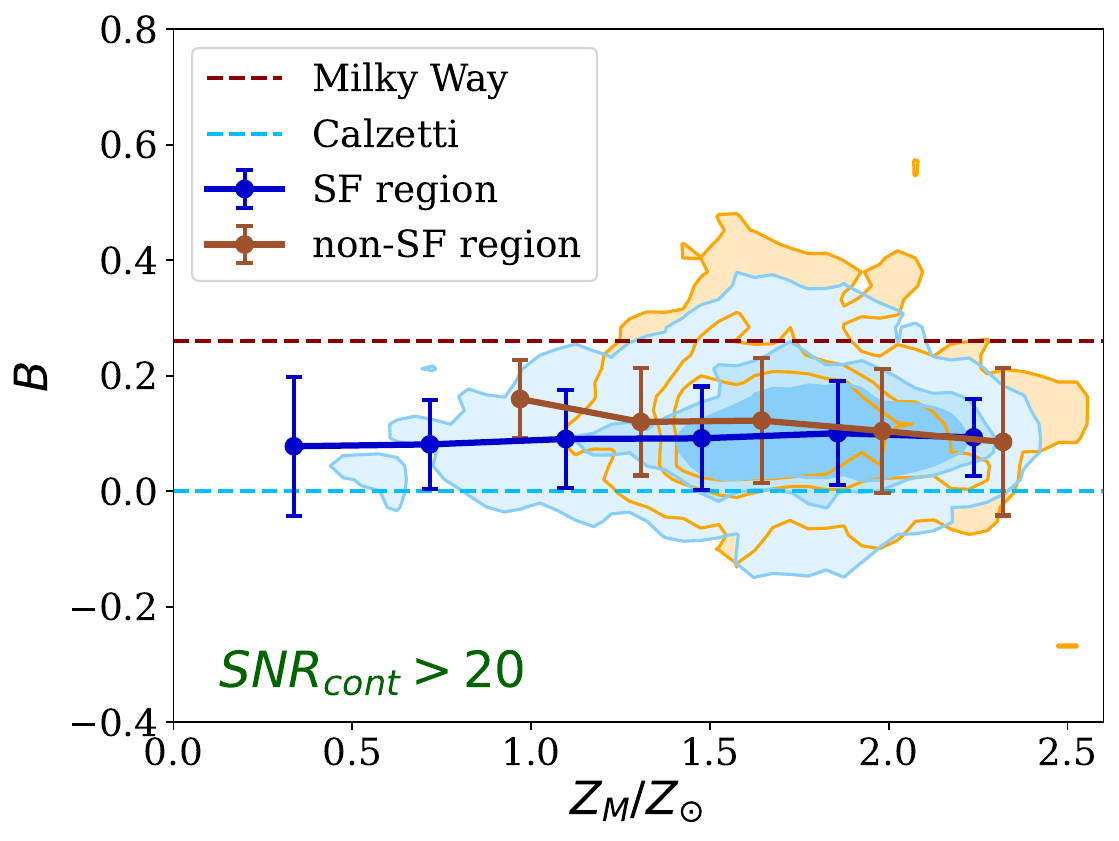}
    \includegraphics[width=0.275\textwidth]{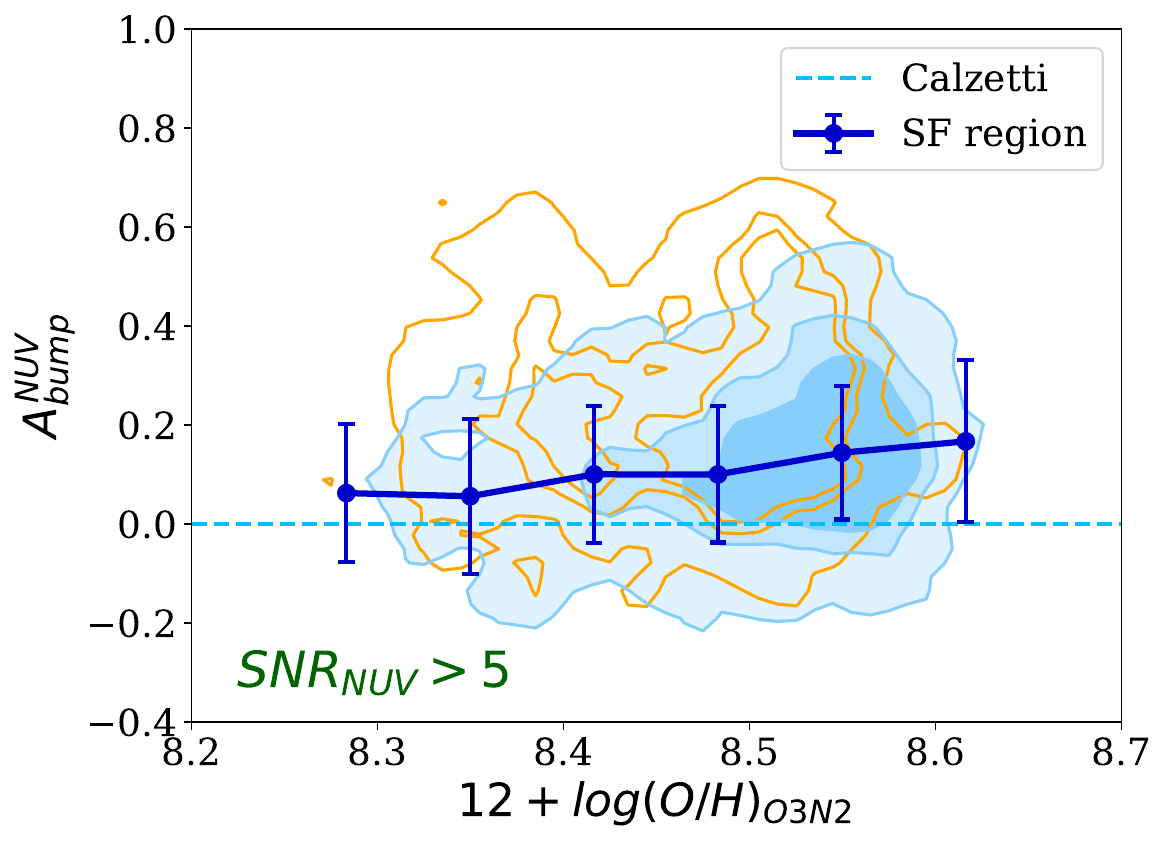}
    \includegraphics[width=0.275\textwidth]{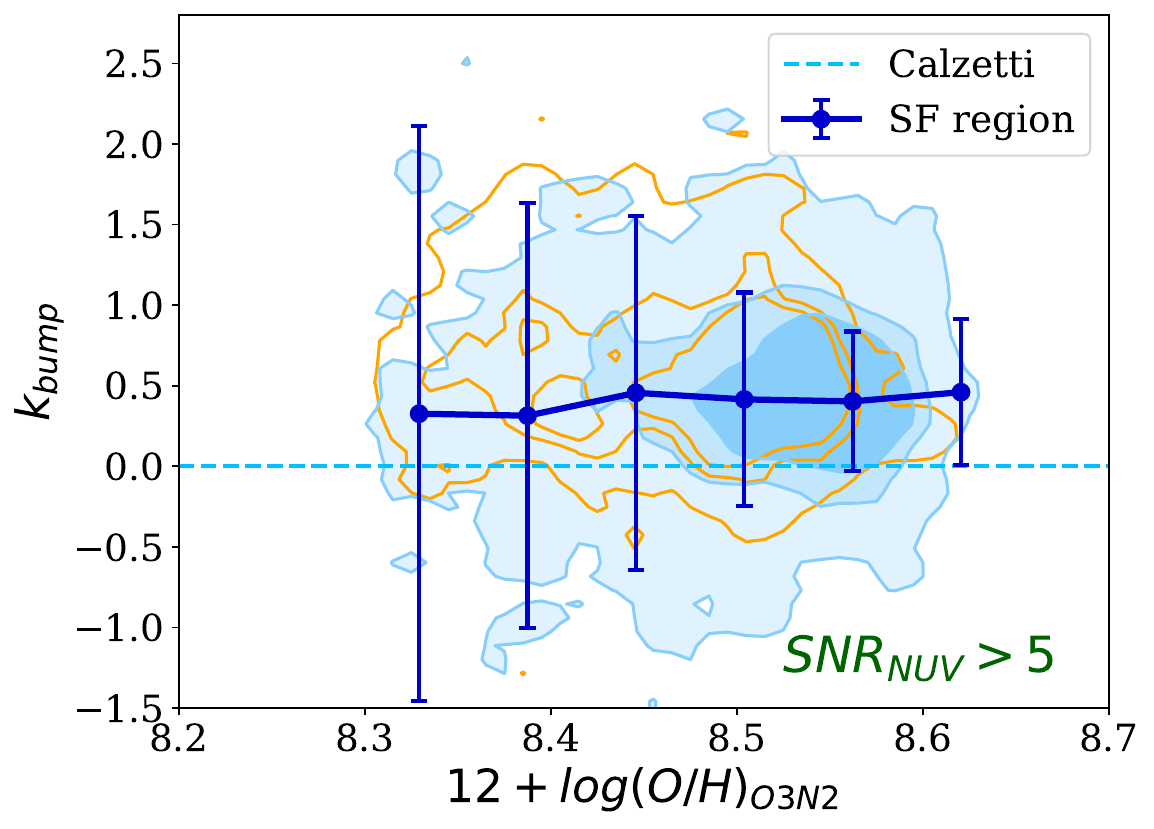}
    \includegraphics[width=0.275\textwidth]{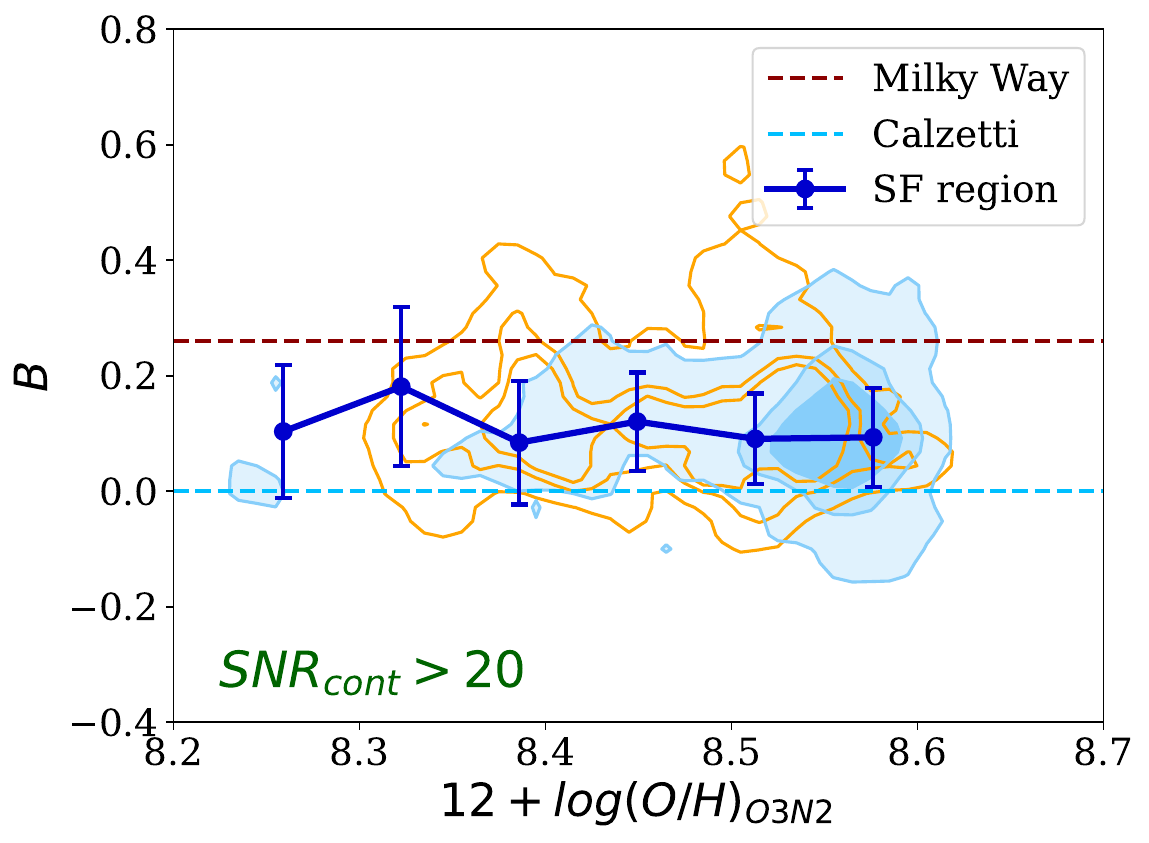}
    \includegraphics[width=0.275\textwidth]{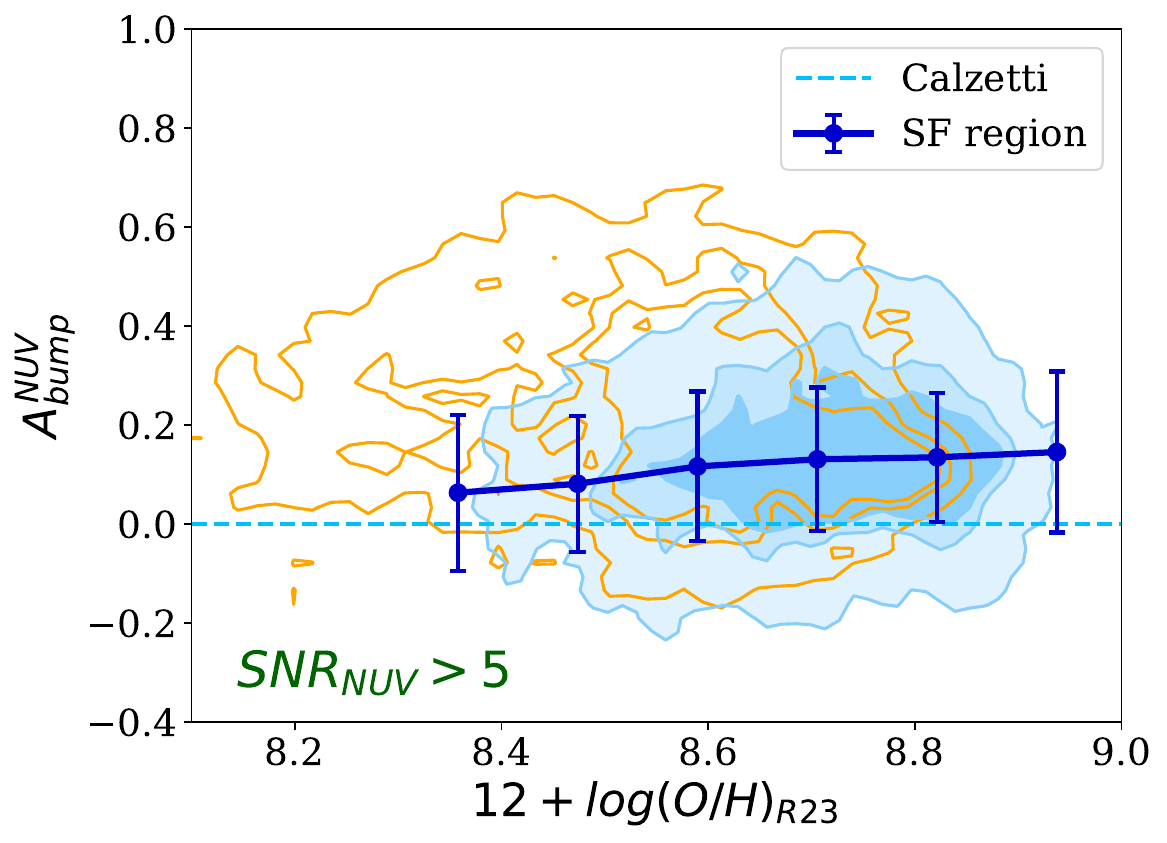}
    \includegraphics[width=0.275\textwidth]{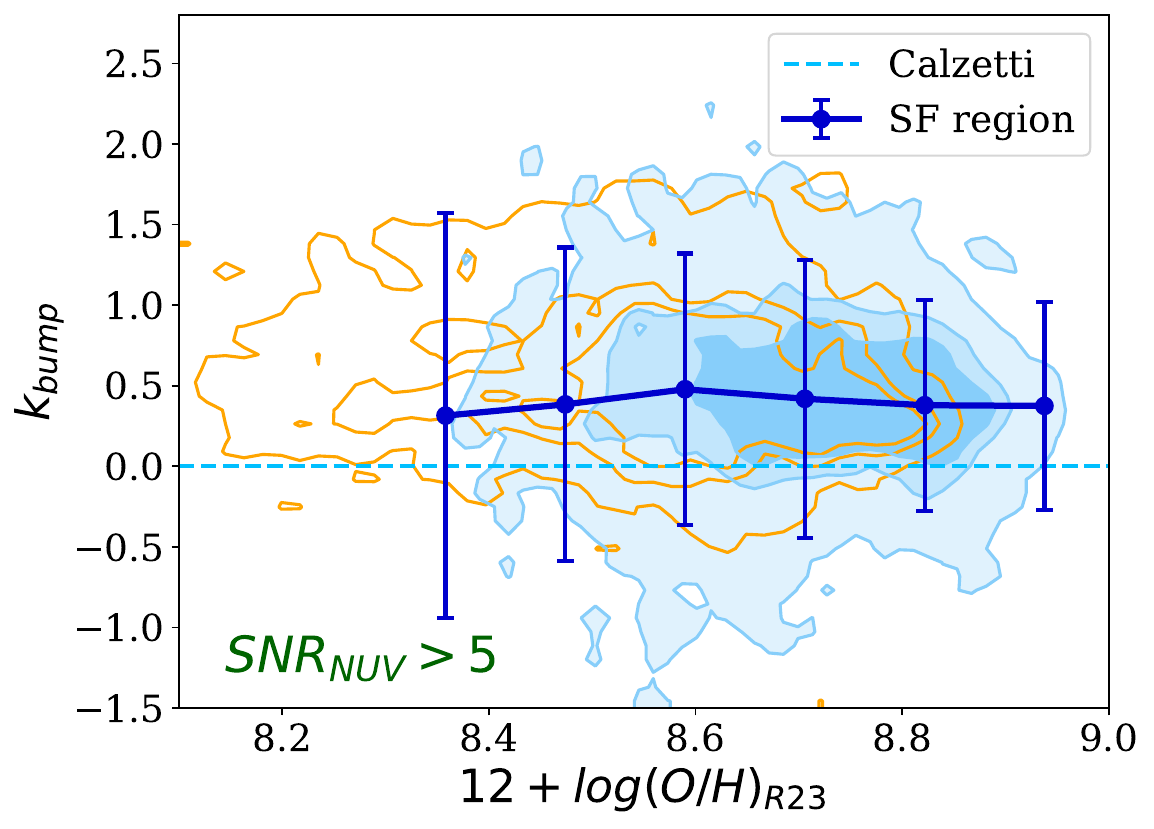}
    \includegraphics[width=0.275\textwidth]{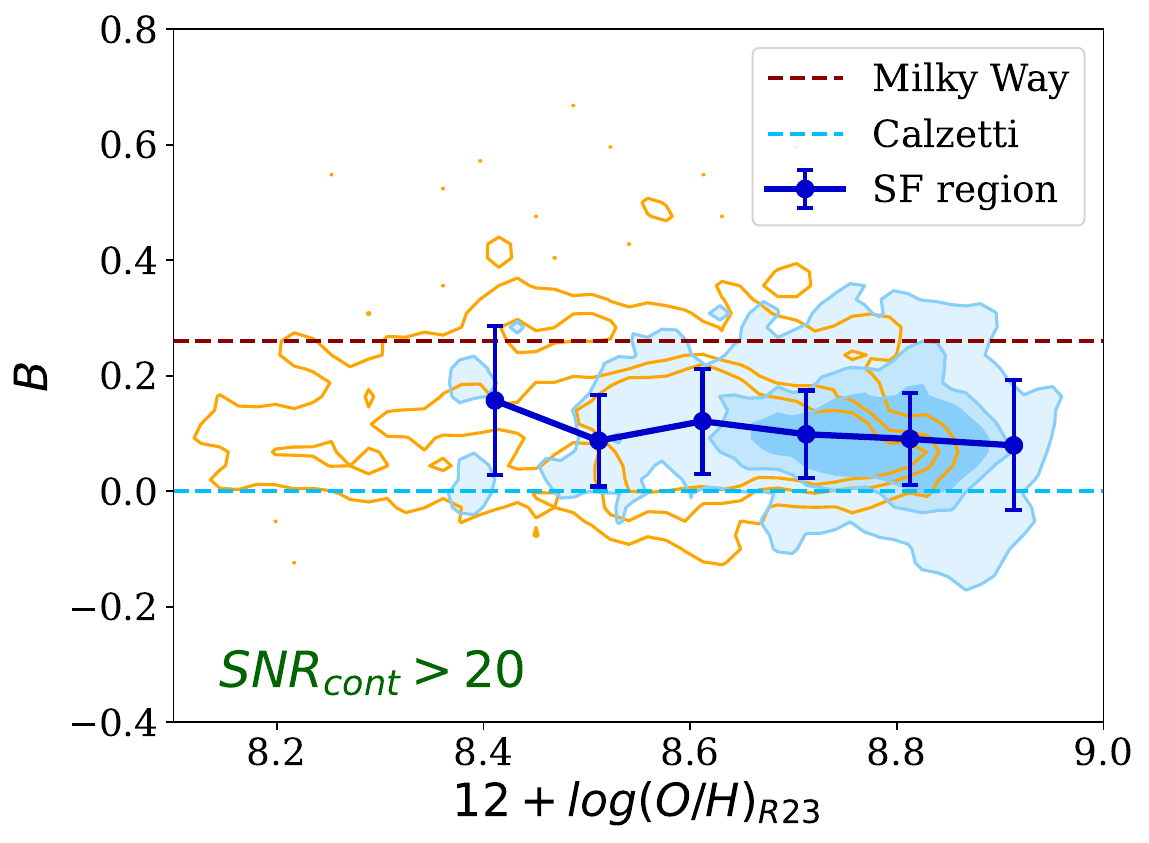}
    \includegraphics[width=0.275\textwidth]{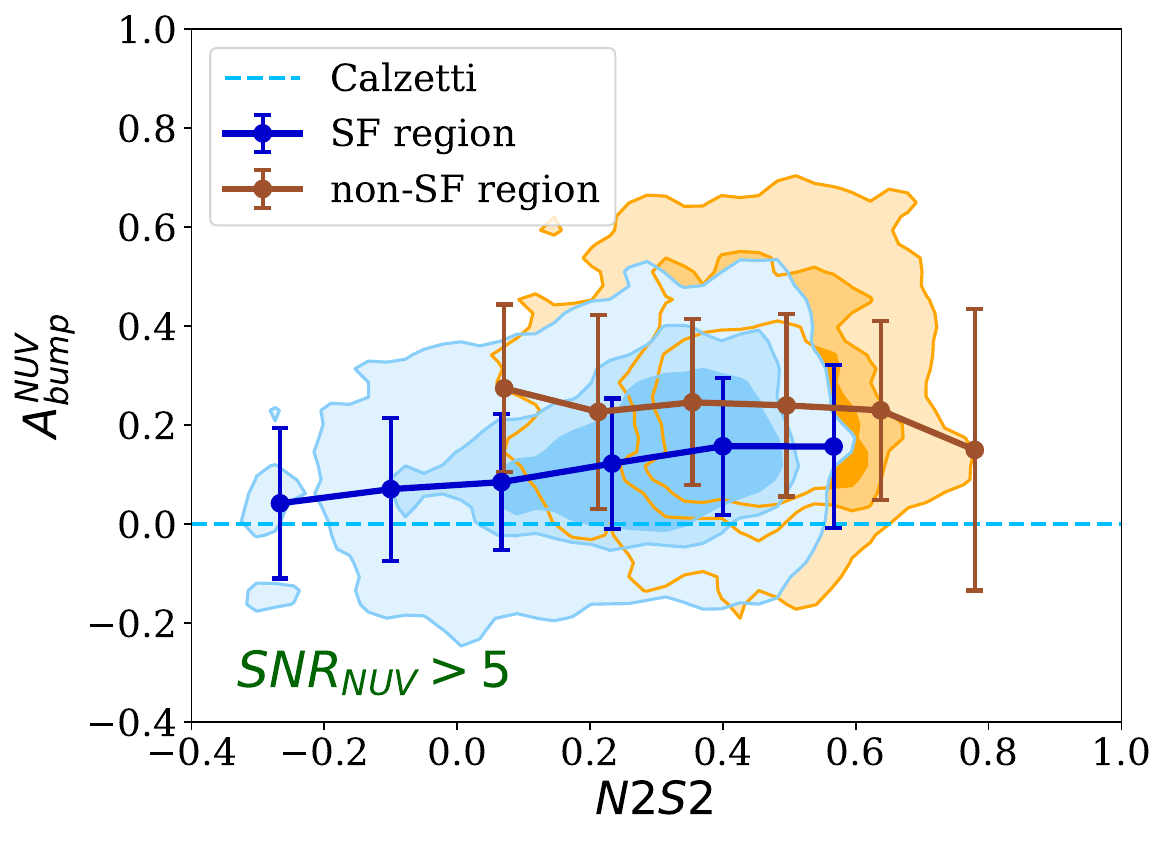}
    \includegraphics[width=0.275\textwidth]{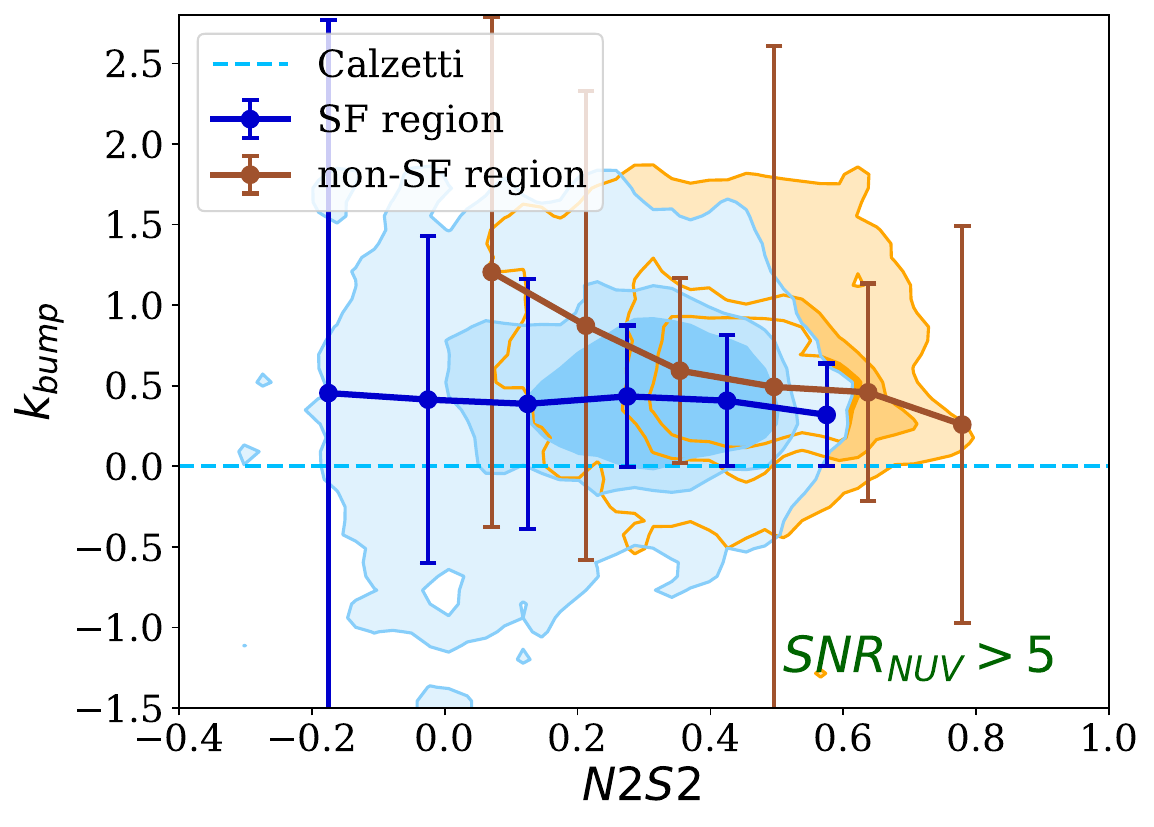}
    \includegraphics[width=0.275\textwidth]{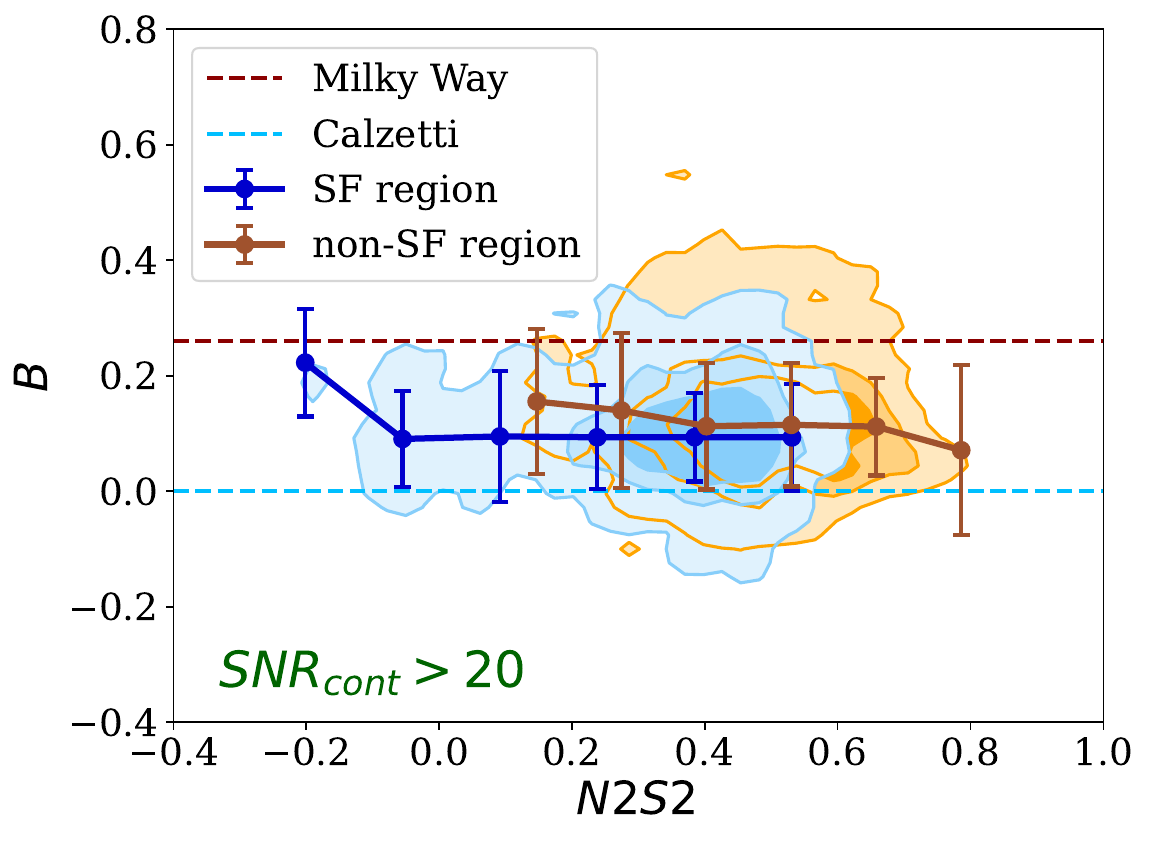}
    \caption{Bump strength versus luminosity-weighted stellar metallicity, mass-weighted stellar metallicity, gas-phase oxygen abundance from O3N2, gas-phase oxygen abundance from R23, and N2S2. Columns show $A_{bump}^{NUV}$, $k_{bump}$, and $B$. For oxygen-abundance panels, median relations are shown only for SF regions, where the calibrations are most reliable.\label{fig:stellar_metal_bump}}
\end{figure}

\begin{figure}
\centering
    \includegraphics[width=0.275\textwidth]{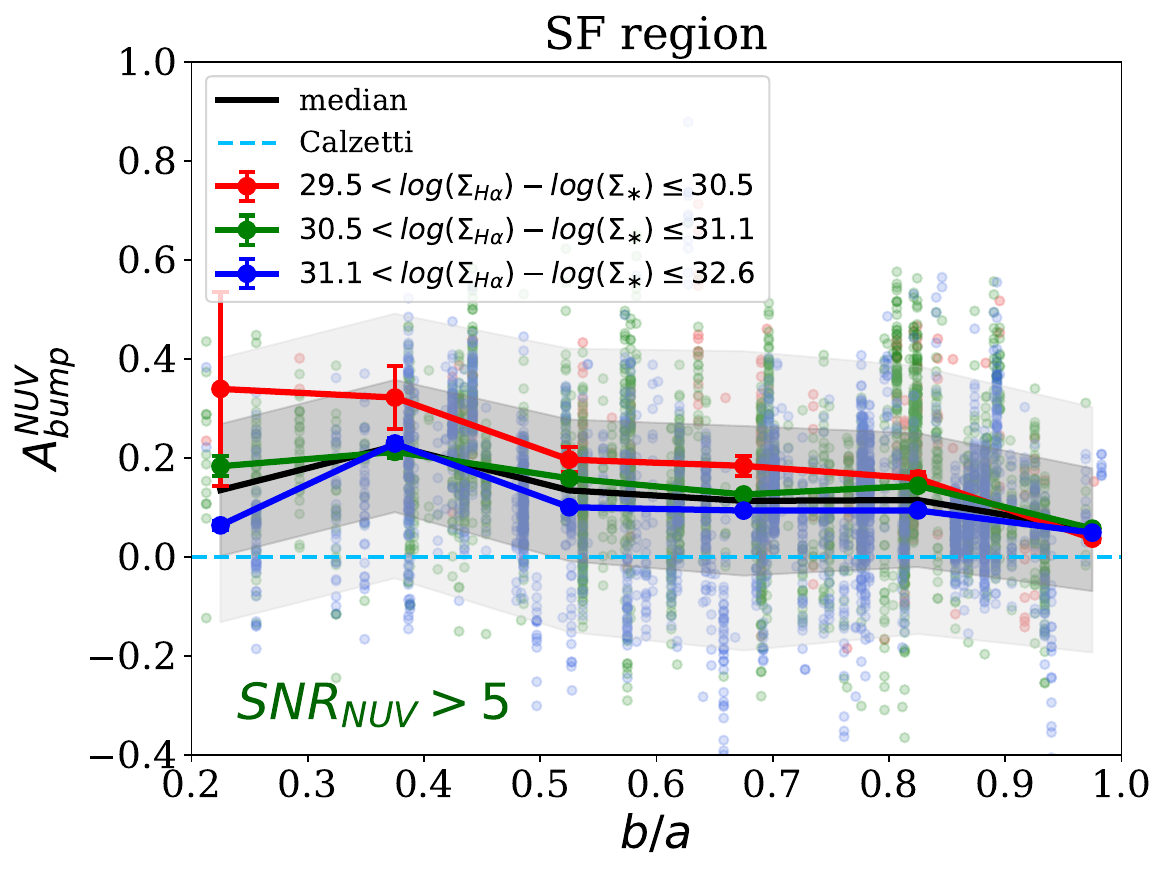}
    \includegraphics[width=0.275\textwidth]{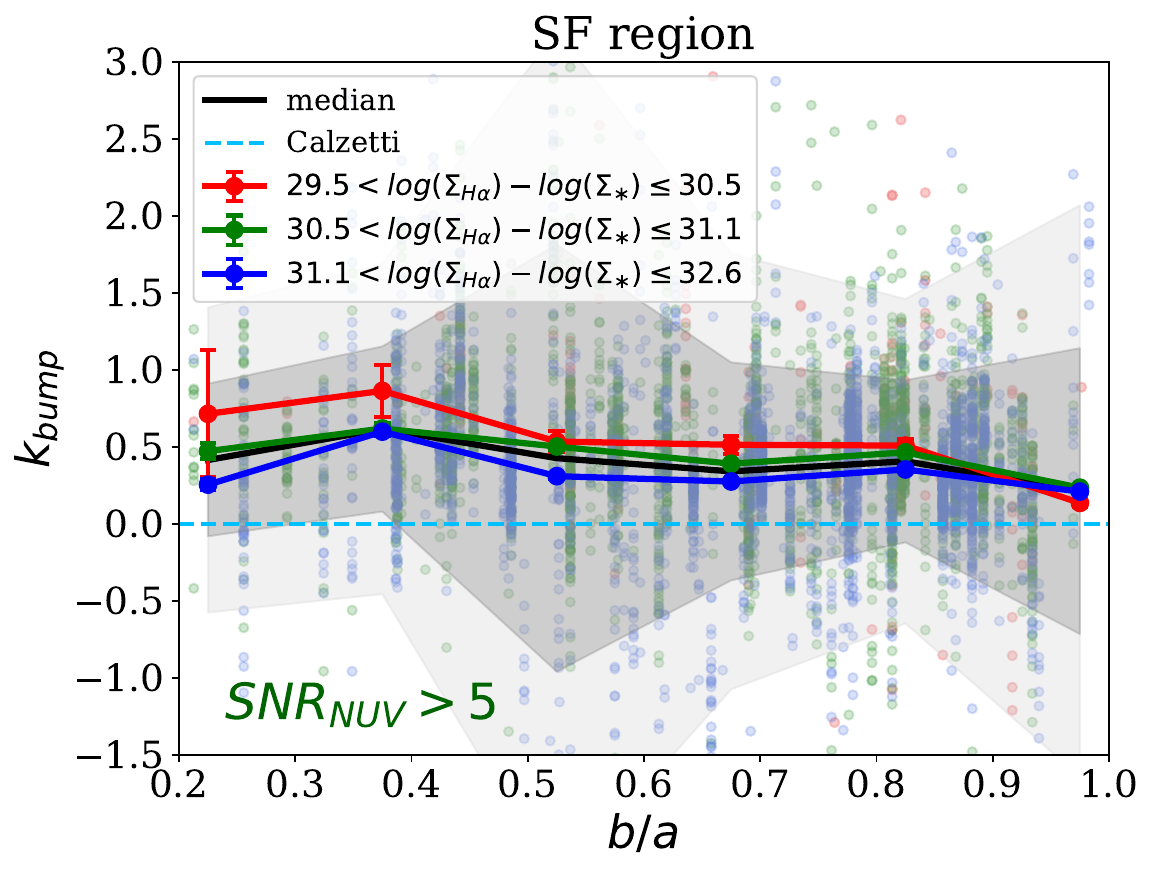}
    \includegraphics[width=0.275\textwidth]{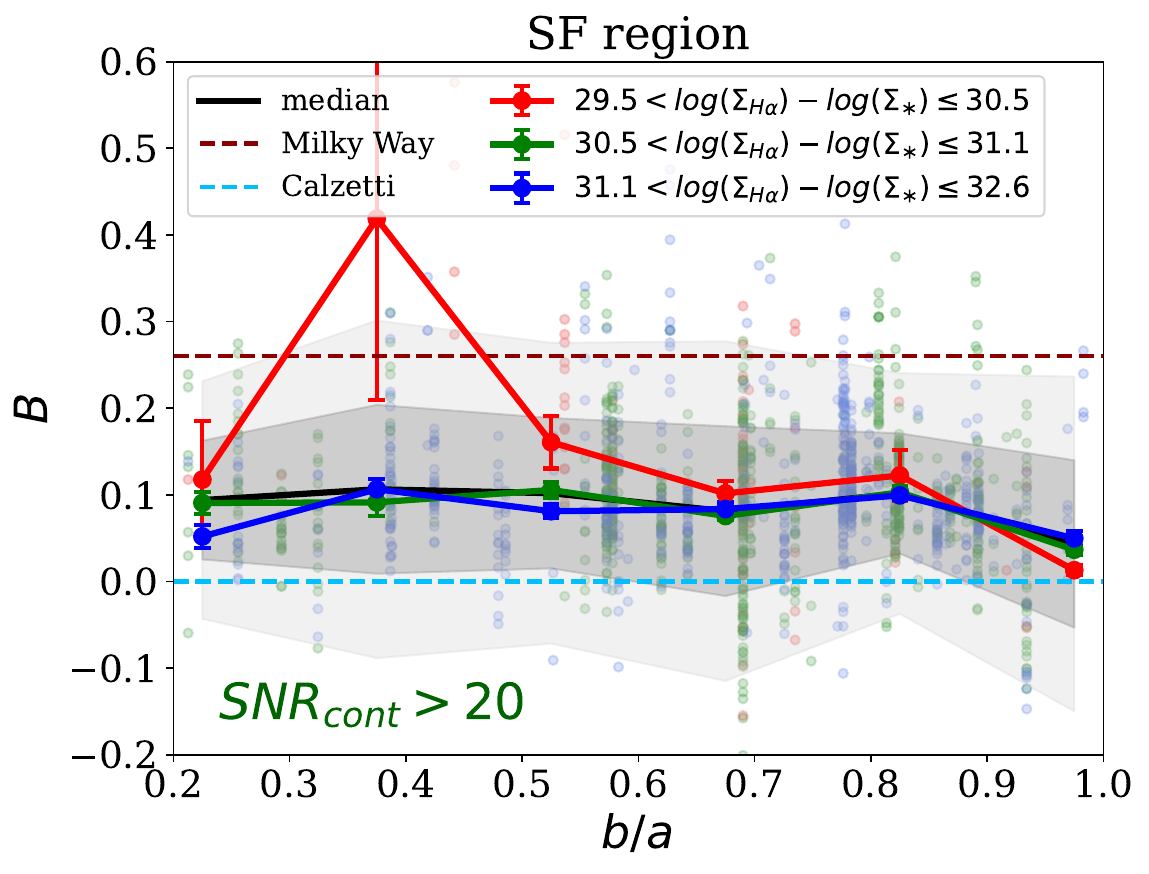}
    \includegraphics[width=0.275\textwidth]{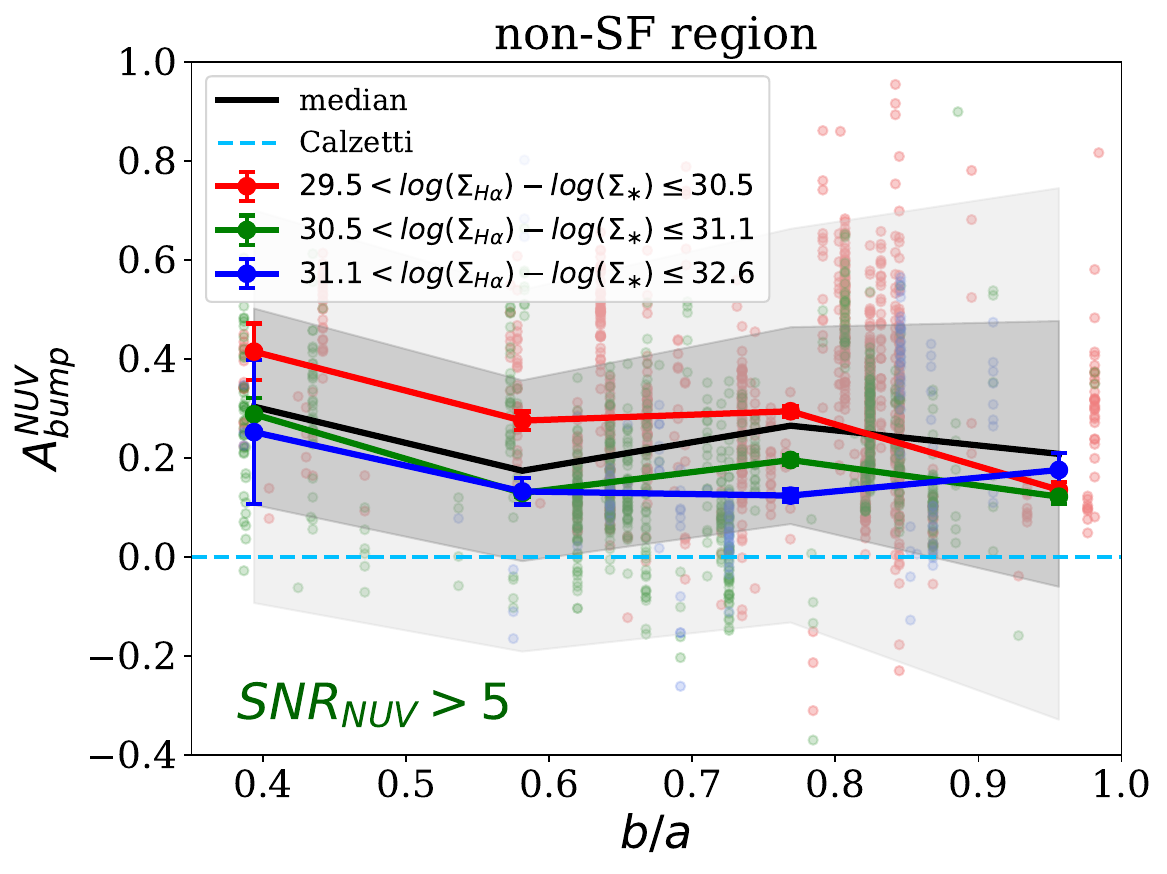}
    \includegraphics[width=0.275\textwidth]{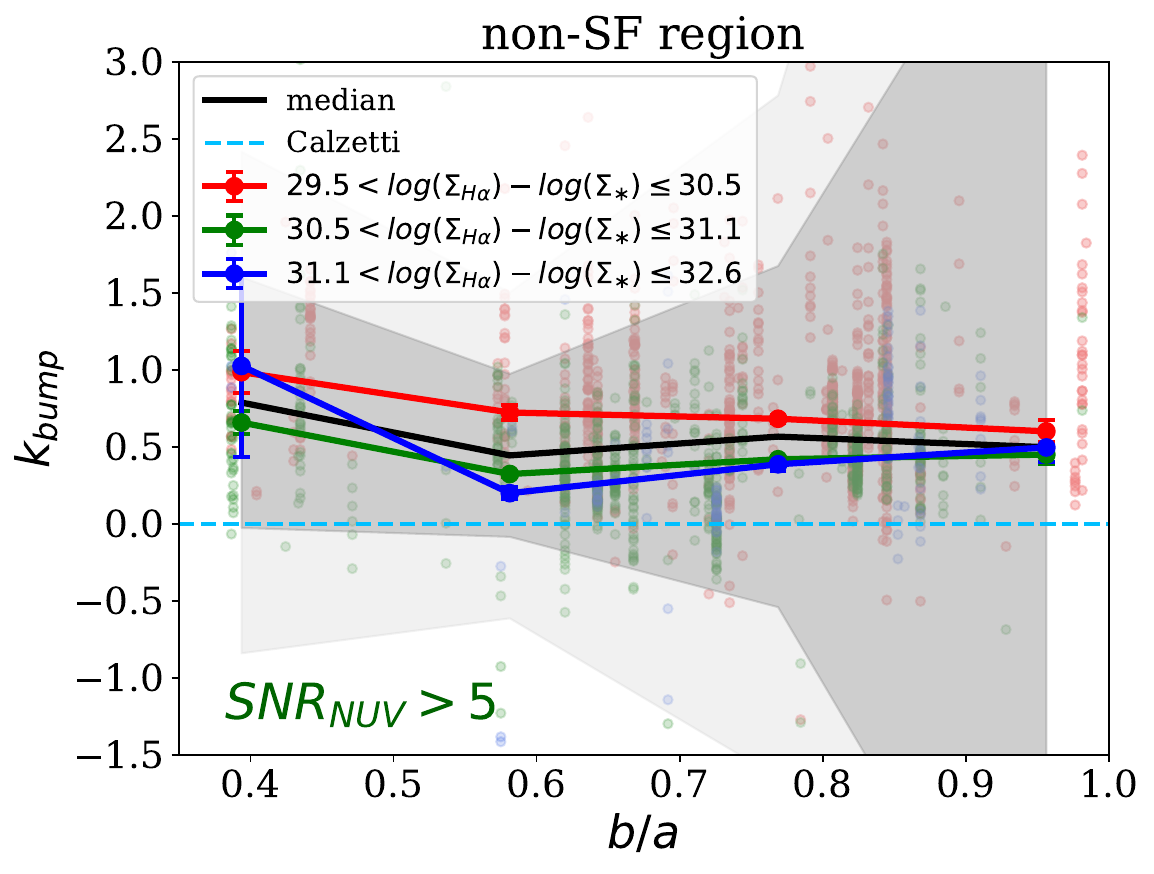}
    \includegraphics[width=0.275\textwidth]{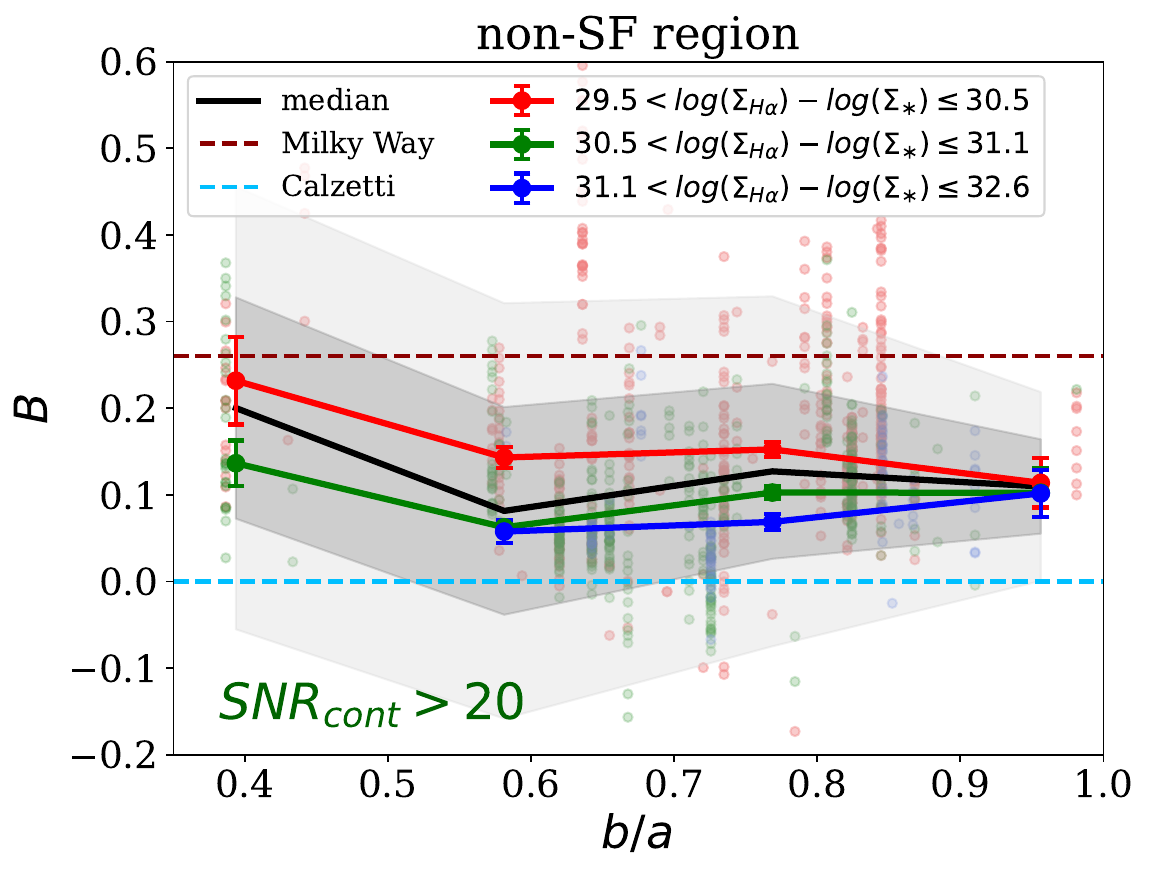}
    \includegraphics[width=0.275\textwidth]{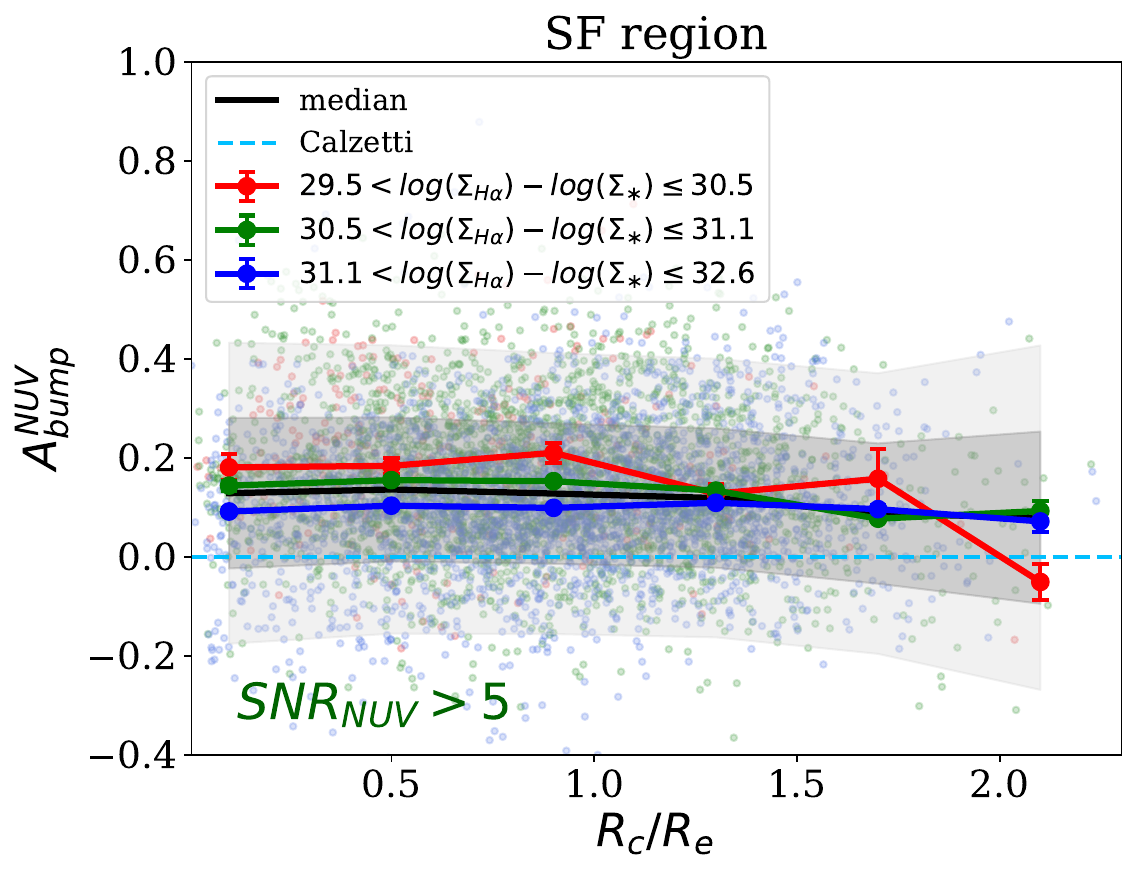}
    \includegraphics[width=0.275\textwidth]{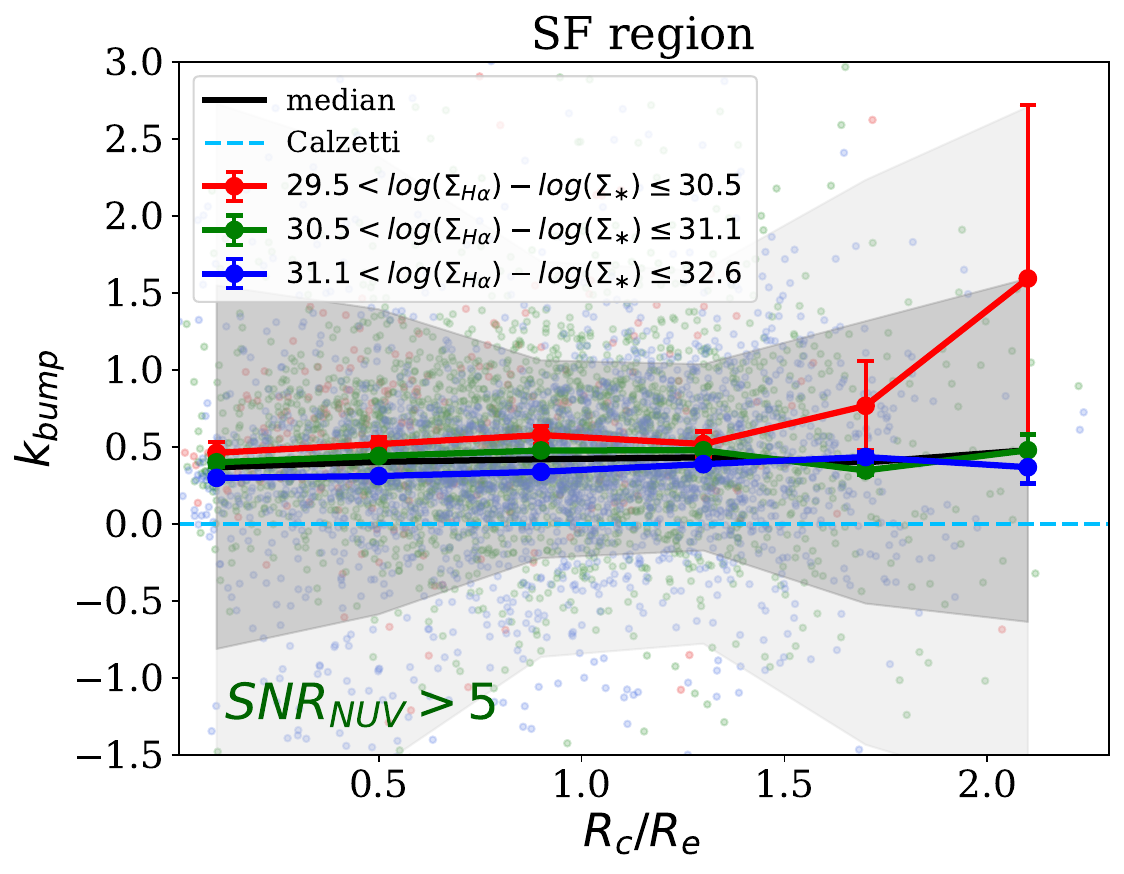}
    \includegraphics[width=0.275\textwidth]{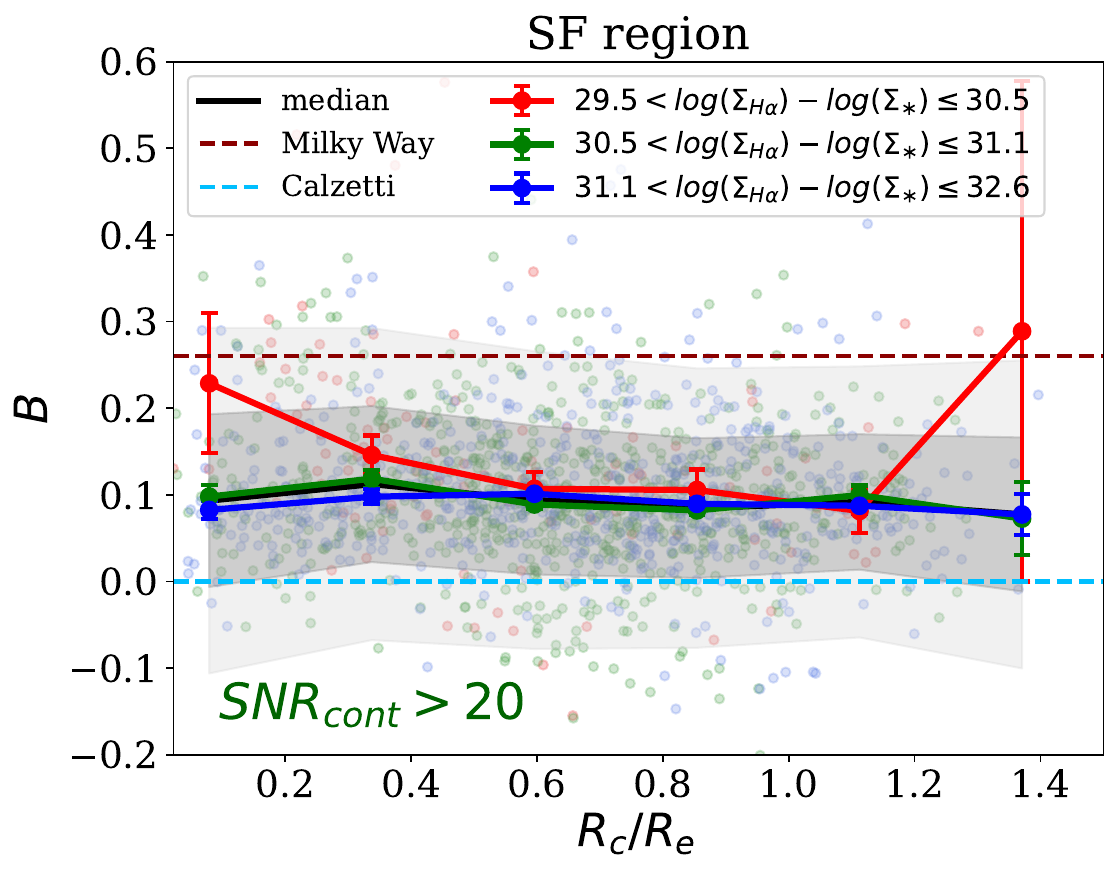}
    \includegraphics[width=0.275\textwidth]{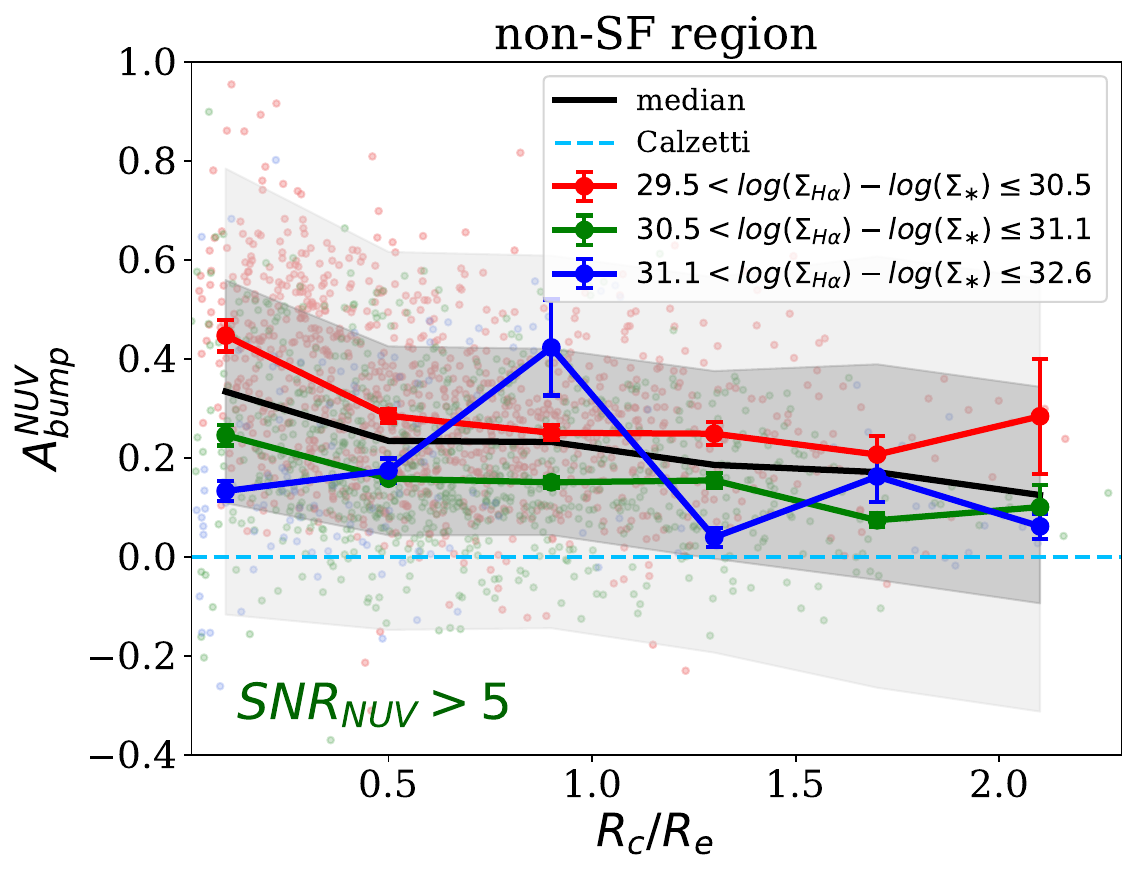}
    \includegraphics[width=0.275\textwidth]{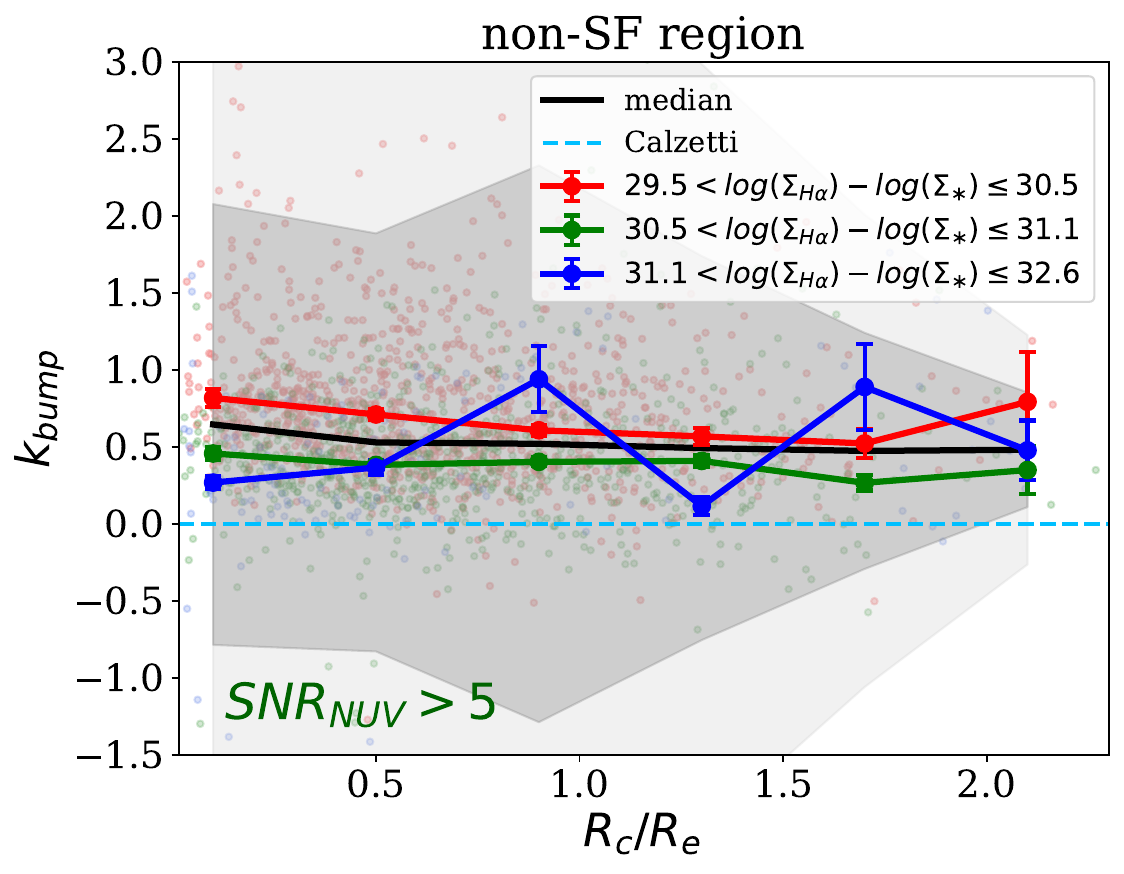}
    \includegraphics[width=0.275\textwidth]{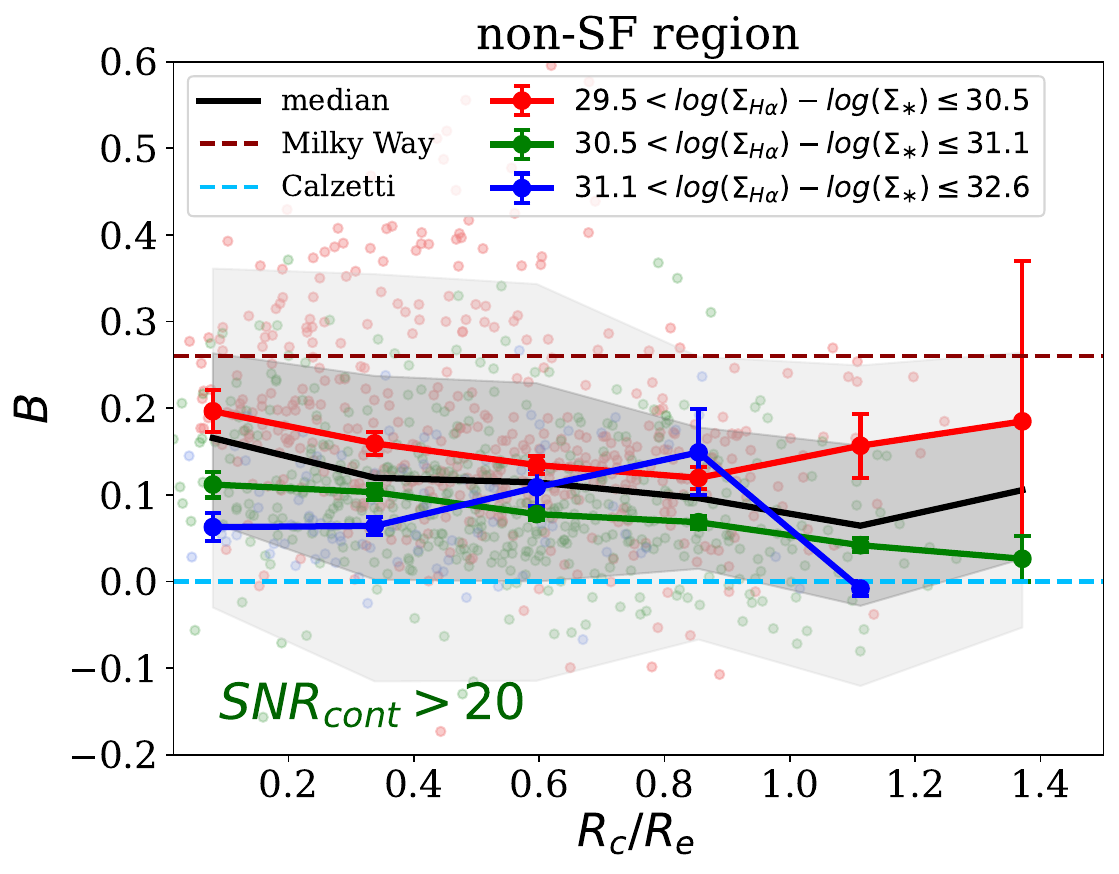}
    \caption{Bump strength versus minor-to-major axis ratio $b/a$ (top two rows) and normalized galactocentric radius $R_c/R_e$ (bottom two rows), shown separately for SF and non-SF regions. Columns show $A_{bump}^{NUV}$, $k_{bump}$, and $B$. Colored medians mark bins of $\log_{10}(\Sigma_{\text{H}\alpha}/\Sigma_\ast)$.\label{fig:geometry_bump}}
\end{figure}

\begin{figure}
\centering
    \includegraphics[width=0.3\textwidth]{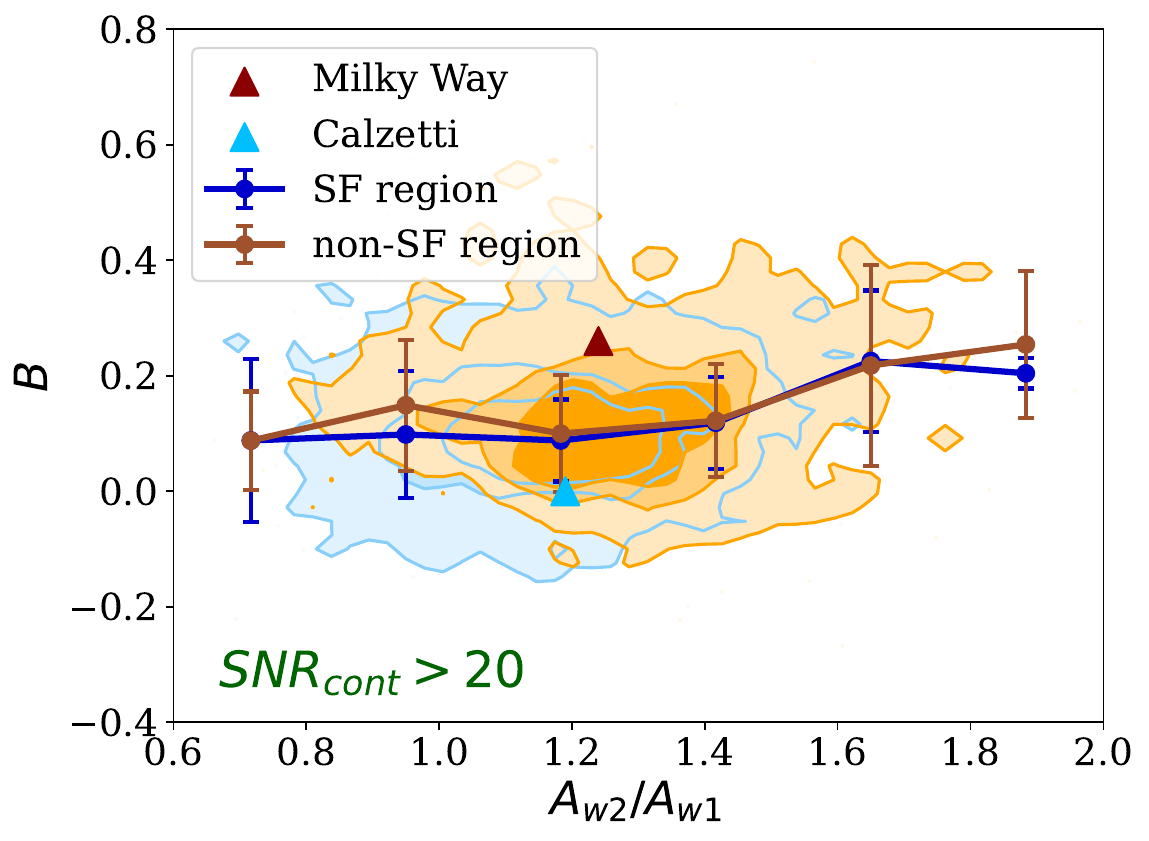}
    \includegraphics[width=0.3\textwidth]{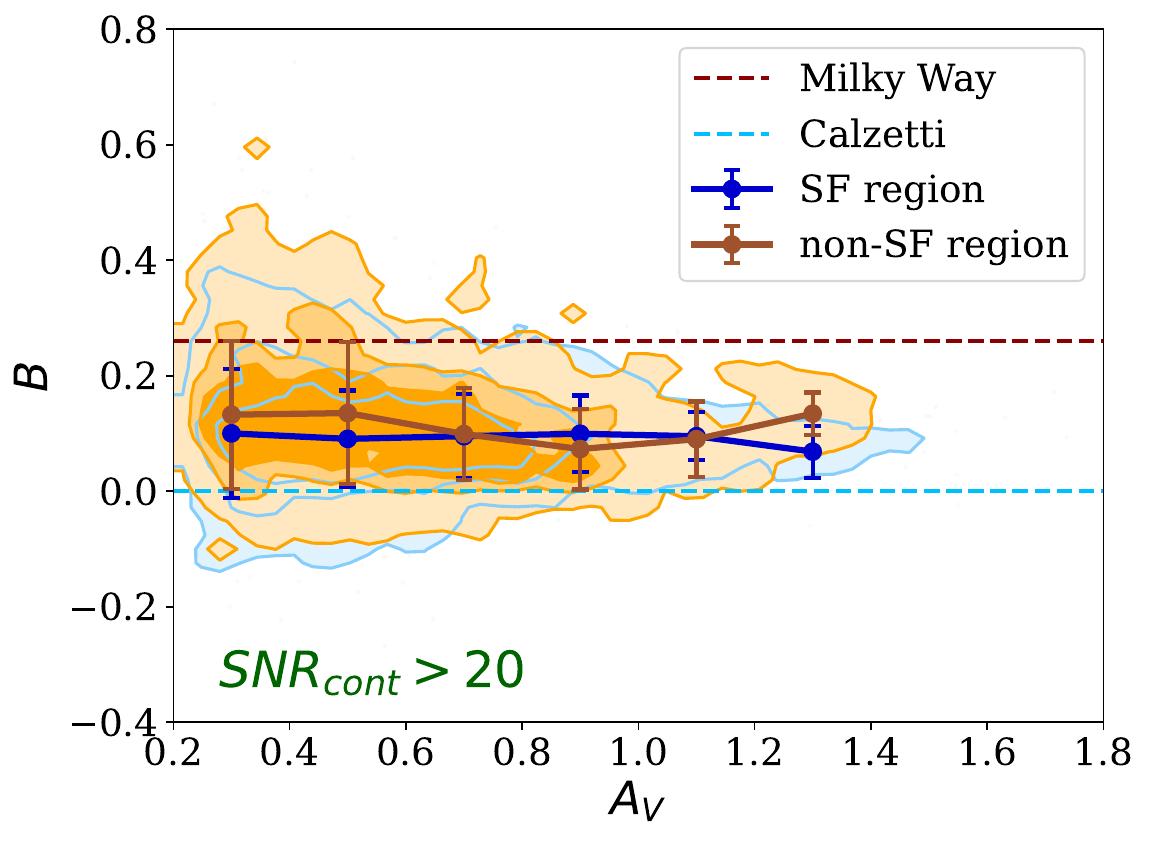}
    \includegraphics[width=0.3\textwidth]{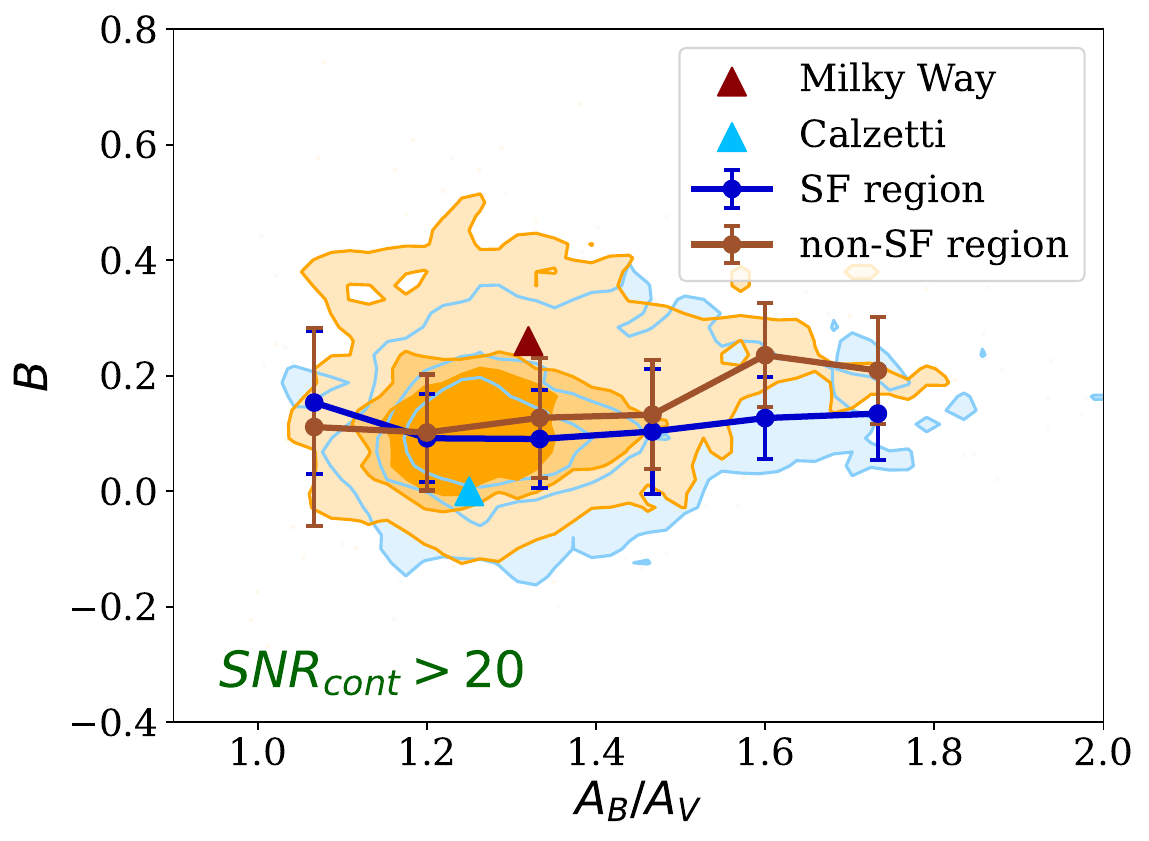}
    \caption{Relative bump strength $B$ versus other attenuation-curve properties in the $SNR_{cont}>20$ sample: NUV slope $A_{w2}/A_{w1}$, optical opacity $A_V$, and optical slope $A_B/A_V$.\label{fig:B_each_other}}
\end{figure}

\twocolumngrid

\bibliographystyle{aasjournal}
\bibliography{bibtex}

@ARTICLE{2025PASA...42...22B,
       author = {{Battisti}, A. and {Shivaei}, I. and {Park}, H. -J. and {Decleir}, M. and {Calzetti}, D. and {Mathew}, J. and {Wisnioski}, E. and {da Cunha}, E.},
        title = "{Constraining the link between the 2175{\r{A}} dust absorption feature and PAHs in Nearby Star-Forming Galaxies using Swift/UVOT and JWST/MIRI}",
      journal = {\pasa},
         year = 2025,
        month = feb,
       volume = {42},
          eid = {e022},
        pages = {e022},
          doi = {10.1017/pasa.2024.129},
archivePrefix = {arXiv},
       eprint = {2412.03690},
 primaryClass = {astro-ph.GA},
       adsurl = {https://ui.adsabs.harvard.edu/abs/2025PASA...42...22B}
}

@ARTICLE{2013A&A...559A.114M,
       author = {{Marino}, R.~A. and {Rosales-Ortega}, F.~F. and {S{\'a}nchez}, S.~F. and {Gil de Paz}, A. and {V{\'\i}lchez}, J. and {Miralles-Caballero}, D. and {Kehrig}, C. and {P{\'e}rez-Montero}, E. and {Stanishev}, V. and {Iglesias-P{\'a}ramo}, J. and {D{\'\i}az}, A.~I. and {Castillo-Morales}, A. and {Kennicutt}, R. and {L{\'o}pez-S{\'a}nchez}, A.~R. and {Galbany}, L. and {Garc{\'\i}a-Benito}, R. and {Mast}, D. and {Mendez-Abreu}, J. and {Monreal-Ibero}, A. and {Husemann}, B. and {Walcher}, C.~J. and {Garc{\'\i}a-Lorenzo}, B. and {Masegosa}, J. and {Del Olmo Orozco}, A. and {Mour{\~a}o}, A.~M. and {Ziegler}, B. and {Moll{\'a}}, M. and {Papaderos}, P. and {S{\'a}nchez-Bl{\'a}zquez}, P. and {Gonz{\'a}lez Delgado}, R.~M. and {Falc{\'o}n-Barroso}, J. and {Roth}, M.~M. and {van de Ven}, G. and {CALIFA Team}},
        title = "{The O3N2 and N2 abundance indicators revisited: improved calibrations based on CALIFA and T$_{e}$-based literature data}",
      journal = {\aap},
         year = 2013,
        month = nov,
       volume = {559},
          eid = {A114},
        pages = {A114},
          doi = {10.1051/0004-6361/201321956},
archivePrefix = {arXiv},
       eprint = {1307.5316},
 primaryClass = {astro-ph.CO},
       adsurl = {https://ui.adsabs.harvard.edu/abs/2013A&A...559A.114M}
}

@ARTICLE{2020ARA&A..58..529S,
       author = {{Salim}, Samir and {Narayanan}, Desika},
        title = "{The Dust Attenuation Law in Galaxies}",
      journal = {\araa},
         year = 2020,
        month = aug,
       volume = {58},
        pages = {529-575},
          doi = {10.1146/annurev-astro-032620-021933},
archivePrefix = {arXiv},
       eprint = {2001.03181},
 primaryClass = {astro-ph.GA},
       adsurl = {https://ui.adsabs.harvard.edu/abs/2020ARA&A..58..529S}
}

@ARTICLE{2010ApJ...718..184C,
       author = {{Conroy}, Charlie and {Schiminovich}, David and {Blanton}, Michael R.},
        title = "{Dust Attenuation in Disk-dominated Galaxies: Evidence for the 2175 {\r{A}} Dust Feature}",
      journal = {\apj},
         year = 2010,
        month = jul,
       volume = {718},
       number = {1},
        pages = {184-198},
          doi = {10.1088/0004-637X/718/1/184},
archivePrefix = {arXiv},
       eprint = {1003.2202},
 primaryClass = {astro-ph.CO},
       adsurl = {https://ui.adsabs.harvard.edu/abs/2010ApJ...718..184C}
}

@ARTICLE{2003ARA&A..41..241D,
       author = {{Draine}, B.~T.},
        title = "{Interstellar Dust Grains}",
      journal = {\araa},
         year = 2003,
        month = jan,
       volume = {41},
        pages = {241-289},
          doi = {10.1146/annurev.astro.41.011802.094840},
archivePrefix = {arXiv},
       eprint = {astro-ph/0304489},
 primaryClass = {astro-ph},
       adsurl = {https://ui.adsabs.harvard.edu/abs/2003ARA&A..41..241D}
}

@ARTICLE{2001ApJ...554..778L,
       author = {{Li}, Aigen and {Draine}, B.~T.},
        title = "{Infrared Emission from Interstellar Dust. II. The Diffuse Interstellar Medium}",
      journal = {\apj},
         year = 2001,
        month = jun,
       volume = {554},
       number = {2},
        pages = {778-802},
          doi = {10.1086/323147},
archivePrefix = {arXiv},
       eprint = {astro-ph/0011319},
 primaryClass = {astro-ph},
       adsurl = {https://ui.adsabs.harvard.edu/abs/2001ApJ...554..778L}
}

@ARTICLE{2023MNRAS.525.2380L,
       author = {{Lin}, Qi and {Yang}, X.~J. and {Li}, Aigen},
        title = "{Polycyclic aromatic hydrocarbon molecules and the 2175{\r{A}} interstellar extinction bump}",
      journal = {\mnras},
         year = 2023,
        month = oct,
       volume = {525},
       number = {2},
        pages = {2380-2387},
          doi = {10.1093/mnras/stad2405},
archivePrefix = {arXiv},
       eprint = {2308.04084},
 primaryClass = {astro-ph.GA},
       adsurl = {https://ui.adsabs.harvard.edu/abs/2023MNRAS.525.2380L}
}

@ARTICLE{2025A&A...694A..84L,
       author = {{Lin}, Qi and {Yang}, Xuejuan and {Li}, Aigen and {Witstok}, Joris},
        title = "{Polycyclic Aromatic Hydrocarbon and the Ultraviolet Extinction Bump at the Cosmic Dawn}",
      journal = {\aap},
         year = 2025,
        month = feb,
	       volume = {694},
	          eid = {A84},
	        pages = {A84},
	          doi = {10.1051/0004-6361/202452372},
	archivePrefix = {arXiv},
	       eprint = {2502.08113},
	 primaryClass = {astro-ph.GA},
	       adsurl = {https://ui.adsabs.harvard.edu/abs/2025A&A...694A..84L}
}

@ARTICLE{2018ApJ...859...11S,
       author = {{Salim}, Samir and {Boquien}, M{\'e}d{\'e}ric and {Lee}, Janice C.},
        title = "{Dust Attenuation Curves in the Local Universe: Demographics and New Laws for Star-forming Galaxies and High-redshift Analogs}",
      journal = {\apj},
         year = 2018,
        month = may,
       volume = {859},
       number = {1},
          eid = {11},
        pages = {11},
          doi = {10.3847/1538-4357/aabf3c},
archivePrefix = {arXiv},
       eprint = {1804.05850},
 primaryClass = {astro-ph.GA},
       adsurl = {https://ui.adsabs.harvard.edu/abs/2018ApJ...859...11S}
}

@ARTICLE{2023ApJ...953...54B,
       author = {{Belles}, Alexander and {Decleir}, Marjorie and {Bowman}, William P. and {Hagen}, Lea M.~Z. and {Gronwall}, Caryl and {Siegel}, Michael H.},
        title = "{Measuring the Dust Attenuation Curves of SINGS/KINGFISH Galaxies Using Swift/UVOT Photometry}",
      journal = {\apj},
         year = 2023,
        month = aug,
       volume = {953},
       number = {1},
          eid = {54},
        pages = {54},
          doi = {10.3847/1538-4357/acd332},
archivePrefix = {arXiv},
       eprint = {2305.05650},
 primaryClass = {astro-ph.GA},
       adsurl = {https://ui.adsabs.harvard.edu/abs/2023ApJ...953...54B}
}

@ARTICLE{1994ApJ...429..582C,
       author = {{Calzetti}, Daniela and {Kinney}, Anne L. and {Storchi-Bergmann}, Thaisa},
        title = "{Dust Extinction of the Stellar Continua in Starburst Galaxies: The Ultraviolet and Optical Extinction Law}",
      journal = {\apj},
         year = 1994,
        month = jul,
       volume = {429},
        pages = {582},
          doi = {10.1086/174346},
       adsurl = {https://ui.adsabs.harvard.edu/abs/1994ApJ...429..582C}
}

@ARTICLE{2011MNRAS.417.1760W,
       author = {{Wild}, Vivienne and {Charlot}, St{\'e}phane and {Brinchmann}, Jarle and {Heckman}, Timothy and {Vince}, Oliver and {Pacifici}, Camilla and {Chevallard}, Jacopo},
        title = "{Empirical determination of the shape of dust attenuation curves in star-forming galaxies}",
      journal = {\mnras},
         year = 2011,
        month = nov,
       volume = {417},
       number = {3},
        pages = {1760-1786},
          doi = {10.1111/j.1365-2966.2011.19367.x},
archivePrefix = {arXiv},
       eprint = {1106.1646},
 primaryClass = {astro-ph.CO},
       adsurl = {https://ui.adsabs.harvard.edu/abs/2011MNRAS.417.1760W}
}

@ARTICLE{2017ApJ...851...90B,
       author = {{Battisti}, A.~J. and {Calzetti}, D. and {Chary}, R. -R.},
        title = "{Characterizing Dust Attenuation in Local Star-forming Galaxies: Inclination Effects and the 2175 {\r{A}} Feature}",
      journal = {\apj},
         year = 2017,
        month = dec,
       volume = {851},
       number = {2},
          eid = {90},
        pages = {90},
          doi = {10.3847/1538-4357/aa9a43},
archivePrefix = {arXiv},
       eprint = {1711.04814},
 primaryClass = {astro-ph.GA},
       adsurl = {https://ui.adsabs.harvard.edu/abs/2017ApJ...851...90B}
}

@ARTICLE{2000ApJ...528..799W,
       author = {{Witt}, Adolf N. and {Gordon}, Karl D.},
        title = "{Multiple Scattering in Clumpy Media. II. Galactic Environments}",
      journal = {\apj},
         year = 2000,
        month = jan,
       volume = {528},
       number = {2},
        pages = {799-816},
          doi = {10.1086/308197},
archivePrefix = {arXiv},
       eprint = {astro-ph/9907342},
 primaryClass = {astro-ph},
       adsurl = {https://ui.adsabs.harvard.edu/abs/2000ApJ...528..799W}
}

@ARTICLE{2007MNRAS.375..640P,
       author = {{Panuzzo}, P. and {Granato}, G.~L. and {Buat}, V. and {Inoue}, A.~K. and {Silva}, L. and {Iglesias-P{\'a}ramo}, J. and {Bressan}, A.},
        title = "{Ultraviolet dust attenuation in spiral galaxies: the role of age-dependent extinction and the initial mass function}",
      journal = {\mnras},
         year = 2007,
        month = feb,
       volume = {375},
       number = {2},
        pages = {640-648},
          doi = {10.1111/j.1365-2966.2006.11337.x},
archivePrefix = {arXiv},
       eprint = {astro-ph/0612087},
 primaryClass = {astro-ph},
       adsurl = {https://ui.adsabs.harvard.edu/abs/2007MNRAS.375..640P}
}

@ARTICLE{2016ApJ...833..201S,
       author = {{Seon}, Kwang-Il and {Draine}, Bruce T.},
        title = "{Radiative Transfer Model of Dust Attenuation Curves in Clumpy, Galactic Environments}",
      journal = {\apj},
         year = 2016,
        month = dec,
       volume = {833},
       number = {2},
          eid = {201},
        pages = {201},
          doi = {10.3847/1538-4357/833/2/201},
archivePrefix = {arXiv},
       eprint = {1606.02030},
 primaryClass = {astro-ph.GA},
       adsurl = {https://ui.adsabs.harvard.edu/abs/2016ApJ...833..201S}
}

@ARTICLE{1986ApJ...307..286F,
       author = {{Fitzpatrick}, E.~L. and {Massa}, D.},
        title = "{An Analysis of the Shapes of Ultraviolet Extinction Curves. I. The 2175 Angstrom Bump}",
      journal = {\apj},
         year = 1986,
        month = aug,
       volume = {307},
        pages = {286},
          doi = {10.1086/164415},
       adsurl = {https://ui.adsabs.harvard.edu/abs/1986ApJ...307..286F}
}

@ARTICLE{1989ApJ...345..245C,
       author = {{Cardelli}, Jason A. and {Clayton}, Geoffrey C. and {Mathis}, John S.},
        title = "{The Relationship between Infrared, Optical, and Ultraviolet Extinction}",
      journal = {\apj},
         year = 1989,
        month = oct,
       volume = {345},
        pages = {245},
          doi = {10.1086/167900},
       adsurl = {https://ui.adsabs.harvard.edu/abs/1989ApJ...345..245C}
}

@ARTICLE{2003ApJ...594..279G,
       author = {{Gordon}, Karl D. and {Clayton}, Geoffrey C. and {Misselt}, K.~A. and {Landolt}, Arlo U. and {Wolff}, Michael J.},
        title = "{A Quantitative Comparison of the Small Magellanic Cloud, Large Magellanic Cloud, and Milky Way Ultraviolet to Near-Infrared Extinction Curves}",
      journal = {\apj},
         year = 2003,
        month = sep,
       volume = {594},
       number = {1},
        pages = {279-293},
          doi = {10.1086/376774},
archivePrefix = {arXiv},
       eprint = {astro-ph/0305257},
 primaryClass = {astro-ph},
       adsurl = {https://ui.adsabs.harvard.edu/abs/2003ApJ...594..279G}
}

@ARTICLE{2000ApJ...533..682C,
       author = {{Calzetti}, Daniela and {Armus}, Lee and {Bohlin}, Ralph C. and {Kinney}, Anne L. and {Koornneef}, Jan and {Storchi-Bergmann}, Thaisa},
        title = "{The Dust Content and Opacity of Actively Star-forming Galaxies}",
      journal = {\apj},
         year = 2000,
        month = apr,
       volume = {533},
       number = {2},
        pages = {682-695},
          doi = {10.1086/308692},
archivePrefix = {arXiv},
       eprint = {astro-ph/9911459},
 primaryClass = {astro-ph},
       adsurl = {https://ui.adsabs.harvard.edu/abs/2000ApJ...533..682C}
}

@ARTICLE{2013ApJ...775L..16K,
       author = {{Kriek}, Mariska and {Conroy}, Charlie},
        title = "{The Dust Attenuation Law in Distant Galaxies: Evidence for Variation with Spectral Type}",
      journal = {\apjl},
         year = 2013,
        month = sep,
       volume = {775},
       number = {1},
          eid = {L16},
        pages = {L16},
          doi = {10.1088/2041-8205/775/1/L16},
archivePrefix = {arXiv},
       eprint = {1308.1099},
 primaryClass = {astro-ph.CO},
       adsurl = {https://ui.adsabs.harvard.edu/abs/2013ApJ...775L..16K}
}

@ARTICLE{2015ApJ...798....7B,
       author = {{Bundy}, Kevin and {Bershady}, Matthew A. and {Law}, David R. and {Yan}, Renbin and {Drory}, Niv and {MacDonald}, Nicholas and {Wake}, David A. and {Cherinka}, Brian and {S{\'a}nchez-Gallego}, Jos{\'e} R. and {Weijmans}, Anne-Marie and {Thomas}, Daniel and {Tremonti}, Christy and {Masters}, Karen and {Coccato}, Lodovico and {Diamond-Stanic}, Aleksandar M. and {Arag{\'o}n-Salamanca}, Alfonso and {Avila-Reese}, Vladimir and {Badenes}, Carles and {Falc{\'o}n-Barroso}, J{\'e}sus and {Belfiore}, Francesco and {Bizyaev}, Dmitry and {Blanc}, Guillermo A. and {Bland-Hawthorn}, Joss and {Blanton}, Michael R. and {Brownstein}, Joel R. and {Byler}, Nell and {Cappellari}, Michele and {Conroy}, Charlie and {Dutton}, Aaron A. and {Emsellem}, Eric and {Etherington}, James and {Frinchaboy}, Peter M. and {Fu}, Hai and {Gunn}, James E. and {Harding}, Paul and {Johnston}, Evelyn J. and {Kauffmann}, Guinevere and {Kinemuchi}, Karen and {Klaene}, Mark A. and {Knapen}, Johan H. and {Leauthaud}, Alexie and {Li}, Cheng and {Lin}, Lihwai and {Maiolino}, Roberto and {Malanushenko}, Viktor and {Malanushenko}, Elena and {Mao}, Shude and {Maraston}, Claudia and {McDermid}, Richard M. and {Merrifield}, Michael R. and {Nichol}, Robert C. and {Oravetz}, Daniel and {Pan}, Kaike and {Parejko}, John K. and {Sanchez}, Sebastian F. and {Schlegel}, David and {Simmons}, Audrey and {Steele}, Oliver and {Steinmetz}, Matthias and {Thanjavur}, Karun and {Thompson}, Benjamin A. and {Tinker}, Jeremy L. and {van den Bosch}, Remco C.~E. and {Westfall}, Kyle B. and {Wilkinson}, David and {Wright}, Shelley and {Xiao}, Ting and {Zhang}, Kai},
        title = "{Overview of the SDSS-IV MaNGA Survey: Mapping nearby Galaxies at Apache Point Observatory}",
      journal = {\apj},
         year = 2015,
        month = jan,
       volume = {798},
       number = {1},
          eid = {7},
        pages = {7},
          doi = {10.1088/0004-637X/798/1/7},
archivePrefix = {arXiv},
       eprint = {1412.1482},
 primaryClass = {astro-ph.GA},
       adsurl = {https://ui.adsabs.harvard.edu/abs/2015ApJ...798....7B}
}

@ARTICLE{2006AJ....131.1163S,
       author = {{Skrutskie}, M.~F. and {Cutri}, R.~M. and {Stiening}, R. and {Weinberg}, M.~D. and {Schneider}, S. and {Carpenter}, J.~M. and {Beichman}, C. and {Capps}, R. and {Chester}, T. and {Elias}, J. and {Huchra}, J. and {Liebert}, J. and {Lonsdale}, C. and {Monet}, D.~G. and {Price}, S. and {Seitzer}, P. and {Jarrett}, T. and {Kirkpatrick}, J.~D. and {Gizis}, J.~E. and {Howard}, E. and {Evans}, T. and {Fowler}, J. and {Fullmer}, L. and {Hurt}, R. and {Light}, R. and {Kopan}, E.~L. and {Marsh}, K.~A. and {McCallon}, H.~L. and {Tam}, R. and {Van Dyk}, S. and {Wheelock}, S.},
        title = "{The Two Micron All Sky Survey (2MASS)}",
      journal = {\aj},
         year = 2006,
        month = feb,
       volume = {131},
       number = {2},
        pages = {1163-1183},
          doi = {10.1086/498708},
       adsurl = {https://ui.adsabs.harvard.edu/abs/2006AJ....131.1163S}
}

@ARTICLE{2023Natur.621..267W,
       author = {{Witstok}, Joris and {Shivaei}, Irene and {Smit}, Renske and {Maiolino}, Roberto and {Carniani}, Stefano and {Curtis-Lake}, Emma and {Ferruit}, Pierre and {Arribas}, Santiago and {Bunker}, Andrew J. and {Cameron}, Alex J. and others},
        title = "{Carbonaceous dust grains seen in the first billion years of cosmic time}",
      journal = {Nature},
         year = 2023,
        month = sep,
       volume = {621},
        pages = {267-270},
          doi = {10.1038/s41586-023-06413-w},
archivePrefix = {arXiv},
       eprint = {2302.05468},
 primaryClass = {astro-ph.GA},
       adsurl = {https://ui.adsabs.harvard.edu/abs/2023Natur.621..267W}
}

@ARTICLE{2025arXiv250221119O,
       author = {{Ormerod}, Katherine and {Witstok}, Joris and {Smit}, Renske and {de Graaff}, Anna and {Helton}, Jakob M. and {Maseda}, Michael V. and {Shivaei}, Irene and {Bunker}, Andrew J. and {Carniani}, Stefano and {D'Eugenio}, Francesco and others},
        title = "{Detection of the 2175{\r{A}} UV Bump at z>7: Evidence for Rapid Dust Evolution in a Merging Reionisation-Era Galaxy}",
	      journal = {\mnras},
	         year = 2025,
	        month = aug,
	       volume = {542},
	        pages = {1136-1154},
	          doi = {10.1093/mnras/staf1228},
	archivePrefix = {arXiv},
	       eprint = {2502.21119},
	 primaryClass = {astro-ph.GA},
	       adsurl = {https://ui.adsabs.harvard.edu/abs/2025MNRAS.542.1136O}
}

@ARTICLE{2023ApJ...957...75Z,
       author = {{Zhou}, Shuang and {Li}, Cheng and {Li}, Niu and {Mo}, Houjun and {Yan}, Renbin and {Eracleous}, Michael and {Molina}, Mallory and {Gronwall}, Caryl and {Ajgaonkar}, Nikhil and {Cheng}, Zhuo and {Guo}, Ruonan},
        title = "{Mapping Dust Attenuation and the 2175 {\r{A}} Bump at Kiloparsec Scales in Nearby Galaxies}",
      journal = {\apj},
         year = 2023,
        month = nov,
       volume = {957},
       number = {2},
          eid = {75},
        pages = {75},
          doi = {10.3847/1538-4357/acfb80},
archivePrefix = {arXiv},
       eprint = {2212.01918},
 primaryClass = {astro-ph.GA},
       adsurl = {https://ui.adsabs.harvard.edu/abs/2023ApJ...957...75Z}
}

@ARTICLE{2025arXiv250314028G,
       author = {{Guo}, Ruonan and {Li}, Cheng and {Zhou}, Shuang and {Li}, Niu and {Jing}, Tao and {Cheng}, Zhuo},
        title = "{Mapping Dust Attenuation at Kiloparsec Scales. II. Attenuation Curves from Near-ultraviolet to Near-infrared}",
      journal = {Research in Astronomy and Astrophysics},
         year = 2025,
        month = jun,
       volume = {25},
       number = {6},
          eid = {065017},
        pages = {065017},
          doi = {10.1088/1674-4527/add673}
}

@ARTICLE{2026arXiv260713868G,
       author = {{Guo}, Ruonan and {Li}, Cheng and {Jing}, Tao and {Zhou}, Shuang and {Li}, Niu and {Cheng}, Zhuo},
        title = "{Mapping Dust Attenuation at Kiloparsec Scales. IV. A Dust-model Interpretation of Attenuation Curves in Nearby Galaxies}",
      journal = {arXiv e-prints},
         year = 2026,
        month = jul,
          eid = {arXiv:2607.13868},
        pages = {arXiv:2607.13868},
archivePrefix = {arXiv},
       eprint = {2607.13868},
 primaryClass = {astro-ph.GA},
          doi = {10.48550/arXiv.2607.13868},
       adsurl = {https://ui.adsabs.harvard.edu/abs/2026arXiv260713868G}
}

@ARTICLE{2020ApJ...896...38L,
       author = {{Li}, Niu and {Li}, Cheng and {Mo}, Houjun and {Hu}, Jian and {Zhou}, Shuang and {Du}, Cheng},
        title = "{Estimating Dust Attenuation from Galactic Spectra. I. Methodology and Tests}",
      journal = {\apj},
         year = 2020,
        month = jun,
       volume = {896},
       number = {1},
          eid = {38},
        pages = {38},
          doi = {10.3847/1538-4357/ab92a1},
archivePrefix = {arXiv},
       eprint = {2001.02815},
 primaryClass = {astro-ph.GA},
       adsurl = {https://ui.adsabs.harvard.edu/abs/2020ApJ...896...38L}
}

@ARTICLE{2023ApJS..268...63M,
       author = {{Molina}, Mallory and {Duffy}, Laura and {Eracleous}, Michael and {Ogborn}, Mary and {Kaldor}, Mary E. and {Yan}, Renbin and {Gronwall}, Caryl and {Ciardullo}, Robin and {Ajgaonkar}, Nikhil},
        title = "{The New Swift/UVOT+MaNGA (SwiM) Value-added Catalog}",
      journal = {\apjs},
         year = 2023,
        month = oct,
       volume = {268},
       number = {2},
          eid = {63},
        pages = {63},
          doi = {10.3847/1538-4365/acf578},
archivePrefix = {arXiv},
       eprint = {2308.15551},
 primaryClass = {astro-ph.GA},
       adsurl = {https://ui.adsabs.harvard.edu/abs/2023ApJS..268...63M}
}

@ARTICLE{2017AJ....154...28B,
       author = {{Blanton}, Michael R. and {Bershady}, Matthew A. and {Abolfathi}, Bela and {Albareti}, Franco D. and {Allende Prieto}, Carlos and {Almeida}, Andres and {Alonso-Garc{\'\i}a}, Javier and {Anders}, Friedrich and {Anderson}, Scott F. and {Andrews}, Brett and {Aquino-Ort{\'\i}z}, Erik and {Arag{\'o}n-Salamanca}, Alfonso and {Argudo-Fern{\'a}ndez}, Maria and {Armengaud}, Eric and {Aubourg}, Eric and {Avila-Reese}, Vladimir and {Badenes}, Carles and {Bailey}, Stephen and {Barger}, Kathleen A. and {Barrera-Ballesteros}, Jorge and {Bartosz}, Curtis and {Bates}, Dominic and {Baumgarten}, Falk and {Bautista}, Julian and {Beaton}, Rachael and {Beers}, Timothy C. and {Belfiore}, Francesco and {Bender}, Chad F. and {Berlind}, Andreas A. and {Bernardi}, Mariangela and {Beutler}, Florian and {Bird}, Jonathan C. and {Bizyaev}, Dmitry and {Blanc}, Guillermo A. and {Blomqvist}, Michael and {Bolton}, Adam S. and {Boquien}, M{\'e}d{\'e}ric and {Borissova}, Jura and {van den Bosch}, Remco and {Bovy}, Jo and {Brandt}, William N. and {Brinkmann}, Jonathan and {Brownstein}, Joel R. and {Bundy}, Kevin and {Burgasser}, Adam J. and {Burtin}, Etienne and {Busca}, Nicol{\'a}s G. and {Cappellari}, Michele and {Delgado Carigi}, Maria Leticia and {Carlberg}, Joleen K. and {Carnero Rosell}, Aurelio and {Carrera}, Ricardo and {Chanover}, Nancy J. and {Cherinka}, Brian and {Cheung}, Edmond and {G{\'o}mez Maqueo Chew}, Yilen and {Chiappini}, Cristina and {Choi}, Peter Doohyun and {Chojnowski}, Drew and {Chuang}, Chia-Hsun and {Chung}, Haeun and {Cirolini}, Rafael Fernando and {Clerc}, Nicolas and {Cohen}, Roger E. and {Comparat}, Johan and {da Costa}, Luiz and {Cousinou}, Marie-Claude and {Covey}, Kevin and {Crane}, Jeffrey D. and {Croft}, Rupert A.~C. and {Cruz-Gonzalez}, Irene and {Garrido Cuadra}, Daniel and {Cunha}, Katia and {Damke}, Guillermo J. and {Darling}, Jeremy and {Davies}, Roger and {Dawson}, Kyle and {de la Macorra}, Axel and {Dell'Agli}, Flavia and {De Lee}, Nathan and {Delubac}, Timoth{\'e}e and {Di Mille}, Francesco and {Diamond-Stanic}, Aleks and {Cano-D{\'\i}az}, Mariana and {Donor}, John and {Downes}, Juan Jos{\'e} and {Drory}, Niv and {du Mas des Bourboux}, H{\'e}lion and {Duckworth}, Christopher J. and {Dwelly}, Tom and {Dyer}, Jamie and {Ebelke}, Garrett and {Eigenbrot}, Arthur D. and {Eisenstein}, Daniel J. and {Emsellem}, Eric and {Eracleous}, Mike and {Escoffier}, Stephanie and {Evans}, Michael L. and {Fan}, Xiaohui and {Fern{\'a}ndez-Alvar}, Emma and {Fernandez-Trincado}, J.~G. and {Feuillet}, Diane K. and {Finoguenov}, Alexis and {Fleming}, Scott W. and {Font-Ribera}, Andreu and {Fredrickson}, Alexander and {Freischlad}, Gordon and {Frinchaboy}, Peter M. and {Fuentes}, Carla E. and {Galbany}, Llu{\'\i}s and {Garcia-Dias}, R. and {Garc{\'\i}a-Hern{\'a}ndez}, D.~A. and {Gaulme}, Patrick and {Geisler}, Doug and {Gelfand}, Joseph D. and {Gil-Mar{\'\i}n}, H{\'e}ctor and {Gillespie}, Bruce A. and {Goddard}, Daniel and {Gonzalez-Perez}, Violeta and {Grabowski}, Kathleen and {Green}, Paul J. and {Grier}, Catherine J. and {Gunn}, James E. and {Guo}, Hong and {Guy}, Julien and {Hagen}, Alex and {Hahn}, ChangHoon and {Hall}, Matthew and {Harding}, Paul and {Hasselquist}, Sten and {Hawley}, Suzanne L. and {Hearty}, Fred and {Gonzalez Hern{\'a}ndez}, Jonay I. and {Ho}, Shirley and {Hogg}, David W. and {Holley-Bockelmann}, Kelly and {Holtzman}, Jon A. and {Holzer}, Parker H. and {Huehnerhoff}, Joseph and {Hutchinson}, Timothy A. and {Hwang}, Ho Seong and {Ibarra-Medel}, H{\'e}ctor J. and {da Silva Ilha}, Gabriele and {Ivans}, Inese I. and {Ivory}, KeShawn and {Jackson}, Kelly and {Jensen}, Trey W. and {Johnson}, Jennifer A. and {Jones}, Amy and {J{\"o}nsson}, Henrik and {Jullo}, Eric and {Kamble}, Vikrant and {Kinemuchi}, Karen and {Kirkby}, David and {Kitaura}, Francisco-Shu and {Klaene}, Mark and {Knapp}, Gillian R. and {Kneib}, Jean-Paul and {Kollmeier}, Juna A. and {Lacerna}, Ivan and {Lane}, Richard R. and {Lang}, Dustin and {Law}, David R. and {Lazarz}, Daniel and {Lee}, Youngbae and {Le Goff}, Jean-Marc and {Liang}, Fu-Heng and {Li}, Cheng and {Li}, Hongyu and {Lian}, Jianhui and {Lima}, Marcos and {Lin}, Lihwai and {Lin}, Yen-Ting and {Bertran de Lis}, Sara and {Liu}, Chao and {de Icaza Lizaola}, Miguel Angel C. and {Long}, Dan and {Lucatello}, Sara and {Lundgren}, Britt and {MacDonald}, Nicholas K. and {Deconto Machado}, Alice and {MacLeod}, Chelsea L. and {Mahadevan}, Suvrath and {Geimba Maia}, Marcio Antonio and {Maiolino}, Roberto and {Majewski}, Steven R. and {Malanushenko}, Elena and {Malanushenko}, Viktor and {Manchado}, Arturo and {Mao}, Shude and {Maraston}, Claudia and {Marques-Chaves}, Rui and {Masseron}, Thomas and {Masters}, Karen L. and {McBride}, Cameron K. and {McDermid}, Richard M. and {McGrath}, Brianne and {McGreer}, Ian D. and {Medina Pe{\~n}a}, Nicol{\'a}s and {Melendez}, Matthew},
        title = "{Sloan Digital Sky Survey IV: Mapping the Milky Way, Nearby Galaxies, and the Distant Universe}",
      journal = {\aj},
         year = 2017,
        month = jul,
       volume = {154},
       number = {1},
          eid = {28},
        pages = {28},
          doi = {10.3847/1538-3881/aa7567},
archivePrefix = {arXiv},
       eprint = {1703.00052},
 primaryClass = {astro-ph.GA},
       adsurl = {https://ui.adsabs.harvard.edu/abs/2017AJ....154...28B}
}

@ARTICLE{2022ApJS..259...35A,
       author = {{Abdurro'uf} and {Accetta}, Katherine and {Aerts}, Conny and {Silva Aguirre}, V{\'\i}ctor and {Ahumada}, Romina and {Ajgaonkar}, Nikhil and {Filiz Ak}, N. and {Alam}, Shadab and {Allende Prieto}, Carlos and {Almeida}, Andr{\'e}s and {Anders}, Friedrich and {Anderson}, Scott F. and {Andrews}, Brett H. and {Anguiano}, Borja and {Aquino-Ort{\'\i}z}, Erik and {Arag{\'o}n-Salamanca}, Alfonso and {Argudo-Fern{\'a}ndez}, Maria and {Ata}, Metin and {Aubert}, Marie and {Avila-Reese}, Vladimir and {Badenes}, Carles and {Barb{\'a}}, Rodolfo H. and {Barger}, Kat and {Barrera-Ballesteros}, Jorge K. and {Beaton}, Rachael L. and {Beers}, Timothy C. and {Belfiore}, Francesco and {Bender}, Chad F. and {Bernardi}, Mariangela and {Bershady}, Matthew A. and {Beutler}, Florian and {Bidin}, Christian Moni and {Bird}, Jonathan C. and {Bizyaev}, Dmitry and {Blanc}, Guillermo A. and {Blanton}, Michael R. and {Boardman}, Nicholas Fraser and {Bolton}, Adam S. and {Boquien}, M{\'e}d{\'e}ric and {Borissova}, Jura and {Bovy}, Jo and {Brandt}, W.~N. and {Brown}, Jordan and {Brownstein}, Joel R. and {Brusa}, Marcella and {Buchner}, Johannes and {Bundy}, Kevin and {Burchett}, Joseph N. and {Bureau}, Martin and {Burgasser}, Adam and {Cabang}, Tuesday K. and {Campbell}, Stephanie and {Cappellari}, Michele and {Carlberg}, Joleen K. and {Wanderley}, F{\'a}bio Carneiro and {Carrera}, Ricardo and {Cash}, Jennifer and {Chen}, Yan-Ping and {Chen}, Wei-Huai and {Cherinka}, Brian and {Chiappini}, Cristina and {Choi}, Peter Doohyun and {Chojnowski}, S. Drew and {Chung}, Haeun and {Clerc}, Nicolas and {Cohen}, Roger E. and {Comerford}, Julia M. and {Comparat}, Johan and {da Costa}, Luiz and {Covey}, Kevin and {Crane}, Jeffrey D. and {Cruz-Gonzalez}, Irene and {Culhane}, Connor and {Cunha}, Katia and {Dai}, Y. Sophia and {Damke}, Guillermo and {Darling}, Jeremy and {Davidson}, Jr., James W. and {Davies}, Roger and {Dawson}, Kyle and {De Lee}, Nathan and {Diamond-Stanic}, Aleksandar M. and {Cano-D{\'\i}az}, Mariana and {S{\'a}nchez}, Helena Dom{\'\i}nguez and {Donor}, John and {Duckworth}, Chris and {Dwelly}, Tom and {Eisenstein}, Daniel J. and {Elsworth}, Yvonne P. and {Emsellem}, Eric and {Eracleous}, Mike and {Escoffier}, Stephanie and {Fan}, Xiaohui and {Farr}, Emily and {Feng}, Shuai and {Fern{\'a}ndez-Trincado}, Jos{\'e} G. and {Feuillet}, Diane and {Filipp}, Andreas and {Fillingham}, Sean P. and {Frinchaboy}, Peter M. and {Fromenteau}, Sebastien and {Galbany}, Llu{\'\i}s and {Garc{\'\i}a}, Rafael A. and {Garc{\'\i}a-Hern{\'a}ndez}, D.~A. and {Ge}, Junqiang and {Geisler}, Doug and {Gelfand}, Joseph and {G{\'e}ron}, Tobias and {Gibson}, Benjamin J. and {Goddy}, Julian and {Godoy-Rivera}, Diego and {Grabowski}, Kathleen and {Green}, Paul J. and {Greener}, Michael and {Grier}, Catherine J. and {Griffith}, Emily and {Guo}, Hong and {Guy}, Julien and {Hadjara}, Massinissa and {Harding}, Paul and {Hasselquist}, Sten and {Hayes}, Christian R. and {Hearty}, Fred and {Hern{\'a}ndez}, Jes{\'u}s and {Hill}, Lewis and {Hogg}, David W. and {Holtzman}, Jon A. and {Horta}, Danny and {Hsieh}, Bau-Ching and {Hsu}, Chin-Hao and {Hsu}, Yun-Hsin and {Huber}, Daniel and {Huertas-Company}, Marc and {Hutchinson}, Brian and {Hwang}, Ho Seong and {Ibarra-Medel}, H{\'e}ctor J. and {Chitham}, Jacob Ider and {Ilha}, Gabriele S. and {Imig}, Julie and {Jaekle}, Will and {Jayasinghe}, Tharindu and {Ji}, Xihan and {Johnson}, Jennifer A. and {Jones}, Amy and {J{\"o}nsson}, Henrik and {Katkov}, Ivan and {Khalatyan}, Dr., Arman and {Kinemuchi}, Karen and {Kisku}, Shobhit and {Knapen}, Johan H. and {Kneib}, Jean-Paul and {Kollmeier}, Juna A. and {Kong}, Miranda and {Kounkel}, Marina and {Kreckel}, Kathryn and {Krishnarao}, Dhanesh and {Lacerna}, Ivan and {Lane}, Richard R. and {Langgin}, Rachel and {Lavender}, Ramon and {Law}, David R. and {Lazarz}, Daniel and {Leung}, Henry W. and {Leung}, Ho-Hin and {Lewis}, Hannah M. and {Li}, Cheng and {Li}, Ran and {Lian}, Jianhui and {Liang}, Fu-Heng and {Lin}, Lihwai and {Lin}, Yen-Ting and {Lin}, Sicheng and {Lintott}, Chris and {Long}, Dan and {Longa-Pe{\~n}a}, Pen{\'e}lope and {L{\'o}pez-Cob{\'a}}, Carlos and {Lu}, Shengdong and {Lundgren}, Britt F. and {Luo}, Yuanze and {Mackereth}, J. Ted and {de la Macorra}, Axel and {Mahadevan}, Suvrath and {Majewski}, Steven R. and {Manchado}, Arturo and {Mandeville}, Travis and {Maraston}, Claudia and {Margalef-Bentabol}, Berta and {Masseron}, Thomas and {Masters}, Karen L. and {Mathur}, Savita and {McDermid}, Richard M. and {Mckay}, Myles and {Merloni}, Andrea and {Merrifield}, Michael and {Meszaros}, Szabolcs and {Miglio}, Andrea and {Di Mille}, Francesco and {Minniti}, Dante and {Minsley}, Rebecca and {Monachesi}, Antonela},
        title = "{The Seventeenth Data Release of the Sloan Digital Sky Surveys: Complete Release of MaNGA, MaStar, and APOGEE-2 Data}",
      journal = {\apjs},
         year = 2022,
        month = apr,
       volume = {259},
       number = {2},
          eid = {35},
        pages = {35},
          doi = {10.3847/1538-4365/ac4414},
archivePrefix = {arXiv},
       eprint = {2112.02026},
 primaryClass = {astro-ph.GA},
       adsurl = {https://ui.adsabs.harvard.edu/abs/2022ApJS..259...35A}
}

@ARTICLE{2016AJ....152...83L,
       author = {{Law}, David R. and {Cherinka}, Brian and {Yan}, Renbin and {Andrews}, Brett H. and {Bershady}, Matthew A. and {Bizyaev}, Dmitry and {Blanc}, Guillermo A. and {Blanton}, Michael R. and {Bolton}, Adam S. and {Brownstein}, Joel R. and {Bundy}, Kevin and {Chen}, Yanmei and {Drory}, Niv and {D'Souza}, Richard and {Fu}, Hai and {Jones}, Amy and {Kauffmann}, Guinevere and {MacDonald}, Nicholas and {Masters}, Karen L. and {Newman}, Jeffrey A. and {Parejko}, John K. and {S{\'a}nchez-Gallego}, Jos{\'e} R. and {S{\'a}nchez}, Sebastian F. and {Schlegel}, David J. and {Thomas}, Daniel and {Wake}, David A. and {Weijmans}, Anne-Marie and {Westfall}, Kyle B. and {Zhang}, Kai},
        title = "{The Data Reduction Pipeline for the SDSS-IV MaNGA IFU Galaxy Survey}",
      journal = {\aj},
         year = 2016,
        month = oct,
       volume = {152},
       number = {4},
          eid = {83},
        pages = {83},
          doi = {10.3847/0004-6256/152/4/83},
archivePrefix = {arXiv},
       eprint = {1607.08619},
 primaryClass = {astro-ph.IM},
       adsurl = {https://ui.adsabs.harvard.edu/abs/2016AJ....152...83L}
}

@ARTICLE{2019AJ....158..231W,
       author = {{Westfall}, Kyle B. and {Cappellari}, Michele and {Bershady}, Matthew A. and {Bundy}, Kevin and {Belfiore}, Francesco and {Ji}, Xihan and {Law}, David R. and {Schaefer}, Adam and {Shetty}, Shravan and {Tremonti}, Christy A. and {Yan}, Renbin and {Andrews}, Brett H. and {Brownstein}, Joel R. and {Cherinka}, Brian and {Coccato}, Lodovico and {Drory}, Niv and {Maraston}, Claudia and {Parikh}, Taniya and {S{\'a}nchez-Gallego}, Jos{\'e} R. and {Thomas}, Daniel and {Weijmans}, Anne-Marie and {Barrera-Ballesteros}, Jorge and {Du}, Cheng and {Goddard}, Daniel and {Li}, Niu and {Masters}, Karen and {Ibarra Medel}, H{\'e}ctor Javier and {S{\'a}nchez}, Sebasti{\'a}n F. and {Yang}, Meng and {Zheng}, Zheng and {Zhou}, Shuang},
        title = "{The Data Analysis Pipeline for the SDSS-IV MaNGA IFU Galaxy Survey: Overview}",
      journal = {\aj},
         year = 2019,
        month = dec,
       volume = {158},
       number = {6},
          eid = {231},
        pages = {231},
          doi = {10.3847/1538-3881/ab44a2},
archivePrefix = {arXiv},
       eprint = {1901.00856},
 primaryClass = {astro-ph.GA},
       adsurl = {https://ui.adsabs.harvard.edu/abs/2019AJ....158..231W}
}

@ARTICLE{2019AJ....158..160B,
       author = {{Belfiore}, Francesco and {Westfall}, Kyle B. and {Schaefer}, Adam and {Cappellari}, Michele and {Ji}, Xihan and {Bershady}, Matthew A. and {Tremonti}, Christy and {Law}, David R. and {Yan}, Renbin and {Bundy}, Kevin and {Shetty}, Shravan and {Drory}, Niv and {Thomas}, Daniel and {Emsellem}, Eric and {S{\'a}nchez}, Sebasti{\'a}n F.},
        title = "{The Data Analysis Pipeline for the SDSS-IV MaNGA IFU Galaxy Survey: Emission-line Modeling}",
      journal = {\aj},
         year = 2019,
        month = oct,
       volume = {158},
       number = {4},
          eid = {160},
        pages = {160},
          doi = {10.3847/1538-3881/ab3e4e},
archivePrefix = {arXiv},
       eprint = {1901.00866},
 primaryClass = {astro-ph.GA},
       adsurl = {https://ui.adsabs.harvard.edu/abs/2019AJ....158..160B}
}

@ARTICLE{2019MNRAS.485.5256Z,
       author = {{Zhou}, Shuang and {Mo}, H.~J. and {Li}, Cheng and {Zheng}, Zheng and {Li}, Niu and {Du}, Cheng and {Mao}, Shude and {Parikh}, Taniya and {Lane}, Richard R. and {Thomas}, Daniel},
        title = "{SDSS-IV MaNGA: stellar initial mass function variation inferred from Bayesian analysis of the integral field spectroscopy of early-type galaxies}",
      journal = {\mnras},
         year = 2019,
        month = jun,
       volume = {485},
       number = {4},
        pages = {5256-5275},
          doi = {10.1093/mnras/stz764},
archivePrefix = {arXiv},
       eprint = {1811.09799},
 primaryClass = {astro-ph.GA},
       adsurl = {https://ui.adsabs.harvard.edu/abs/2019MNRAS.485.5256Z}
}

@ARTICLE{2020ApJ...888...88L,
       author = {{Lin}, Zesen and {Kong}, Xu},
        title = "{A Variant Stellar-to-nebular Dust Attenuation Ratio on Subgalactic and Galactic Scales}",
      journal = {\apj},
         year = 2020,
        month = jan,
       volume = {888},
       number = {2},
          eid = {88},
        pages = {88},
          doi = {10.3847/1538-4357/ab5f0e},
archivePrefix = {arXiv},
       eprint = {1912.01851},
 primaryClass = {astro-ph.GA},
       adsurl = {https://ui.adsabs.harvard.edu/abs/2020ApJ...888...88L}
}

@ARTICLE{2020MNRAS.499.5749J,
       author = {{Ji}, Xihan and {Yan}, Renbin},
        title = "{Constraining photoionization models with a reprojected optical diagnostic diagram}",
      journal = {\mnras},
         year = 2020,
        month = dec,
       volume = {499},
       number = {4},
        pages = {5749-5764},
          doi = {10.1093/mnras/staa3259},
archivePrefix = {arXiv},
       eprint = {2007.09159},
 primaryClass = {astro-ph.GA},
       adsurl = {https://ui.adsabs.harvard.edu/abs/2020MNRAS.499.5749J}
}

@ARTICLE{2022ApJS..262....3N,
       author = {{Nakajima}, Kimihiko and {Ouchi}, Masami and {Xu}, Yi and {Rauch}, Michael and {Harikane}, Yuichi and {Nishigaki}, Moka and {Isobe}, Yuki and {Kusakabe}, Haruka and {Nagao}, Tohru and {Ono}, Yoshiaki and {Onodera}, Masato and {Sugahara}, Yuma and {Kim}, Ji Hoon and {Komiyama}, Yutaka and {Lee}, Chien-Hsiu and {Zahedy}, Fakhri S.},
        title = "{EMPRESS. V. Metallicity Diagnostics of Galaxies over 12 + log(O/H) ≃ 6.9-8.9 Established by a Local Galaxy Census: Preparing for JWST Spectroscopy}",
      journal = {\apjs},
         year = 2022,
        month = sep,
       volume = {262},
       number = {1},
          eid = {3},
        pages = {3},
          doi = {10.3847/1538-4365/ac7710},
archivePrefix = {arXiv},
       eprint = {2206.02824},
 primaryClass = {astro-ph.GA},
       adsurl = {https://ui.adsabs.harvard.edu/abs/2022ApJS..262....3N}
}

@ARTICLE{2020ApJ...903..146B,
       author = {{Bari{\v{s}}i{\'c}}, Ivana and {Pacifici}, Camilla and {van der Wel}, Arjen and {Straatman}, Caroline and {Bell}, Eric F. and {Bezanson}, Rachel and {Brammer}, Gabriel and {D'Eugenio}, Francesco and {Franx}, Marijn and {van Houdt}, Josha and {Maseda}, Michael V. and {Muzzin}, Adam and {Sobral}, David and {Wu}, Po-Feng},
        title = "{Dust Attenuation Curves at z {\ensuremath{\sim}} 0.8 from LEGA-C: Precise Constraints on the Slope and 2175{\v{S}}Bump Strength}",
      journal = {\apj},
         year = 2020,
        month = nov,
       volume = {903},
       number = {2},
          eid = {146},
        pages = {146},
          doi = {10.3847/1538-4357/abba37},
archivePrefix = {arXiv},
       eprint = {2010.01147},
 primaryClass = {astro-ph.GA},
       adsurl = {https://ui.adsabs.harvard.edu/abs/2020ApJ...903..146B}
}

@ARTICLE{2021ApJ...909..213K,
       author = {{Kashino}, Daichi and {Lilly}, Simon J. and {Silverman}, John D. and {Renzini}, Alvio and {Daddi}, Emanuele and {Bardelli}, Sandro and {Cucciati}, Olga and {Kartaltepe}, Jeyhan S. and {Mainieri}, Vincenzo and {Pell{\'o}}, Roser and {Peng}, Ying-jie and {Sanders}, David B. and {Zucca}, Elena},
        title = "{The 2175 {\r{A}} Dust Feature in Star-forming Galaxies at 1.3 {\ensuremath{\leq}} z {\ensuremath{\leq}} 1.8: The Dependence on Stellar Mass and Specific Star Formation Rate}",
      journal = {\apj},
         year = 2021,
        month = mar,
       volume = {909},
       number = {2},
          eid = {213},
        pages = {213},
          doi = {10.3847/1538-4357/abdf62},
archivePrefix = {arXiv},
       eprint = {2009.13582},
 primaryClass = {astro-ph.GA},
       adsurl = {https://ui.adsabs.harvard.edu/abs/2021ApJ...909..213K}
}

@ARTICLE{2019MNRAS.486..743D,
       author = {{Decleir}, Marjorie and {De Looze}, Ilse and {Boquien}, M{\'e}d{\'e}ric and {Baes}, Maarten and {Verstocken}, Sam and {Calzetti}, Daniela and {Ciesla}, Laure and {Fritz}, Jacopo and {Kennicutt}, Rob and {Nersesian}, Angelos and {Page}, Mathew},
        title = "{Revealing the dust attenuation properties on resolved scales in NGC 628 with SWIFT UVOT data}",
      journal = {\mnras},
         year = 2019,
        month = jun,
       volume = {486},
       number = {1},
        pages = {743-767},
          doi = {10.1093/mnras/stz805},
archivePrefix = {arXiv},
       eprint = {1903.06715},
 primaryClass = {astro-ph.GA},
       adsurl = {https://ui.adsabs.harvard.edu/abs/2019MNRAS.486..743D}
}

@ARTICLE{2021ApJ...917...72L,
       author = {{Li}, Niu and {Li}, Cheng and {Mo}, Houjun and {Zhou}, Shuang and {Liang}, Fu-heng and {Boquien}, M{\'e}d{\'e}ric and {Drory}, Niv and {Fern{\'a}ndez-Trincado}, Jos{\'e} G. and {Greener}, Michael and {Riffel}, Rog{\'e}rio},
        title = "{Estimating Dust Attenuation From Galactic Spectra. II. Stellar and Gas Attenuation in Star-forming and Diffuse Ionized Gas Regions in MaNGA}",
      journal = {\apj},
         year = 2021,
        month = aug,
       volume = {917},
       number = {2},
          eid = {72},
        pages = {72},
          doi = {10.3847/1538-4357/ac0973},
archivePrefix = {arXiv},
       eprint = {2103.00666},
 primaryClass = {astro-ph.GA},
       adsurl = {https://ui.adsabs.harvard.edu/abs/2021ApJ...917...72L}
}

@ARTICLE{2011A&A...533A.117F,
       author = {{Fischera}, J. and {Dopita}, M.},
        title = "{On the missing 2175 {\r{A}}-bump in the Calzetti extinction curve}",
      journal = {\aap},
         year = 2011,
        month = sep,
       volume = {533},
          eid = {A117},
        pages = {A117},
          doi = {10.1051/0004-6361/201116644},
archivePrefix = {arXiv},
       eprint = {1108.1971},
 primaryClass = {astro-ph.GA},
       adsurl = {https://ui.adsabs.harvard.edu/abs/2011A&A...533A.117F}
}

@ARTICLE{2020MNRAS.492.3779H,
       author = {{Hirashita}, Hiroyuki and {Murga}, Maria S.},
        title = "{Self-consistent modelling of aromatic dust species and extinction curves in galaxy evolution}",
      journal = {\mnras},
         year = 2020,
        month = mar,
       volume = {492},
       number = {3},
        pages = {3779-3793},
          doi = {10.1093/mnras/stz3640},
archivePrefix = {arXiv},
       eprint = {2001.01968},
 primaryClass = {astro-ph.GA},
       adsurl = {https://ui.adsabs.harvard.edu/abs/2020MNRAS.492.3779H}
}

@ARTICLE{2022MNRAS.514.1886S,
       author = {{Shivaei}, Irene and {Boogaard}, Leindert and {D{\'\i}az-Santos}, Tanio and {Battisti}, Andrew and {da Cunha}, Elisabete and {Brinchmann}, Jarle and {Maseda}, Michael and {Matthee}, Jorryt and {Monreal-Ibero}, Ana and {Nanayakkara}, Themiya and {Popping}, Gerg{\"o} and {Vidal-Garc{\'\i}a}, Alba and {Weilbacher}, Peter M.},
        title = "{The UV 2175{\r{A}} attenuation bump and its correlation with PAH emission at z   2}",
      journal = {\mnras},
         year = 2022,
        month = aug,
       volume = {514},
       number = {2},
        pages = {1886-1894},
          doi = {10.1093/mnras/stac1313},
archivePrefix = {arXiv},
       eprint = {2203.09153},
 primaryClass = {astro-ph.GA},
       adsurl = {https://ui.adsabs.harvard.edu/abs/2022MNRAS.514.1886S}
}

@ARTICLE{2023ApJ...944L..16E,
       author = {{Egorov}, Oleg V. and {Kreckel}, Kathryn and {Sandstrom}, Karin M. and {Leroy}, Adam K. and {Glover}, Simon C.~O. and {Groves}, Brent and {Kruijssen}, J.~M. Diederik and {Barnes}, Ashley. T. and {Belfiore}, Francesco and {Bigiel}, F. and {Blanc}, Guillermo A. and {Boquien}, M{\'e}d{\'e}ric and {Cao}, Yixian and {Chastenet}, J{\'e}r{\'e}my and {Chevance}, M{\'e}lanie and {Congiu}, Enrico and {Dale}, Daniel A. and {Emsellem}, Eric and {Grasha}, Kathryn and {Klessen}, Ralf S. and {Larson}, Kirsten L. and {Liu}, Daizhong and {Murphy}, Eric J. and {Pan}, Hsi-An and {Pessa}, Ismael and {Pety}, J{\'e}r{\^o}me and {Rosolowsky}, Erik and {Scheuermann}, Fabian and {Schinnerer}, Eva and {Sutter}, Jessica and {Thilker}, David A. and {Watkins}, Elizabeth J. and {Williams}, Thomas G.},
        title = "{PHANGS-JWST First Results: Destruction of the PAH Molecules in H II Regions Probed by JWST and MUSE}",
      journal = {\apjl},
         year = 2023,
        month = feb,
       volume = {944},
       number = {2},
          eid = {L16},
        pages = {L16},
          doi = {10.3847/2041-8213/acac92},
archivePrefix = {arXiv},
       eprint = {2212.09159},
 primaryClass = {astro-ph.GA},
       adsurl = {https://ui.adsabs.harvard.edu/abs/2023ApJ...944L..16E}
}

@ARTICLE{1999ApJ...527...54B,
       author = {{Balogh}, Michael L. and {Morris}, Simon L. and {Yee}, H.~K.~C. and {Carlberg}, R.~G. and {Ellingson}, Erica},
        title = "{Differential Galaxy Evolution in Cluster and Field Galaxies at z\raisebox{-0.5ex}\textasciitilde0.3}",
      journal = {\apj},
         year = 1999,
        month = dec,
       volume = {527},
       number = {1},
        pages = {54-79},
          doi = {10.1086/308056},
archivePrefix = {arXiv},
       eprint = {astro-ph/9906470},
 primaryClass = {astro-ph},
       adsurl = {https://ui.adsabs.harvard.edu/abs/1999ApJ...527...54B}
}

@ARTICLE{2015AJ....150...19L,
       author = {{Law}, David R. and {Yan}, Renbin and {Bershady}, Matthew A. and {Bundy}, Kevin and {Cherinka}, Brian and {Drory}, Niv and {MacDonald}, Nicholas and {S{\'a}nchez-Gallego}, Jos{\'e} R. and {Wake}, David A. and {Weijmans}, Anne-Marie and {Blanton}, Michael R. and {Klaene}, Mark A. and {Moran}, Sean M. and {Sanchez}, Sebastian F. and {Zhang}, Kai},
        title = "{Observing Strategy for the SDSS-IV/MaNGA IFU Galaxy Survey}",
      journal = {\aj},
         year = 2015,
        month = jul,
       volume = {150},
       number = {1},
          eid = {19},
        pages = {19},
          doi = {10.1088/0004-6256/150/1/19},
archivePrefix = {arXiv},
       eprint = {1505.04285},
 primaryClass = {astro-ph.IM},
       adsurl = {https://ui.adsabs.harvard.edu/abs/2015AJ....150...19L}
}

@ARTICLE{2017AJ....154...86W,
       author = {{Wake}, David A. and {Bundy}, Kevin and {Diamond-Stanic}, Aleksandar M. and {Yan}, Renbin and {Blanton}, Michael R. and {Bershady}, Matthew A. and {S{\'a}nchez-Gallego}, Jos{\'e} R. and {Drory}, Niv and {Jones}, Amy and {Kauffmann}, Guinevere and {Law}, David R. and {Li}, Cheng and {MacDonald}, Nicholas and {Masters}, Karen and {Thomas}, Daniel and {Tinker}, Jeremy and {Weijmans}, Anne-Marie and {Brownstein}, Joel R.},
        title = "{The SDSS-IV MaNGA Sample: Design, Optimization, and Usage Considerations}",
      journal = {\aj},
         year = 2017,
        month = sep,
       volume = {154},
       number = {3},
          eid = {86},
        pages = {86},
          doi = {10.3847/1538-3881/aa7ecc},
archivePrefix = {arXiv},
       eprint = {1707.02989},
 primaryClass = {astro-ph.GA},
       adsurl = {https://ui.adsabs.harvard.edu/abs/2017AJ....154...86W}
}

@ARTICLE{1997ApJ...487..625G,
       author = {{Gordon}, Karl D. and {Calzetti}, Daniela and {Witt}, Adolf N.},
        title = "{Dust in Starburst Galaxies}",
      journal = {\apj},
         year = 1997,
        month = oct,
       volume = {487},
       number = {2},
        pages = {625-635},
          doi = {10.1086/304654},
archivePrefix = {arXiv},
       eprint = {astro-ph/9705043},
 primaryClass = {astro-ph},
       adsurl = {https://ui.adsabs.harvard.edu/abs/1997ApJ...487..625G}
}

@ARTICLE{2024ApJ...970...51G,
       author = {{Gordon}, Karl D. and {Fitzpatrick}, E.~L. and {Massa}, Derck and {Bohlin}, Ralph and {Chastenet}, J{\'e}r{\'e}my and {Murray}, Claire E. and {Clayton}, Geoffrey C. and {Lennon}, Daniel J. and {Misselt}, Karl A. and {Sandstrom}, Karin},
        title = "{Expanded Sample of Small Magellanic Cloud Ultraviolet Dust Extinction Curves: Correlations between the 2175 {\r{A}} Bump, $q_{\rm PAH}$, Ultraviolet Extinction Shape, and N(HI)/A(V)}",
      journal = {\apj},
         year = 2024,
        month = jul,
       volume = {970},
       number = {1},
          eid = {51},
        pages = {51},
          doi = {10.3847/1538-4357/ad4be1},
archivePrefix = {arXiv},
       eprint = {2405.12792},
 primaryClass = {astro-ph.GA},
       adsurl = {https://ui.adsabs.harvard.edu/abs/2024ApJ...970...51G}
}

@ARTICLE{2025ApJ...989...61C,
       author = {{Clayton}, Geoffrey C. and {Yanchulova Merica-Jones}, Petia and {Gordon}, Karl D. and {Decleir}, Marjorie and {Murray}, Claire E. and {Bohlin}, Ralph and {Bianchi}, Luciana and {Massey}, Philip and {Wolff}, Michael J.},
        title = "{Nature or Nurture: LMC-like Dust in the Solar Metallicity Galaxy M31}",
      journal = {\apj},
         year = 2025,
        month = aug,
       volume = {989},
       number = {1},
          eid = {61},
        pages = {61},
          doi = {10.3847/1538-4357/adef45},
archivePrefix = {arXiv},
       eprint = {2507.08612},
 primaryClass = {astro-ph.GA},
       adsurl = {https://ui.adsabs.harvard.edu/abs/2025ApJ...989...61C}
}

@ARTICLE{2026ApJ..1001..221G,
       author = {{Gordon}, Karl D. and {Yanchulova Merica-Jones}, Petia and {Clayton}, Geoffrey C. and {Bohlin}, Ralph and {Decleir}, Marjorie and {Murray}, Claire E. and {Bianchi}, Luciana},
        title = "{Large Variations Seen in First Ultraviolet Spectroscopic M33 Dust Extinction Curves}",
      journal = {\apj},
         year = 2026,
        month = apr,
       volume = {1001},
       number = {2},
          eid = {221},
        pages = {221},
          doi = {10.3847/1538-4357/ae5702},
archivePrefix = {arXiv},
       eprint = {2603.23662},
 primaryClass = {astro-ph.GA},
       adsurl = {https://ui.adsabs.harvard.edu/abs/2026ApJ..1001..221G}
}

\end{document}